\documentclass[letterpaper,titlepage,12pt]{article}
\usepackage[colorlinks=true,urlcolor=blue,citecolor=magenta]{hyperref}
\usepackage{amssymb,amsmath,amsfonts}
\usepackage{epsfig}
\usepackage{epsfig}
\usepackage{graphicx}
\usepackage{epstopdf}
\usepackage{caption}
\usepackage{subcaption}
\usepackage{amscd}
\usepackage{amsthm}
\usepackage{latexsym}
\usepackage{amsbsy}
\usepackage{bbm}
\usepackage[english]{babel}
\usepackage{psfrag}
\usepackage{tabularx}

\allowdisplaybreaks

\def\clock{{\count0=\time
           \divide\count0 60
           \ifnum\count0<10 0\fi\the\count0
           \multiply\count0 -60 \advance\count0 \time
           :\ifnum\count0<10 0\fi \the\count0
         }}
\newcommand{\timestamp}{{\small\vbox{\hbox{\tt\jobname.tex}
\hbox{\the\day/\the\month/\the\year, \clock}}}}

\numberwithin{equation}{section}

\begin{document}

\begin{titlepage}

\rightline{\vbox{\hfill  CCTP-2015-05
}}
\rightline{\vbox{\hfill  CCQCN-2015-64}}

 \vskip 1.4 cm

\centerline{\LARGE \bf Field Theory on Newton--Cartan Backgrounds}
\vskip .5cm

\centerline{\LARGE \bf and Symmetries of the Lifshitz Vacuum}

\vskip 1.5cm

\centerline{\large {{\bf Jelle Hartong$^1$, Elias Kiritsis$^{2,3}$, Niels A. Obers$^4$}}}

\vskip .8cm

\begin{center}

\sl $^1$ Physique Th\'eorique et Math\'ematique and International Solvay Institutes,\\
Universit\'e Libre de Bruxelles, C.P. 231, 1050 Brussels, Belgium.\\
\sl $^2$ Crete Center for Theoretical Physics, Department of Physics, University of Crete,\\
 71003 Heraklion, Greece.\\
\sl $^3$ APC, Universit\'e Paris 7,
CNRS/IN2P3, CEA/IRFU, Obs. de Paris,
Sorbonne Paris\\ Cit\'e, B\^atiment Condorcet, F-75205, Paris Cedex 13, France,
(UMR du CNRS 7164).\\
\sl $^4$ The Niels Bohr Institute, Copenhagen University,\\
\sl  Blegdamsvej 17, DK-2100 Copenhagen \O , Denmark.
\vskip 0.4cm

\end{center}
\vskip 0.6cm


\vskip .8cm \centerline{\bf Abstract} \vskip 0.2cm \noindent

Holography for Lifshitz space-times corresponds to dual field theories on a fixed torsional Newton--Cartan (TNC) background. We examine the coupling of non-relativistic field theories to TNC backgrounds and uncover a novel mechanism  by which a global $U(1)$ can become local.  This involves the TNC vector $M_\mu$ which sources a particle number current, and which for flat NC space-time satisfies $M_{\mu}=\partial_{\mu}M$ with a Schr\"odinger symmetry realized on $M$. We discuss various toy model field theories on flat NC space-time for which the new mechanism leads to extra global space-time symmetries beyond the generic Lifshitz symmetry, allowing for an enhancement to Schr\"odinger symmetry. On the holographic side, the source $M$ also appears in the Lifshitz vacuum with exactly the same properties as for flat NC space-time. In particular, the bulk diffeomorphisms that preserve the boundary conditions realize a Schr\"odinger algebra on $M$, allowing for a conserved particle number current. Finally, we present a probe action for a complex scalar field on the Lifshitz vacuum, which exhibits Schr\"odinger invariance in the same manner as seen in the field theory models.

\end{titlepage}


\tableofcontents


\section{Introduction}

Extending holography to settings that go beyond
the original AdS-setup has received considerable attention in recent years.
This has been motivated in part by
applying holographic ideas to the study of strongly coupled condensed matter
systems, which often exhibit non-relativistic scaling, and thus necessitate the
consideration of bulk space-times with asymptotics different from AdS \cite{Son:2008ye,Balasubramanian:2008dm,Kachru:2008yh,Taylor:2008tg}.
These include in particular
 Schr\"odinger, Lifshitz and hyperscaling violating geometries, which have in common that they exhibit
  a dynamical exponent $z$ characterizing the anisotropic scaling ratio between time and space on the boundary. 
 
 Besides the interest in such space-times in view of their application to non-relativistic field theories,
 examining to what extent holography is applicable in spaces with
different asymptotics is also of intrinsic importance. It may provide hints towards flat-space holography and, more generally, 
shed light on the nature of quantum gravity and elucidate puzzles in black hole physics.  Moreover, generalizing holographic techniques to non-AdS settings has the potential to reveal novel geometric structures on the boundary, which are interesting in their own right and at the same time present new perspectives on field theories
when coupling to these structures.

There is thus an extra, perhaps unusual, but rather important motivation for studying exotic theories for gravity, including those that we consider in this paper.
This stems from the fact that  such theories can be viewed as the Schwinger source functionals of non-relativistic quantum field  theories
(e.g. those used in condensed matter systems). The sources in question are the various components of
the metric and the relevant operator is the (non-relativistic) stress tensor. Once the symmetries of the quantum field theory above are specified, the symmetries of the relevant gravitational theory follow, and constraint the form of such source functionals. The usefulness of this procedure, beyond the realm of holography (which is a concrete realization of this idea), has been recently emphasized also in \cite{Son:2013rqa}.
This paper is a direct implementation of these ideas in a specific class of examples characterized by Lifshitz scaling symmetry and extended Schr\"odinger symmetry.

In particular, it was recently found that the boundary geometry for Lifshitz space-times is described by a new extension of Newton-Cartan (NC) geometry%
\footnote{We refer to \cite{Dautcourt,Eisenhart,Trautman,Kuenzle:1972zw,Duval:1984cj,Duval:1990hj,Julia:1994bs} for earlier work on Newton--Cartan geometry.}
with a specific torsion tensor, called torsional Newton-Cartan (TNC) geometry. This was first observed \cite{Christensen:2013lma,Christensen:2013rfa} for a specific action supporting $z=2$ Lifshitz geometries, and generalized to a large class of Lifshitz models for arbitrary value of $z$
 \cite{Hartong:2014oma,Hartong:2014}.
These works identified the Lifshitz UV completion and resulting boundary geometry by solving for the most general solution near the Lifshitz boundary using a vielbein formalism along with well-chosen linear combinations of the timelike vielbein and bulk gauge field.  By considering the coupling of this geometry to the boundary field
theory  the  vevs dual to the sources were computed, and moreover their Ward identities were written down in a TNC
covariant form.  In parallel, in \cite{Bergshoeff:2014uea} it was shown in detail how TNC geometry arises by gauging the Schr\"odinger algebra.
The coupling of non-relativistic field theories to TNC geometry  was also considered in 
\cite{Jensen:2014aia} from non-holographic perspective. 

The work of  \cite{Bergshoeff:2014uea} was used in the holographic context  to show  \cite{Hartong:2014oma,Hartong:2014}  that for Lifshitz space-times  there is an 
 underlying Schr\"odinger symmetry that acts on the sources and vevs, strongly suggesting that the boundary theory has a Schr\"odinger invariance. 
This observation was supported in the Letter \cite{Hartong:2014pma} by a complimentary analysis of  bulk versus boundary Killing symmetries
(employing the  TNC analogue of a conformal Killing vector \cite{Christensen:2013rfa}), by considering the  conditions for the boundary theory to admit conserved currents. Crucially, it was shown  that for field theories on a TNC background the interplay between conserved currents and space-time isometries is markedly different from the relativistic case.
The purpose of the present paper is to provide an in-depth analysis and discussion of this new mechanism, which, in its most general sense,
shows that Lifshitz holography describes a dual version of field theories on TNC backgrounds. 

Our results are of relevance to understanding the holographic dictionary in case of tractable examples of non-AdS space-times,
first and foremost for Lifshitz space-times
\cite{Ross:2009ar,Ross:2011gu,Baggio:2011cp,Mann:2011hg,Griffin:2011xs,Korovin:2013bua,Christensen:2013lma,Christensen:2013rfa,Chemissany:2014xpa,Chemissany:2014xsa,Hartong:2014oma,Hartong:2014pma,Hartong:2014},  but possibly also for other cases,
e.g.   Schr\"odinger and warped AdS$_3$ space-times \cite{Guica:2010sw,Hartong:2010ec,Guica:2011ia,Hartong:2013cba,Compere:2014bia,Andrade:2014iia,Andrade:2014kba}.
While this is interesting in its own right, there are also concrete direct applications to condensed matter type systems.
In particular, there is a growing body of recent work on using TNC geometry  in relation to field theory analyses of problems with 
strongly correlated  electrons, such as the quantum Hall effect (see e.g. \cite{Gromov:2014vla,Geracie:2014nka,Brauner:2014jaa,Geracie:2014zha,Wu:2014osa,Geracie:2014mta} following the earlier work \cite{Son:2013rqa} that introduced NC geometry to this problem).

\subsection{Outline and summary}

An outline and summary of the present paper is as follows. Our presentation below alternates between short summaries of the sections
and putting the results in context along with presenting the main conclusions. 

One of our key points is that in order to understand holography for Lifshitz space-times one must understand field theories on torsional Newton--Cartan (TNC) geometries. This is one of the reasons we spend a large fraction of this paper, sections \ref{sec:FT} and \ref{sec:flatspace}, entirely on that subject. The evidence for this is by now rather substantial. We have the null reductions on the AdS boundary of \cite{Christensen:2013lma,Christensen:2013rfa}, the general structure of the sources for asymptotically Lifshitz space-times as discussed in \cite{Hartong:2014oma,Hartong:2014} and in section \ref{sec:holoLif} of this paper and finally we have the discussion of exact (vacuum) Lifshitz space-times given in \cite{Hartong:2014pma} and section \ref{sec:Lifspace} of this paper that in the appropriate coordinates reflects all properties of flat NC space-times from a bulk point of view.

{\bf Summary of section \ref{sec:holoLif}.}
We start in section  \ref{sec:holoLif} with a brief review of the definition of the sources for asymptotically Lifshitz space-times in the Einstein--Proca dilaton model. This includes a derivation of the action of local bulk transformations such as diffeomorphisms, etc.  on the sources. The resulting local transformations of the sources are given in \eqref{eq:trafosources} which is in agreement with the way background fields transform in TNC geometry \cite{Bergshoeff:2014uea}. In \cite{Bergshoeff:2014uea} it shown that the transformations \eqref{eq:trafosources} can be written such that they make a local Schr\"odinger algebra acting on the sources manifest. In order to do this one must choose certain Schr\"odinger covariant curvature constraints that make local time and space translations equivalent to diffeomorphisms. The resulting TNC geometry on the boundary is discussed in subsection \ref{subsec:TNCgeometry} and readers who are not interested in the holographic origin of this geometry may immediately jump to this subsection.

{\bf TNC geometry.}
TNC geometry was found for the first time in \cite{Christensen:2013lma,Christensen:2013rfa} and a geometrical foundation for it has been given in \cite{Hartong:2014oma,Hartong:2014pma,Bergshoeff:2014uea} which appeared simultaneously with \cite{Jensen:2014aia}%
\footnote{See also the recent work \cite{Jensen:2014wha}, where the relation with relativistic field theories was revisited using non-relativistic limits.}  (how \cite{Jensen:2014aia}
fits into our framework will be commented on below).  It is well-known that the geometrical framework on which general relativity is based can be obtained by gauging the Poincar\'e algebra and imposing so-called curvature constraints to make local space-time translations equivalent to diffeomorphisms. In much the same way  it is shown in \cite{Bergshoeff:2014uea} that TNC geometry can be seen as arising from gauging the Schr\"odinger algebra and imposing suitable curvature constraints,  following the earlier work \cite{Andringa:2010it} that showed how to get NC geometry from gauging the
Bargmann algebra. The resulting geometrical framework provides us with various connections such as the affine connection
 which carries torsion and is TNC metric compatible (see eq.~\eqref{eq:GammaTNC}), but also for example the spin connections for local rotations 
 and Galilean boosts (see section \ref{sec:spincon}) and finally a dilatation connection (see section \ref{subsec:scaleandCKVs}). As an aid to our discussion below, we remark here that
 the relevant geometric structures in TNC are a time-like vielbein $\tau_\mu$, space-like vielbeins $e^a_\mu$ and a vector field $M_\mu$, that
 will play an important role. The fields transform under local tangent space transformations, namely local spatial rotations and Galilean boosts (Milne transformations in \cite{Jensen:2014aia}) and diffeomorphisms and local scale transformations. Crudely speaking we need the vector field $M_\mu$ because mass and energy are not equivalent and $M_\mu$ can be thought as the source for the mass current while $\tau_\mu$ sources the energy current. The precise definition of the energy-momentum tensor which contains the energy and momentum currents and the definition of the mass current will be given in section \ref{sec:FT}.

{\bf TNC geometry and its coupling to non-relativistic field theories.}
The natural framework to consider the covariant coupling of non-relativistic field theories to a space-time background is TNC geometry. 
Thus, armed with these geometrical tools we can write down actions for  field theories that are coupled to a TNC background
(section \ref{sec:FT}) and in particular study their global space-time symmetry properties on a flat NC background (section  \ref{sec:flatspace}). 
 The coupling should be done with respect to the so-called geometric invariants that are invariant under the local tangent space transformations\footnote{These are called Milne boost invariants in \cite{Jensen:2014aia}.} as discussed in \cite{Hartong:2014pma,Jensen:2014aia} and further elaborated on in section \ref{sec:FT}. These are certain combinations of $\tau_\mu$, $e^a_\mu$ and $M_\mu$ that are invariant under the local tangent space transformations. Essentially all of $M_\mu$ disappears into these geometric invariants with the exception of one scalar combination that we denote by $\tilde\Phi$, which is closely related to the Newtonian potential.
 
 {\bf Summary of section \ref{sec:FT}.}
In section \ref{sec:FT} we first  discuss the definition of the energy-momentum tensor $T^\mu{}_\nu$ and mass current $T^\mu$ that result from our coupling prescriptions and derive various Ward identities such as local scale and diffeomorphism Ward identities. There are also Ward identities for the local tangent space transformations, i.e. the local spatial rotations and Galilean boosts. These reduce the number of independent components of $T^\mu{}_\nu$ and $T^\mu$, e.g. the only independent component in $T^\mu$ is the mass density $\tau_\mu T^\mu$ that couples to $\tilde\Phi$ whereas $T^\mu{}_\nu$ contains the energy and momentum currents as well as the symmetric spatial stress tensor.

The diffeomorphism Ward identity  will also enables to define the notion of TNC Killing vectors $K^\nu$ by demanding that $K^\nu T^\mu{}_\nu$ is a conserved current. For scale invariant theories this leads to the notion of a TNC conformal Killing vector. In order to gain some intuition about field theories on TNC geometries, in particular with regards to global space-time symmetries, we introduce a number of field theory toy models. The lessons learned from these toy models  will be insightful when discussing global space-time symmetries in the holography setting. In particular, we introduce  the $z=2$ Schr\"odinger model (see section \ref{sec:Schmodel}) and a deformation of it (see section \ref{sec:defSchmodel}). Then we will show that these models realize some global space-time symmetries in a manner that has no relativistic counterpart and that crucially depends on the coupling to the background field $M_\mu$. We show that $M_\mu$ can become a gauge connection making a global $U(1)$ invariance into a local symmetry, and we discuss how this is done in the deformed Schr\"odinger model (see section \ref{sec:locU}) and how this allows for global space-time symmetries. The important role of $M_\mu$ is further commented on in section \ref{sec:comM} (see also below).  We also show in section \ref{subsec:notildePhicoupling} that, again by choosing the coupling to $M_\mu$ in a special way, namely such that we do not couple to the invariant $\tilde\Phi$, one can couple the $z=2$ Lifshitz scalar field model to TNC geometry, which is interesting to contrast to the Schr\"odinger model. We also comment there on how TNC structures can be used
to describe other situations, including the case considered in \cite{Hoyos:2013qna} as well as actions that only couple to a Lorentzian metric.

 {\bf Summary of section \ref{sec:flatspace}.}
Section \ref{sec:flatspace} specializes to the case of flat NC space-time.  We start by defining what we mean by flat NC space-time in section \ref{subsec:flatNC} and show in particular that this implies that the vector field $M_\mu $ is a total derivative of a function $M$. We will define the notion of a flat NC space-time in what are called global inertial coordinate systems. We subsequently compute the residual coordinate transformations that preserve the choice of global inertial coordinates up to local scale transformations in section \ref{sec:residualtrafos}. The analogous calculation for a flat Minkowski space-time would give us the conformal group. Here we show that the resulting set of transformations forms a realization of the Schr\"odinger algebra on $M$. The flat NC space-time conformal Killing vectors are computed in the later section \ref{subsec:flatNCKillingvectors} and shown to agree with those residual transformations that leave $M$ invariant. We show that there are three different functions $M$ for which the conformal Killing vectors span the Lifshitz algebra and that there does not exist an $M$ for which they generate the Schr\"odinger algebra. These three families of $M$ are related by the action of the Schr\"odinger group on $M$.

{\bf Global symmetries in non-relativistic field theories.}
We study scale invariant field theories on a flat NC space-time and the role played by $M$ in section \ref{subsec:FTonflatNC}. The two toy models that we consider are: i). the deformed Schr\"odinger model and ii). the Lifshitz model. Both these models are scale invariant but due to the way $M$ appears in these models they have various degrees of additional global space-time symmetries. The deformed Schr\"odiger model comes with two parameters $a$ and $b$ and we show that on a flat NC space-time with $a=b=0$ the model has full $z=2$ Schr\"odinger symmetry which for $a\neq 0$ and $b=0$ gets broken to Lifshitz plus Galilean boosts and when $b\neq 0$ it gets broken to Lifshitz. The real scalar Lifshitz model on the other hand is just Lifshitz invariant and differs from the deformed Schr\"odinger model in that it is higher order in derivatives (2nd order time derivatives as opposed to 1st order ones). Another important difference between the Lifshitz model and the deformed Schr\"odinger model with $b\neq 0$ is that the former has no notion of particle number, i.e. $\tau_\mu T^\mu=0$, whereas the latter has a particle number current $T^\mu$ whose conservation is explicitly broken by the $b$ term.

{\bf Elimination of $M$.} The different amounts of global space-time symmetries thus ranges for scale invariant from Lifshitz to Schr\"odinger and this is controlled by $M$. In section \ref{subsec:orbitsM} we define the notion of the orbit of $M$. This is defined to be all $M$ related to $M=\text{cst}$ that upon some $M$-dependent field redefinition lead to the same action. These field redefinitions `eat up M' in that they remove $M$ from the action, so that it is no longer a background field. For example we will see that for the scalar Lifshitz model all $M$ lead to inequivalent actions while for the Schr\"odinger model any $M$ related to $M=\text{cst}$ by a Schr\"odinger transformation leads to the same action. In general, the size of the orbit of $M$ depends on the couplings to the background fields.

As remarked above, one cannot view all elements of the Schr\"odinger group as conformal Killing vectors of a flat NC space-time. The global space-time symmetries that are outside a Lifshitz subalgebra of the Schr\"odinger group become global symmetries only in situations where we have a non-trivial orbit of $M$. This is because there are space-time diffeomorphisms that act on $M_\mu$ as a gauge transformation, i.e. $\delta M_\mu=\partial_\mu\tilde\sigma$ which takes one from element of the orbit to another one and this transformation gets compensated by a local phase rotation of some complex scalar field, say. This is the basic mechanism by which field theories are Galilean boost and/or special conformal invariant. The special conformal symmetry requires also a local scale transformation of the scalar field.

{\bf Lifshitz vacuum as the holographic dual of flat NC space-time.}
After this long detour on field theory on TNC geometries we return to the subject of holography for Lifshitz geometries in section \ref{sec:Lifspace}. We first show in section \ref{subsec:examples} that the Lifshitz vacuum in a coordinate system such that the boundary geometry is a flat NC space-time also comes with a function $M$ which on the boundary corresponds to $M_\mu=\partial_\mu M$. The $M$ dependent Lifshitz metric is given in equation \eqref{eq:LifshitzallM}. This is not written in the same gauge in which we defined the boundary conditions (sources) in section \ref{subsec:bdryconditions}. We show that one can perform a coordinate transformation that does not affect the sources which brings \eqref{eq:LifshitzallM} into radial gauge. In deriving these coordinate transformations we use a coordinate independent definition of a Lifshitz space-time given in appendix \ref{app:coordinateindependentLif}. We then continue to show in section \ref{subsec:symmetriesLifspace} that the bulk  Penrose--Brown--Henneaux (PBH) diffeomorphisms are exactly the same as those of section \ref{sec:residualtrafos}, i.e. the bulk PBH transformations realize the Schr\"odinger algebra on $M$. Hence the Lifshitz vacuum is the holographic dual of flat NC space-time.

We therefore have the right structure for the dual field theory to show global Schr\"odinger invariance. Just like in the toy models of section \ref{subsec:FTonflatNC} this requires fields living on a Lifshitz space-time to have a local symmetry that can be used to remove $M$ from the equation of motion. We will demonstrate in section \ref{subsec:probes} that one can indeed construct such probes on a Lifshitz space-time. 

{\bf Particle number symmetry.}
The existence of a local Schr\"odinger symmetry by which $M$ gets shifted, as it does under the bulk PBH transformations, can correspond to a particle number symmetry of the dual field theory. This is shown in section \ref{subsec:particlenumbercurrent}. Put another way we show that the residual bulk diffeomorphisms that realize a Schr\"odinger algebra on $M$ can lead to a conserved particle number current that relates to $T^\mu$ by an improvement. This is quite an uncommon feature. The bulk Einstein--Proca-dilaton theory has no local gauge symmetry, still the dual field theory can have a conserved particle number current. This can happen because $M_\mu$ plays a double role: it is part of the geometry through its appearance in the geometric invariants but it also sources the particle number current. Hence it can happen that bulk PBH transformations act non-trivially on $M_\mu$ which in turn has implications for the properties of $T^\mu$. 

{\bf On the role of the St\"uckelberg scalar $\chi$.}
We stress that in our formulation of TNC geometry the massive vector $M_\mu$ does not by itself have any gauge transformations under particle number. This only happens when we choose our couplings to the TNC geometry appropriately. Formulating the construction this way is forced upon us by the holographic dual model we are using which contains a massive vector field so that there is no local $U(1)$ in the bulk. One can go to a formulation with an internal particle number transformation by making a St\"uckelberg decomposition of $M_\mu$ via $M_\mu=\tilde m_\mu-\partial_\chi$ where $\chi$ is a St\"uckelberg scalar and where $\tilde m_\mu$ can be related to the gauge connection $m_\mu$ of the particle number symmetry inside the local Schr\"odinger algebra under which the background TNC fields transform \cite{Bergshoeff:2014uea} (for $z=2$ we have $\tilde m_\mu=m_\mu$). In cases where the coupling to the TNC background fields is chosen such that there is an additional local symmetry acting on $M_\mu$ of the form $\delta M_\mu=\partial_\mu\alpha$ (combined with some local transformation in field space) we can fix the $\alpha$ gauge transformation to remove $\chi$ and  doing so our formalism becomes identical to that of  \cite{Jensen:2014aia}. However we would like to emphasize that, independent of the holographic setup, our way of describing TNC geometries allows for much more general field theories than discussed in \cite{Jensen:2014aia}. As discussed above it also allows us to study cases such as Lifshitz invariant theories. In fact the conformal Killing vectors of flat NC space-time span just the Lifshitz algebra and nothing more. 

The enhancement to Galilean boost invariance is a property of the model just like it is for the case of scale symmetries. Not every field theory on a Minkowski space-time is scale invariant. In much the same way we see that not every theory on flat NC space-time is Galilean boost invariant (scale invariance is likewise not guaranteed). If we restrict to the class of scale invariant theories, TNC geometries form the natural habitat of Lifshitz invariant field theories. The geometrical framework then must include $\chi$ because there is no notion of particle number and $\chi$ allows us to deal with that kind of situations (see sections \ref{sec:comM} and \ref{subsec:notildePhicoupling}). One should not conclude that when $\chi$ appears in the formalism that this implies absence of particle number symmetries as we can have either an extra local shift symmetry that allows us to remove $\chi$ or because we can perform an improvement of the current $T^\mu$ sourced by $M_\mu$ such that we get a conserved particle number current (see sections \ref{subsec:particlenumbercurrent} and \ref{subsec:comments}).

In summary, our main results and findings are as follows
\begin{itemize}
\item Non-relativistic field theories coupled to TNC geometry, depending on the couplings of the field theory, exhibit a new mechanism, tied to the TNC vector  $M_\mu$, by which a global $U(1)$ becomes local with gauge connection $M_\mu$. 
\item  We elucidate the role of the mass current $T^\mu$ that couples to $M_\mu$ and its relation to a conserved particle number current, in different field theory setups. 
\item We provide a characterization of flat NC space in global inertial coordinates that emphasizes the relevance of the free function $M$ (in $M_\mu = \partial_\mu M$).
\item We work out the residual transformations preserving our notion of flat NC space-time and show that these realize a Schr\"odinger algebra on $M$.
\item When coupling a field theory to a flat NC space-time there can be a non-trivial orbit of $M$, i.e. a set of $M$'s related to $M=\text{cst}$ by the action of the Schr\"odinger group, such that for each of these $M$ we can write down the same action. This involves an $M$-dependent field redefinition of the mater fields, i.e. the matter fields eat up $M$ so that it is no longer a source, i.e. background field.
\item When there is a non-trivial orbit the theory exhibits extra global space-time symmetries (Galilean boost and/or special conformal symmetries) beyond the generic Lifshitz symmetries, allowing for an enhancement to Schr\"odinger symmetries. 
\item In the holographic context, we find a general form of the (bulk) Lifshitz metric that exhibits the source $M$. The bulk PBH transformations realize a Schr\"odinger algebra on $M$. Those PBH transformations that leave $M$ invariant form a Lifshitz algebra. This is the same manner in which Schr\"odinger symmetries appear in field theories on a flat NC background. 
\item We construct scalar probes on a bulk Lifshitz background that are invariant under a global Sch\"odinger group, supporting the claim that also in the holographic setup the background field $M$ can be eaten up by the bulk fields. 
\end{itemize}

\section{Holography for Lifshitz space-times}\label{sec:holoLif}

We will be working in the bulk with a gravitational theory containing Einstein gravity and a massive vector field (and possibly a dilaton). In this section we will show that the geometry on the boundary of asymptotically locally Lifshitz space-time is given by Newton--Cartan geometry with torsion. This is essentially a summary of the results found in \cite{Hartong:2014oma} (see also \cite{Hartong:2014}). The main results of this section that will be needed in the other sections are the definitions of the sources (the boundary conditions) and their local transformations (that preserve the boundary conditions). These are given in section \ref{subsec:localtrafosources}. In section \ref{subsec:TNCgeometry} we will review the properties of Newton--Cartan geometry with torsion.

\subsection{The Einstein--Proca-dilaton model}

We will work with a bulk theory consisting of a metric $g_{MN}$, a massive vector field $B_M$ and a scalar $\Phi$ (Einstein--Proca-dilaton (EPD) theory) whose dynamics is governed by the following action%
\footnote{Capital roman indices $M=(r,\mu)$ denote four-dimensional bulk space-time, with boundary space-time indices $\mu$. The boundary tangent space indices will be $0,a$ with $a=1,2$.} 
\begin{equation}\label{eq:action}
 S=\int d^4x\sqrt{-g}\left(R-\frac{1}{4}Z(\Phi)F^2-\frac{1}{2}W(\Phi)B^2-\frac{1}{2}(\partial\Phi)^2-V(\Phi)\right)\,,
\end{equation}
where $F=dB$. The equations of motion are
\begin{eqnarray}
\hspace{-1cm}\frac{1}{\sqrt{-g}}\partial_M\left(\sqrt{-g}ZF^{MN}\right) & = & WB^N\,,\label{eq:vectoreomagain}\\
\hspace{-1cm}\square\Phi & = & \frac{1}{4}Z'F^2+\frac{1}{2}W'B^2+V'\,,\\
\hspace{-1cm}R_{MN} & = & \frac{1}{2}Vg_{MN}+\frac{1}{2}Z\left(F_{MP}F_N{}^P-\frac{1}{4}F^2g_{MN}\right)+\frac{1}{2}WB_M B_N\,.
\end{eqnarray}
The Lagrangian has a broken $U(1)$ gauge symmetry signaled by the mass term of $B_M$. The functions $Z(\Phi)$ and $W(\Phi)$ are positive but otherwise arbitrary functions of the scalar field $\Phi$ and
the potential $V(\Phi)$ is negative close to a Lifshitz solution.

This model admits Lifshitz solutions (with $z>1$)
\begin{equation}\label{eq:solution}
 ds^2  =  -\frac{1}{r^{2z}}dt^2+\frac{1}{r^2}\left(dr^2+dx^2+dy^2\right)\,,\qquad
 B =  A_0\frac{1}{r^{z}}dt\,,\qquad
 \Phi  =  \Phi_\star\,,
 \end{equation} 
where $\Phi_*$ is  constant, $ A_0^2 = 2(z-1)/(z Z_0)$ and 
\begin{equation} \label{eq:conditionV1}
 V_0  =  -(z^2+z+4)\,,\qquad \frac{W_0}{Z_0}  = 2z\,,\qquad V_1  =  (za+2b)(z-1)\,,
\end{equation}
with  $a=Z_1/Z_0$, $b=W_1/W_0$ and $Z_i,W_i,V_i$ the Taylor coefficients of the functions $Z, W, V $ around
$\Phi_*$, the value of which, together with $z$, is determined by the first two equations in \eqref{eq:conditionV1}. 
The third equation in \eqref{eq:conditionV1} is a constraint on the potential making Lifshitz a non-generic solution of \eqref{eq:action}.

\subsection{Boundary conditions}\label{subsec:bdryconditions}

Because of the anisotropy of the Lifshitz metric, which is a property that will be retained in the definition of asymptotically locally Lifshitz space-times, it is very convenient to define the boundary conditions using bulk vielbeins \cite{Ross:2011gu}. Further we define a holographic coordinate $r$ by demanding that the metric is asymptotically (conformally) radial\footnote{The need for this was observed in \cite{Christensen:2013rfa} and will be further elaborated on in \cite{Hartong:2014}. It plays no crucial role in this work. We just keep $R$ for generality.}. We can always write for the metric
\begin{equation}\label{eq:gaugemetric}
ds^2=\frac{dr^2}{Rr^2}-E^0E^0+\delta_{ab}E^aE^b\,,
\end{equation}
where $E^0_r=E^a_r=0$. We will think of $r$ as the holographic coordinate with the boundary at $r=0$. By asymptotically locally Lifshitz we will mean the following metric boundary conditions%
\footnote{We note that these boundary conditions differ from those in \cite{Ross:2011gu}, which employs a radial gauge ($R=1$)
and assumes that $\tau_\mu$ in \eqref{bc1} is hyper surface orthogonal. When requiring the latter condition in our setup, the boundary geometry is called twistless torsional Newton-Cartan (TTNC) \cite{Christensen:2013lma,Christensen:2013rfa}.}
\begin{eqnarray}
R & = & O(1)\,,\\
E^0_\mu & = & O(r^{-z})\,,\\
E^a_\mu & = & O(r^{-1})\,,
\end{eqnarray}
where $z>1$.

For the massive vector field $B=B_r dr+B_\mu dx^\mu$ we have
\begin{eqnarray}
B_\mu & = & O(r^{-z})\,,
\end{eqnarray}
where the leading order behavior of $B_r$ is determined by the metric and $B_\mu$ via the equation 
\begin{equation}
\partial_M\left(\sqrt{-g}WB^M\right)=\partial_r \left(\sqrt{-g}WB^r\right)+\partial_\mu \left(\sqrt{-g}WB^\mu\right)=0
\end{equation}
which follows from \eqref{eq:vectoreomagain}. Integrating over $r$ we see that $B_r$ is determined up to a term of the form $\tfrac{f(x)}{r^2RW\sqrt{-g}}$ where $f$ is an arbitrary function of the boundary coordinates, i.e.
\begin{equation}\label{eq:Br}
B_r=g_{rr}B^r=\frac{f(x)}{r^2RW\sqrt{-g}}+\frac{1}{r^2RW\sqrt{-g}}\int^r_0 dr'\partial_\mu \left(\sqrt{-g}WB^\mu\right)\,.
\end{equation}
The $f$ term contributes for the first time to the expansion of $rB_r$ at order $r^{z+2}$. The freedom of adding $f(x)$ does not affect the leading order behavior of $B_r$. The boundary condition for $B_\mu$ is not a choice but enforced by the equations of motion. It is necessary in order to support the leading order behavior of $E^0_\mu$. We will phrase this by saying that there exists a function $\alpha$ such that
\begin{equation}\label{eq:alpha}
B_\mu-\alpha E^0_\mu=o(r^{-z})\,,
\end{equation}
where $\alpha$ is $O(1)$ near $r=0$. By little $o(1)$ we mean anything that goes to zero as $r$ goes to zero.

The boundary condition for the dilaton will simply be the statement that
\begin{equation}\label{eq:bdryPhi}
\Phi \simeq r^\Delta \phi\,,
\end{equation}
where $\Delta\ge 0$. The symbol $\simeq$ refers to the leading order term in the near-boundary $r$-expansion. Here $\phi$ is the boundary value of the dilaton which is an arbitrary function of $x$.

Going back to the boundary conditions for the metric we will impose
\begin{equation}\label{bc1}
E^0_\mu \simeq r^{-z}  \alpha_{(0)}^{1/3}\tau_\mu\,,\qquad E^a_\mu \simeq r^{-1} \alpha_{(0)}^{-1/3} e^a_\mu\,,\qquad R \simeq R_{(0)}\,,
\end{equation}
where $\alpha_{(0)}$ is the leading term in the expansion of $\alpha$ which is defined in \eqref{eq:alpha}, an equation that will be made more precise later in equation \eqref{eq:defMmu}. As derived in \cite{Hartong:2014} for $1<z\le 2$, which is the range we will work with from now on, it turns out that the equations of motion fix the form of $R_{(0)}$ and $\alpha_{(0)}$ either by fixing them to be specific constants or as certain functions of the boundary field $\phi$ (this depends on $z$ and the functions $Z$, $W$ and $V$), so these are not independent sources\footnote{The way in which $\alpha_{(0)}$ appears in \eqref{bc1} is explained in \cite{Hartong:2014} and is not essential for what follows. They are nothing but convenient rescalings of the boundary vielbeins that enable us to write expressions (see the next subsection) for the transformations of the sources, that preserve the boundary conditions and are independent of $\alpha_{(0)}$.}. We will treat the functions $R$ and $\alpha$ as scalars depending on $\Phi$.

For the inverse vielbeins the boundary conditions read
\begin{equation}
E_0^\mu \simeq - r^z \alpha_{(0)}^{-1/3} v^\mu\,,\qquad E_a^\mu \simeq r \alpha_{(0)}^{1/3}e_a^\mu\,,
\end{equation}
where we have the orthogonality relations
\begin{equation}
\label{eq:ortho}
v^\mu\tau_\mu=-1\,,\qquad v^\mu e_\mu^a=0\,,\qquad e^\mu_a\tau_\mu=0\,,\qquad e^\mu_a e_\mu^b=\delta^b_a\,.
\end{equation}
The completeness relation is $e^\mu_a e^a_\nu = \delta^\mu_\nu + v^\mu \tau_\nu$.

The boundary conditions for the vielbeins \eqref{bc1} tell us that the light cones flatten out as we approach the boundary. The bulk vielbeins $E^0_\mu$ and $E^a_\mu$ transform under local Lorentz transformations. If we ask that these respect our boundary conditions we find that the boundary vielbeins transform as
\begin{eqnarray}
\delta\tau_\mu  & = &  0\,,\\
\delta e^a_\mu  & = &  \lambda^a \tau_\mu + \lambda^a{}_b e^b_\mu\,.\label{eq:trafoe_mu^a}
\end{eqnarray}
This has been shown in \cite{Christensen:2013rfa} for $z=2$ and is easily generalized to any value of $z$ (see also \cite{Hartong:2014}). These transformations will be referred to as local Galilean boosts ($\lambda^a$) and local rotations ($\lambda^a{}_b$). The boundary values of the inverse vielbeins transform as
\begin{eqnarray}
\delta v^\mu  & = & \lambda^a e^\mu_a \,,\\
\delta e^\mu_a  & = & \lambda_a{}^b e_b^\mu\,,
\end{eqnarray}
as follows from \eqref{eq:ortho}. All terms in the near boundary expansion of the metric when expressed in terms of the boundary vielbeins should be invariant under these transformations. If we look at the expansion of \eqref{eq:gaugemetric} at order $r^{-2}$ we see that we get 
\begin{equation}
\alpha_{(0)}^{-2/3}\delta_{ab}e^a_\mu e^b_\nu+\ldots
\end{equation}
where the dots denote contributions from the expansion of $E^0_\mu E^0_\nu$. The complete term at order $r^{-2}$ should be Galilean invariant. However the first term coming from the leading term of $\delta_{ab}E^a_\mu E^b_\nu$ is not invariant because
\begin{equation}
\delta \left(\delta_{ab}e^a_\mu e^b_\nu\right)=\lambda_a \tau_\mu e^a_\nu+\lambda_a \tau_\nu e^a_\mu\,,
\end{equation}
under \eqref{eq:trafoe_mu^a}. Hence it must be that there is a contribution coming from $E^0_\mu E^0_\nu$ at order $r^{-2}$ that compensates for this non-invariance. In other words it must be that
\begin{equation}
E^0_\mu=r^{-z}\alpha_{(0)}^{1/3}\tau_\mu+\ldots+r^{z-2}\alpha_{(0)}^{-1}X_\mu+\ldots
\end{equation}
so that the complete order $r^{-2}$ term in the metric reads
\begin{equation}\label{eq:orderr-2}
\alpha_{(0)}^{-2/3}\left(\delta_{ab}e^a_\mu e^b_\nu-\tau_\mu X_\nu-\tau_\nu X_\mu\right)\,,
\end{equation}
with $X_\mu$ transforming as
\begin{equation}
\delta X_\mu=e^a_\mu\lambda_a\,,
\end{equation}
under local Galilean boosts\footnote{We thank Matthias Blau for useful discussions on this point.}. 

What we are asking for is that for any configuration of sources we can write down a vector $X_\mu$ that makes the metric at order $r^{-2}$ Galilean boost invariant. The vector $X_\mu$ must involve a new source because we cannot create such a transformation out of the vielbein sources $\tau_\mu$ and $e^a_\mu$. Thus there must exist a boundary vector field $M_\mu$ such that $X_\mu=M_\mu+I_\mu$ where $I_\mu$ is invariant under local Galilean boosts and local rotations so that $\delta X_\mu=\delta M_\mu$. The invariant part of $X_\mu$ is therefore of no interest to us where it concerns this problem. All we will assume about $I_\mu$ is that it can be written as $I\tau_\mu$ where $I$ is a scalar invariant. We stress that this assumption is not essential as it will not affect the properties of $M_\mu$. Even though we say that the relation $I_\mu=I\tau_\mu$ is an assumption, we not have not managed to find a counterexample using vevs and derivatives of sources that make up a Galilean invariant object that has the right scaling dimension to appear at order $r^{2-z}$. Nevertheless we are not aware of a general proof that it should always be that $I_\mu=I\tau_\mu$. As mentioned already it does not affect the properties of $M_\mu$ which is we are after, it merely changes slightly the presentation of some equations. We will comment on this as we go on. We thus have
\begin{equation}\label{eq:expansionE0}
E^0_\mu=r^{-z}\alpha_{(0)}^{1/3}\tau_\mu+\ldots+r^{z-2}\alpha_{(0)}^{-1}\left(M_\mu+I\tau_\mu\right)+\ldots\,.
\end{equation}
Because the massive vector is Galilean boost invariant at each order in $r$ we can write
\begin{equation}\label{eq:expansionBmu}
B_\mu=r^{-z}\alpha_{(0)}^{4/3}\tau_\mu+\ldots+r^{z-2}\tilde I\tau_\mu+\ldots\,,
\end{equation}
where $\tilde I$ is also rotation and Galilean boost invariant. Here the same comment applies; we could have written $\tilde I_\mu$ but we take it to be $\tilde I\tau_\mu$. For a suitably chosen function $\alpha$ that has an expansion of the form
\begin{equation}
\alpha=\alpha_{(0)}+r^{2z-2}\alpha_{(0)}^{-1/3}\left(\tilde I-I\right)+\ldots\,,
\end{equation}
we can obtain\footnote{If we had not assumed $I_\mu=I\tau_\mu$ and similarly for $\tilde I_\mu$ we would have found
\begin{equation*}
B_\mu - \alpha (\Phi) E^0_\mu \simeq -r^{z-2} (M_\mu+\bar I_\mu) \,,
\end{equation*}
where $\bar I_\mu$ is yet another invariant. We can fix $\alpha$ in the same way by demanding the component of $\bar I_\mu$ along $\tau_\mu$ vanishes. Since $\bar I_\mu$ is an invariant this does not affect the properties of the source $M_\mu$.}
\begin{equation}\label{eq:defMmu}
B_\mu - \alpha (\Phi) E^0_\mu \simeq -r^{z-2} M_\mu \,.
\end{equation}
We think of this as the boundary condition that defines the source $M_\mu$. We note that this definition is intimately
related to what we mean with $\alpha$, as can be seen from \eqref{eq:expansionE0} and \eqref{eq:expansionBmu}.

Using \eqref{eq:defMmu} we find that
\begin{equation}
B_0=E^\mu_0B_\mu=O(1)\,,\qquad B_a=E^\mu_aB_\mu=O(r^{z-1})\,,
\end{equation}
so that 
\begin{equation}
B^\mu=E^\mu_0B^0+E^\mu_aB^a=O(r^z)\,.
\end{equation}
Using \eqref{eq:Br} it follows that
\begin{equation}\label{eq:bdryconditionBr}
B_r=O(r^{z-1})\,.
\end{equation}

One can make a St\"uckelberg decomposition of $B_M$, i.e.
\begin{equation}\label{eq:Stueckelbergdecomposition}
B_M=A_M-\partial_M\Xi\,,
\end{equation}
and we can do the same for $M_\mu$, i.e.
\begin{equation}\label{eq:Mandtildem}
M_\mu=\tilde m_\mu-\partial_\mu\chi\,.
\end{equation}
To this end we need to take for $\Xi$ the boundary condition
\begin{equation}\label{bc2} 
\Xi \simeq -r^{z-2} \chi \,.
\end{equation}
The boundary condition for $\Xi$ is a choice. We fixed the choice by demanding that $\chi$ has the right scaling dimension to combine with $M_\mu$ as in \eqref{eq:Mandtildem}. In general one can put anything for the boundary condition of $\Xi$  since it is just a St\"uckelberg scalar. Since using equation \eqref{eq:bdryconditionBr} we know that
\begin{equation}
B_r=A_r-\partial_r\Xi=O(r^{z-1})\,,
\end{equation}
we get
\begin{equation}\label{eq:Ar}
A_r \simeq -(z-2) r^{z-3} \chi \,.
\end{equation}
This condition for $A_r$ is a necessary condition in order that $B_r=O(r^{z-1})$. One might wonder what about subleading terms. The fact that $A_M$ and $\Xi$ always appear in the combination that gives $B_M$ via \eqref{eq:Stueckelbergdecomposition} guarantees that the subleading orders in $A_r$ and $\Xi$ will cancel such that  $B_r=O(r^{z-1})$. From \eqref{eq:defMmu}, \eqref{eq:Stueckelbergdecomposition} and \eqref{bc2} it follows that we have
\begin{equation}\label{eq:deftildem}
A_\mu - \alpha (\Phi) E^0_\mu \simeq -r^{z-2} \tilde m_\mu \,.
\end{equation}
 
We will formulate the boundary conditions in terms of the metric and the massive vector field, i.e. without reference to vielbeins and St\"uckelberg decompositions, at the end of section \ref{subsec:TNCgeometry}.

\subsection{Local transformations of the sources}\label{subsec:localtrafosources}

We already discussed how the sources transform under local tangent space transformations, i.e. the Galilean boosts and spatial rotations. These transformations are a consequence of us choosing to work with vielbeins. Towards the end of the previous section we introduced yet another local symmetry: the St\"uckelberg $U(1)$ which acts on $A_M$ and $\Xi$ as $\delta A_M=\partial_M\Lambda$ and $\delta\Xi=\Lambda$. The boundary conditions \eqref{bc2} and \eqref{eq:Ar} are preserved by the bulk St\"uckelberg gauge transformations for which $\Lambda\simeq r^{z-2}\sigma$. The sources $\tilde m_\mu$ and $\chi$ defined in \eqref{eq:deftildem} and \eqref{bc2}, respectively, then simply transform as $\delta \tilde m_\mu=\partial_\mu\sigma$ and $\delta \chi=\sigma$. 

However, by far the most relevant local symmetries are the bulk diffeomorphisms that preserve our conformally radial gauge choice made in \eqref{eq:gaugemetric}. These will play a prominent role in this work and we will refer to them as the Penrose--Brown--Henneaux (PBH) transformations \cite{Penrose:1986ca,Brown:1986nw}. They are defined as those transformations that preserve the form of the metric \eqref{eq:gaugemetric} and boundary conditions, i.e. they are such that $Rg_{MN}$ remains in radial gauge after acting on it with a diffeomorphism. From this condition we can conclude that the PBH transformations are generated by a bulk vector $\zeta^M$ which is of the form
\begin{eqnarray}
\zeta^r & = & -r\Lambda_D\,,\label{eq:PBH1}\\
\zeta^\mu & = & \xi^\mu+O(r^2)\,,\label{eq:PBH2}
\end{eqnarray}
where $\Lambda_D$ and $\xi^\mu$ are arbitrary functions of the boundary coordinates. We note that when $\Lambda_D$ is not constant we necessarily need to have a term of order $r^2$ and possibly higher order terms as well in the expansion of $\zeta^\mu$. For later purposes we highlight the fact that for any local rescaling $\Lambda_D$ of $r$ and any boundary diffeomorphism $\xi^\mu$ there exists corrections to $\zeta^\mu$ starting at order $r^2$ such that we maintain a radial gauge. We can think of the PBH transformations as consisting of two parts: 1). the transformations generated by $\zeta^r = -r\Lambda_D$ and $\zeta^\mu = \xi^\mu$ and 2). the transformations generated by $\zeta^r = 0$ and $\zeta^\mu=O(r^2)$. The first transformation takes us possibly out of radial gauge and acts non-trivially on the sources while the second one takes us back to radial gauge and does not act on the sources\footnote{For more background on the role of PBH transformations in AdS/CFT we refer the reader to \cite{Imbimbo:1999bj} (see also \cite{Skenderis:2000in}). In a situation where we have full control over the asymptotic expansion in the sense that the full asymptotic solution space is determined by the sources and the vevs, the knowledge of the PBH transformations together with the Fefferman--Graham expansion is sufficient to compute the asymptotic symmetry algebra.}. We act on all bulk fields such as $E^0_\mu$, $B_\mu-\alpha E^0_\mu$, $\chi$, etc with a bulk diffeomorphism. From this we can read off how the sources transform under $\Lambda_D$ and $\xi^\mu$.

Combining all local transformations we conclude that the sources transform as \cite{Hartong:2014oma}
\begin{equation}\label{eq:trafosources}
\begin{array}{rcl}
\delta \tau_\mu & = & \mathcal{L}_\xi\tau_\mu+z \Lambda_D \tau_\mu \,,  \\
\delta e_\mu^a & = & \mathcal{L}_\xi e^a_\mu+\lambda^a \tau_\mu + \lambda^{a}{}_b e_{\mu}^{b} + \Lambda_D e_\mu^a \,, \\
\delta M_\mu & = & \mathcal{L}_\xi M_\mu+ e_\mu^a\lambda_a+(2-z)\Lambda_D M_\mu\,,\\
\delta\chi & = & \mathcal{L}_\xi\chi+\sigma+(2-z)\Lambda_D\chi\,,\\
\delta v^\mu & = & \mathcal{L}_\xi v^\mu+\lambda^a e_a^\mu -z \Lambda_D v^\mu\,,\\
\delta e^\mu_a & = &  \mathcal{L}_\xi e^\mu_a+\lambda_a{}^b e^\mu_b - \Lambda_D e^\mu_a \,,\\
\delta M_a & = & \mathcal{L}_\xi M_a+\lambda_a{}^b M_b+\lambda_a+(1-z)\Lambda_D M_a\,,
\end{array}
\end{equation}
where $M_a=e^\mu_a M_\mu$. Here $\lambda^a$  correspond to Galilean boosts ($G$), $\lambda_a{}^b$ to spatial rotations ($J$), $\Lambda_D$ to dilatations ($D$) and $\sigma$ to St\"uckelberg gauge transformations ($N$). The fields $M_a$ and $\chi$ undergo shift transformations with respect to Galilean boosts and St\"uckelberg gauge transformations. The fields $M_a$ and $\chi$ play a special role in field theories on TNC backgrounds as we will see in subsection \ref{sec:FT}. 

We emphasize that the transformations \eqref{eq:trafosources} are not special to sources in Lifshitz holography. This is the way in which TNC background fields must transform as shown in \cite{Bergshoeff:2014uea}. In \cite{Bergshoeff:2014uea} it shown that the transformations \eqref{eq:trafosources} can be written such that they make a local Schr\"odinger algebra acting on the sources manifest. In order to do this one must choose certain Schr\"odinger covariant curvature constraints that make local time and space translations equivalent to diffeomorphisms.

\subsection{Torsional Newton--Cartan geometry}\label{subsec:TNCgeometry}

As explained in detail in \cite{Hartong:2014oma,Bergshoeff:2014uea,Hartong:2014} the boundary geometry is described by torsional Newton--Cartan geometry. Here we collect the basic elements of such a geometry that will be needed later when we study symmetries of the Lifshitz vacuum and its Newton--Cartan boundary geometry. We will divide the local symmetries \eqref{eq:trafosources} into two groups. The first contain diffeomorphisms and dilatations and the second what we might call the internal symmetries. The latter are $G$, $J$ and $N$. The local Galilean boosts are what are called Milne boosts in \cite{Jensen:2014aia}. If one wishes to draw an analogy with Lorentzian geometry then the local rotations play the role of the local Lorentz transformations, but there is no relativistic counterpart for the presence of the $G$ and $N$ local shift symmetries that act on the fields $M_a$ and $\chi$. We will further elaborate on this in the next subsection.

It will prove very convenient to define what we call geometric invariants by which we mean tensors that transform covariantly under the local transformations of the first group and that are invariant under the internal symmetries. The invariants one can build out of $\tau_\mu$, $e^a_\mu$ and $M_\mu$ are
\begin{equation}\label{eq:tildePhi}
\begin{array}{rcl}
\hat v^\mu & = & v^\mu - h^{\mu \nu} M_\nu \,, \\ 
\bar h_{\mu\nu} & = & h_{\mu\nu}-\tau_\mu M_\nu-\tau_\nu M_\mu\,,\\
\tilde\Phi & = & -v^\mu M_\mu+\frac{1}{2}h^{\mu\nu}M_\mu M_\nu\,,
\end{array}
\end{equation}
together with the degenerate metric invariants $\tau_\mu$ and $h^{\mu\nu} = e_a^\mu e_b^\nu \delta^{ab}$ and the determinant $e=\text{det}(\tau_\mu\,, e_\mu^a)$. We will also make use of the $G$ and $N$ invariant vielbein $\hat e_\mu^a$ defined as
\begin{equation}\label{eq:hate}
\hat e_\mu^a = e_\mu^a-\tau_\mu M^a\,.
\end{equation}
The objects $\hat e_\mu^a$, $\hat v^\mu$, $\tau_\mu$, $e^\mu_a$ form an orthonormal set. Useful relations are
\begin{equation}\label{eq:relations}
h^{\nu\rho}\bar h_{\rho\mu}=\delta^\nu_\mu+\hat v^\nu\tau_\mu\,,\quad \hat v^\mu\bar h_{\mu\nu}=2\tau_\nu\tilde\Phi\,,\quad\hat e^a_\mu\hat e_{\nu a}=\bar h_{\mu\nu}+2\tilde\Phi\tau_\mu\tau_\nu\,,\quad -\hat v^\nu\tau_\mu+\hat e^a_\mu e_a^\nu=\delta^\nu_\mu\,.
\end{equation}
In section \ref{subsec:flatNC} we will see that $\tilde\Phi$ is closely related to the Newton potential denoted by $\Phi$ when the space-time is flat (see also \cite{Bergshoeff:2014uea}). We use the same symbol for the Newton potential as for the bulk space-time dilaton. We hope that this does not cause any confusion.

There is a unique affine connection that is invariant under the internal symmetries $G$, $J$, $N$ that is metric compatible by which we mean
\begin{equation}\label{eq:TNC}
\nabla_\mu\tau_\nu = 0\,,\hskip 1cm\nabla_\mu h^{\nu\rho} = 0\,,
\end{equation}
and it is given by \cite{Hartong:2014oma,Bergshoeff:2014uea} (see also \cite{Jensen:2014aia,Bekaert:2014bwa})\footnote{To be precise, the uniqueness of this connection requires the additional assumption that it is linear in $M_\mu$ which is a natural property from the point of view of gauging the Schr\"odinger algebra \cite{Bergshoeff:2014uea}. If we drop this condition we can write down a one parameter family of $G$, $J$, $N$ invariant connections that are metric compatible in the sense of \eqref{eq:TNC} that are of the form
\begin{equation*}
\Gamma^{\rho}_{\mu\nu} = -\hat v^\rho\partial_\mu\tau_\nu+\frac{1}{2}h^{\rho\sigma}\left(\partial_\mu X_{\nu\sigma}+\partial_\nu X_{\mu\sigma}-\partial_\sigma X_{\mu\nu}\right)\,,
\end{equation*}
where $X_{\mu\nu}=\bar h_{\mu\nu}+\alpha\tilde\Phi\tau_\mu\tau_\nu$ where $\alpha$ is an arbitrary constant (see section \ref{subsec:notildePhicoupling}).\label{footnote:newconnection}} 
\begin{equation}\label{eq:GammaTNC}
\Gamma^{\rho}_{\mu\nu} = -\hat v^\rho\partial_\mu\tau_\nu+\frac{1}{2}h^{\rho\sigma}\left(\partial_\mu\bar h_{\nu\sigma}+\partial_\nu \bar h_{\mu\sigma}-\partial_\sigma\bar h_{\mu\nu}\right)\,.
\end{equation}
This connection has torsion since the first term is not symmetric in $\mu$ and $\nu$. This is why we call the geometry torsional Newton--Cartan (TNC) \cite{Christensen:2013lma,Christensen:2013rfa}.

\subsubsection{Spin connections for rotations and Galilean boosts \label{sec:spincon}}

Although we will not need them in this work we mention for completeness that one can define spin connections for local rotations and Galilean boosts. This is useful for example when coupling fields with spin to a TNC background.

We define the following covariant derivatives \cite{Bergshoeff:2014uea}
\begin{equation}
\begin{array}{rcl}
\mathcal{D}_\mu\tau_\nu & = &\partial_\mu\tau_\nu-\Gamma^\rho_{\mu\nu}\tau_\rho\,,  \\
\mathcal{D}_\mu e_\nu{}^a & = & \partial_\mu e_\nu{}^a-\Gamma^\rho_{\mu\nu}e_\rho{}^a-\Omega_\mu{}^a\tau_\nu-\Omega_\mu{}^{a}{}_be_{\nu}{}^b \,, \\
\mathcal{D}_\mu v^\nu & = & \partial_\mu v^\nu+\Gamma^\nu_{\mu\rho}v^\rho-\Omega_\mu{}^a e^\nu{}_a\,,  \\
\mathcal{D}_\mu e^\nu{}_a & = & \partial_\mu e^\nu{}_a+\Gamma^\nu_{\mu\rho}e^\rho{}_a+\Omega_\mu{}^{b}{}_ae^\nu{}_b \,, 
\end{array}
\end{equation}
and impose the following vielbein postulates
\begin{equation}
\begin{array}{rclccrcl}
\mathcal{D}_\mu\tau_\nu & = & 0 \,,  &&&\mathcal{D}_\mu e_\nu{}^a & = & 0 \,, \\
\mathcal{D}_\mu v^\nu & = & 0\,,&&&\mathcal{D}_\mu e^\nu{}_a & = & 0\,,
\end{array}
\end{equation}
and take $\Gamma^\rho_{\mu\nu}$ as in \eqref{eq:GammaTNC}. The connections $\Omega_\mu{}^a$ and $\Omega_\mu{}^{ab}$ can be solved for in terms of $\Gamma^\rho_{\mu\nu}$. It can be shown by either using the covariance of the $\mathcal{D}_\mu$ derivative or by solving the vielbein postulates in terms of the vielbeins that the rotation and Galilean boost connections transform as
\begin{eqnarray}
\delta \Omega_\mu{}^{ab} & =& \mathcal{L}_\xi\Omega_\mu{}^{ab}+\partial_\mu \lambda^{ab} + 2 \lambda^{c[a} \Omega_\mu{}^{b]}{}_c\,, \\
\delta \Omega_\mu{}^a & = &\mathcal{L}_\xi \Omega_\mu{}^a+ \partial_\mu \lambda^a + \lambda^a{}_b \Omega_\mu{}^b + \lambda^b \Omega_{\mu b}{}^a \,,
\end{eqnarray}
respectively.

\subsubsection{Bulk metric boundary conditions and TNC invariants}

Now that we have the invariants at our disposal we can formulate the boundary conditions of section \ref{subsec:bdryconditions} in a metric/massive vector field language as follows (use equations \eqref{eq:gaugemetric}, \eqref{eq:Br}, \eqref{bc1}, \eqref{eq:expansionE0}--\eqref{eq:defMmu})
\begin{eqnarray}
ds^2 & = & \frac{dr^2}{Rr^2}-\alpha_{(0)}^{2/3}r^{-2z}\tau_\mu\tau_\nu dx^\mu dx^\nu+\ldots\nonumber\\
&&+\alpha_{(0)}^{-2/3}r^{-2}\left(\bar h_{\mu\nu}+I\tau_\mu\tau_\nu\right)dx^\mu dx^\nu+\ldots\,,\label{eq:metricbdryconditions}\\
B & = & B_rdr+\alpha_{(0)}^{4/3}r^{-z}\tau_\mu dx^\mu+\ldots+r^{z-2}\tilde I\tau_\mu dx^\mu+\ldots\,.\label{eq:massivevectorbdryconditions}
\end{eqnarray}
In here $I$ and $\tilde I$ are invariants defined in \eqref{eq:expansionE0} and \eqref{eq:expansionBmu}, respectively, with dilatation weight\footnote{A field $X$ has dilatation weight $w$ if it transforms as $\delta X=-w\Lambda_D X$ under $\Lambda_D$ transformations.} $2(z-1)$. One such object is $\tilde\Phi$, but it may happen that $I$ and $\tilde I$ also involve certain scalar vevs associated with the presence of the dilaton (see for example appendix D of \cite{Christensen:2013rfa}). The dots on the first line of \eqref{eq:metricbdryconditions} originate from the product $-E^0_\mu E^0_\nu$. The first set of dots of \eqref{eq:massivevectorbdryconditions} allow for the possibility that terms involving derivatives of the sources may appear between the orders $r^{-z}$ and $r^{2-z}$. The structure of the terms on the dots, also those at the end of \eqref{eq:metricbdryconditions} and \eqref{eq:massivevectorbdryconditions}, are determined by the equations of motion. It would be interesting to compute these expansions for an exact Lifshitz background. In appendix \ref{app:coordinateindependentLif} we provide a coordinate independent definition of a Lifshitz space-time, so we could approach this problem by solving equations \eqref{eq:pureLif-I}--\eqref{eq:pureLif-VI} all of whose solutions are locally Lifshitz. We hope to report on such an analysis in the future.

Having introduced our model and setup for Lifshitz holography and the relation of the sources to TNC geometry, we first take
a step back in the coming two sections, where we will present a purely field-theory discussion of properties of non-relativistic
field theories coupled to a TNC background. We return to holography in section \ref{sec:Lifspace}, where we discuss the symmetries
of the Lifshitz vacuum and its implications for the symmetries of the dual field theory,  using the insights gained from
sections \ref{sec:FT} and \ref{sec:flatspace}.

\section{Scale invariant field theories on TNC backgrounds}\label{sec:FT}

In this and the next section we consider scale invariant field theories on TNC backgrounds with particular focus on their symmetries.
Our analysis is at the classical level, and hence we ignore possible quantum anomalies. 
We emphasize that the toy models that we construct and discuss are not expected to be directly related to the dual field theories
that arise in Lifshitz holography, but they will serve as analogue models to illustrate  the symmetry properties that we observe 
in the holographic context. 

We work in this section and onwards with an arbitrary number of spatial dimensions. We couple a field theory to a TNC geometry by writing an action whose background fields are the geometric invariants discussed in section \ref{subsec:TNCgeometry}, i.e. we write
\begin{equation}
S=S[\hat v^\mu, h^{\mu\nu}, \tilde\Phi]\,.
\end{equation}
When varying the background fields we can choose to vary either $\hat v^\mu$, $h^{\mu\nu}$ and $\tilde\Phi$ or the background fields $v^\mu$, $e_a^\mu$ and $M_\mu$ (and even instead of $M_\mu$ the fields $\tilde m_\mu$ and $\chi$ via $M_\mu=\tilde m_\mu-\partial_\mu\chi$). We will discuss below the effect of either of these variations. We can also couple to the invariants $\bar h_{\mu\nu}$ and $\tau_\mu$ but these are not independent 
\begin{eqnarray}
\delta\tau_\mu & = & \tau_\mu\tau_\nu\delta\hat v^\nu-\bar h_{\mu\rho}\tau_\nu\delta h^{\nu\rho}\,,\label{eq:variationidentity1}\\
\delta\bar h_{\mu\nu} & = & -2\tau_\mu\tau_\nu\delta\tilde\Phi+\left(\tau_\mu\bar h_{\nu\rho}+\tau_\nu\bar h_{\mu\rho}\right)\delta\hat v^\rho-\bar h_{\mu\rho}\bar h_{\nu\sigma}\delta h^{\rho\sigma}\,.\label{eq:variationidentity2}
\end{eqnarray}
as follows from \eqref{eq:relations}. 

\subsection{The energy-momentum tensor and mass current}\label{subsec:EMTs}

The variation with respect to the background (bg) fields is written as
\begin{equation}\label{eq:variation}
\delta_{\text{bg}} S = \int d^{d+1}x e\left[-S^0_\mu\delta v^\mu+S^a_\mu\delta e^\mu_a+T^0\delta\tilde m_0+T^a\delta\tilde m_a+\langle O_\chi\rangle\delta\chi\right]\,,
\end{equation}
where $\tilde m_0=-v^\mu\tilde m_\mu$ and $\tilde m_a=e^\mu_a\tilde m_\mu$. Using that $\tilde m_\mu=M_\mu+\partial_\mu\chi$ this can also be written as
\begin{eqnarray}
\delta S & = & \int d^{d+1}x e\left[-\left(S^0_\nu+T^0\partial_\nu\chi\right)\delta v^\mu+\left(S^a_\nu+T^a\partial_\nu\chi\right)\delta e^\mu_a\right.\nonumber\\
&&\left.+T^0\delta M_0+T^a\delta M_a+\left(\langle O_\chi\rangle-\frac{1}{e}\partial_\mu\left(eT^\mu\right)\right)\delta\chi\right]\,,\label{eq:variation1b}
\end{eqnarray}
where $T^\mu$ is given by
\begin{equation}
T^\mu = -T^0 v^\mu+T^a e_a^\mu\,.
\end{equation}
Just like for the TNC geometry it is useful to find invariants, i.e. $G$, $J$, $N$ invariant quantities built out of $S^0_\mu$, $S^a_\mu$, $T^0$, $T^a$ and $\langle O_\chi\rangle$ that transform as tensors. In order to find these we rewrite the variations with respect to $v^\mu$, $e_a^\mu$ and $M_\mu$ by using that \eqref{eq:variation} can, by using the relations of the previous subsection, equivalently be written as
\begin{eqnarray}
\delta_{\text{bg}} S & = & \int d^{d+1}x e\left[-\tau_\nu T^\nu{}_\mu\delta\hat v^\mu-\left(\hat e^a_\nu\hat v^\mu T^\nu{}_\mu\right)\hat e_{\sigma a}\tau_\rho\delta h^{\rho\sigma}\right.\nonumber\\
&&\left.+\frac{1}{2}\left(\hat e^b_\nu e^\mu_a T^\nu{}_\mu\right)\hat e_{\rho b}\hat e_{\sigma}^a\delta h^{\rho\sigma}+\tau_\mu T^\mu\delta\tilde\Phi\right.\nonumber\\
&&\left.+\left(\langle O_\chi\rangle-\frac{1}{e}\partial_\mu\left(eT^\mu\right)\right)\delta\chi+\left(\hat e_\mu^a T^\mu-\tau_\nu e^{\mu a}T^\nu{}_\mu\right)\delta M_a\right.\nonumber\\
&&\left.-\frac{1}{2}\hat e^{[a}_\nu e^{b]\mu}T^\nu{}_\mu\left(\hat e_{\rho a}\delta e^\rho_b-\hat e_{\rho b}\delta e^\rho_a\right)\right]\,,\label{eq:variation2}
\end{eqnarray}
where we defined the energy momentum tensor $T^\mu{}_\nu$ via \cite{Hartong:2014oma,Hartong:2014pma} 
\begin{equation}
T^\mu{}_\nu = -\left(S^0_\nu+T^0\partial_\nu\chi\right)v^\mu+\left(S^a_\nu+T^a\partial_\nu\chi\right)e^\mu_a\,.
\end{equation}
The vielbein components of $ T^\mu{}_\nu$ with respect to $e_\mu^a$, $v^\mu$, $\tau_\mu$, $e^\mu_a$ give us the energy density, energy flux, momentum density and stress, whereas the vielbein component $T^0=\tau_\mu T^\mu$ is the mass density and $T^a=e^a_\mu T^\mu$ the mass flux. The Ward identities for the St\"uckelberg $U(1)$ (the transformation $\delta \tilde m_\mu=\partial_\mu\sigma$ and $\delta\chi=\sigma$) and local Galilean boosts are
\begin{eqnarray}
e^{-1}\partial_\mu\left(e T^\mu\right) & = & \langle O_\chi\rangle\,,\label{eq:WI-N}\\
\hat e_\mu^a T^\mu-\tau_\nu e^{\mu a}T^\nu{}_\mu & = & 0\,.\label{eq:WI-G}
\end{eqnarray}
These are associated with the local shift transformations acting on $M_a$ and $\chi$. Further since we only couple to $\hat v^\mu$, $h^{\mu\nu}$ and $\tilde\Phi$ the last line of \eqref{eq:variation2} should vanish. This gives us the Ward identity associated with local rotational symmetries (and is the non-relativistic analogue of the fact the energy momentum tensor obtained by coupling to a Lorentzian metric is symmetric)
\begin{equation}\label{eq:WI-J}
\hat e^{[a}_\nu e^{b]\mu}T^\nu{}_\mu=0\,.
\end{equation}
Since \eqref{eq:WI-N}--\eqref{eq:WI-J} are satisfied off-shell we can simplify \eqref{eq:variation2} to
\begin{eqnarray}
\delta_{\text{bg}} S & = & \int d^{d+1}x e\left[-\tau_\nu T^\nu{}_\mu\delta\hat v^\mu-\left(\hat e^a_\nu\hat v^\mu T^\nu{}_\mu\right)\hat e_{\sigma a}\tau_\rho\delta h^{\rho\sigma}\right.\nonumber\\
&&\left.+\frac{1}{2}\left(\hat e^b_\nu e^{\mu a} T^\nu{}_\mu\right)\hat e_{\rho b}\hat e_{\sigma a}\delta h^{\rho\sigma}+\tau_\mu T^\mu\delta\tilde\Phi\right]\,,\label{eq:variation3}
\end{eqnarray}
where only the symmetric part of $\hat e^b_\nu e^{\mu a} T^\nu{}_\mu$ features.

For applications to field theory on TNC geometries discussed here it will sometimes prove convenient to treat $S$ as a functional of $v^\mu$, $h^{\mu\nu}$ and $M_\mu$. With respect to these background fields the variation can be written as
\begin{eqnarray}
\delta_{\text{bg}} S & = & \int d^{d+1}x e\left[-\mathcal{T}_\mu\delta v^\mu+\frac{1}{2}\mathcal{T}_{\mu\nu}\delta h^{\mu\nu}+T^\mu\delta M_\mu\right]\,,\label{eq:variation4}
\end{eqnarray}
where $\mathcal{T}_\mu$ and $\mathcal{T}_{\mu\nu}=\mathcal{T}_{\nu\mu}$ are given by
\begin{eqnarray}
\mathcal{T}_\mu & = & \tau_\nu\left(T^\nu{}_\mu+T^\nu M_\mu\right)\,,\label{eq:relationEMTS1}\\
\mathcal{T}_{\mu\nu} & = & -2\left(\hat e^a_\rho\hat v^\sigma T^\rho{}_\sigma\right)\hat e_{a(\mu}\tau_{\nu)}+\left(\hat e^b_\rho e^\sigma_a T^\rho{}_\sigma\right)\hat e_{b(\mu}\hat e_{\nu)}^a\nonumber\\
&&+2\tau_\rho T^\rho{}_{(\mu}M_{\nu)}+\tau_\rho T^\rho M_\mu M_\nu+X\tau_\mu\tau_\nu\,,\label{eq:relationEMTS2}
\end{eqnarray}
where $X$ is undetermined due to the identity $\tau_\mu\tau_\nu\delta h^{\mu\nu}=0$. We can fix $X$ for example by demanding that $v^\mu v^\nu \mathcal{T}_{\mu\nu}=0$. We do not lose information by fixing $X$, since with $X$ fixed there are as many components in $\mathcal{T}_\mu$, $\mathcal{T}_{\mu\nu}$ as there are in $T^\mu{}_\nu$ which obey \eqref{eq:WI-J}. The boost Ward identity relating $\mathcal{T}_\mu$ and $T^\mu$ reads
\begin{equation}\label{eq:boostWI2}
\mathcal{T}_\mu e^\mu_a=T^\mu e_{\mu a}\,.
\end{equation}
Making frequent use of the relations \eqref{eq:relations} and the Ward identities \eqref{eq:WI-G} and \eqref{eq:WI-J} it can be shown that
\begin{equation}\label{eq:scriptT2}
\mathcal{T}_{\mu\nu}=-2\tau_{(\mu}h_{\nu)\rho}v^\sigma\left(T^\rho{}_\sigma+T^\rho M_\sigma\right)+h_{\mu\rho}h^\sigma{}_{\nu}\left(T^\rho{}_\sigma+T^\rho M_\sigma\right)\,,
\end{equation}
where the last term is symmetric due to the Ward identities \eqref{eq:WI-G} and \eqref{eq:WI-J}. This equation together with \eqref{eq:relationEMTS1} shows that $\mathcal{T}_\mu$, $\mathcal{T}_{\mu\nu}$ are fully determined by $T^\nu{}_\mu+T^\nu M_\mu$. Combining \eqref{eq:relationEMTS1} and \eqref{eq:scriptT2} gives
\begin{equation}\label{eq:relationsEMTS3}
h^{\nu\rho}\mathcal{T}_{\rho\mu}-v^\nu\mathcal{T}_\mu=T^\nu{}_\mu+T^\nu M_\mu\,.
\end{equation}
We will study the difference between $T^\mu{}_\nu$ and $\mathcal{T}_\mu$, $\mathcal{T}_{\mu\nu}$ for the case of a point particle in section \ref{subsec:flatNC}.

\subsection{Diffeomorphisms and TNC Killing vectors}\label{subsec:diffsandKVs}

So far we have only looked at general variations of the background fields. We will next discuss two different types of global TNC space-time symmetries. We start with the first set which is the more conventional set of global TNC space-time symmetries in the sense that they have a relativistic counterpart. By this we mean we will look for transformations that leave the background fields invariant so that $\delta_{\text{bg}}S=0$. The most convenient way of writing the variation for this type of question is \eqref{eq:variation3} because it is written in terms of invariants. This means that the quantities $\tau_\nu T^\nu{}_\mu$, $\hat e^a_\nu\hat v^\mu T^\nu{}_\mu$, $\hat e^b_\nu e^{\mu a} T^\nu{}_\mu$ and $\tau_\mu T^\mu$ are not related by any of the Ward identities that are due to local $G$, $J$ or $N$ transformations. The variation of $S$ with respect to diffeomorphisms acting only on the background fields is
\begin{eqnarray}
\delta_{\text{bg}}[\xi] S & = & \int d^{d+1}x e\left[-\tau_\nu T^\nu{}_\mu\mathcal{L}_\xi\hat v^\mu-\left(\hat e^a_\nu\hat v^\mu T^\nu{}_\mu\right)\hat e_{\sigma a}\tau_\rho\mathcal{L}_\xi h^{\rho\sigma}\right.\nonumber\\
&&\left.+\frac{1}{2}\left(\hat e^b_\nu e^{\mu a} T^\nu{}_\mu\right)\hat e_{\rho b}\hat e_{\sigma a}\mathcal{L}_\xi h^{\rho\sigma}+\tau_\mu T^\mu\mathcal{L}_\xi\tilde\Phi\right]\,.\label{eq:difftrafo1}
\end{eqnarray}
Hence demanding that we get zero leads to global symmetries that are determined by the following equations
\begin{equation}\label{eq:TNCKillingvectors}
\mathcal{L}_\xi\hat v^\mu=0\,,\qquad\mathcal{L}_\xi h^{\mu\nu}=0\,,\qquad \mathcal{L}_\xi\tilde\Phi=0\,,
\end{equation}
whose solutions $\xi^\mu=K^\mu$ define the notion of a Killing vector for a TNC geometry. The variation \eqref{eq:difftrafo1} can also be written as
\begin{eqnarray}
\delta_{\text{bg}}[\xi] S & = & -\int d^{d+1}x\partial_\nu\left(e\xi^\mu T^\nu{}_\mu\right)+\int d^{d+1}x e\xi^\rho\left[e^{-1}\partial_\nu\left(e T^\nu{}_\rho\right)+\tau_\mu T^\mu\partial_\rho\tilde\Phi\right.\nonumber\\
&&\left.+T^\nu{}_\mu\left(\hat v^\mu\partial_\rho\tau_\nu-e^\mu_a\partial_\rho\hat e^a_\nu\right)\right]\,.\label{eq:difftrafo2}
\end{eqnarray}
If we include the variation of the fields under a diffeomorphism our action remains invariant. The variation with respect to the fields gives a boundary term plus a variation that is proportional to the equations of motion. Hence on-shell we have the diffeomorphism Ward identity
\begin{equation}\label{eq:diffeomorphismWI}
0 = e^{-1}\partial_\nu\left(eT^\nu{}_\mu\right)+T^\rho{}_\nu\left(\hat v^\nu\partial_\mu\tau_\rho-e_{a}^\nu\partial_\mu\hat e_{\rho}^{a}\right)+\tau_\nu T^\nu\partial_\mu\tilde\Phi\,, 
\end{equation}
where we note the extra force term due to the potential $\tilde\Phi$. 
Since the variation in \eqref{eq:difftrafo2} vanishes for $\xi^\mu=K^\mu$ satisfying \eqref{eq:TNCKillingvectors} it follows that we have the on-shell conserved currents
\begin{equation}
\partial_\nu\left(eK^\mu T^\nu{}_\mu\right)=0\,.
\end{equation}
One can check that this is indeed the case by using \eqref{eq:TNCKillingvectors} and \eqref{eq:diffeomorphismWI}.

\subsection{Local scale transformations: the dilatation connection $b_\mu$}\label{subsec:scaleandCKVs}

We now turn our attention to scale transformations. If we assume that the theory under consideration is scale invariant, we can assign an appropriate set of dilatation weights to the fields such that the combined transformation of the background fields transforming with their canonical weights and fields leaves the action invariant. 

We first briefly recall how one might derive a conserved dilatation current in the case of a relativistic field theory. We assume that the metric $g_{\mu\nu}$ has been introduced following the minimal coupling prescription. Next we introduce a new connection $b_\mu$, the dilatation connection, which transforms as
\begin{equation}
\delta b_\mu=\mathcal{L}_\xi b_\mu+\partial_\mu\Lambda_D\,,
\end{equation}
where the metric $g_{\mu\nu}$ has dilatation weight $-2$ under $\Lambda_D$. We introduce $b_\mu$ by the method of Weyl gauging, i.e. we replace the covariant derivative $\nabla_\mu$ (containing the Levi--Civit\`a connection) acting on some tensor $T^{\rho\cdots}_{\nu\cdots}$ with dilatation weight $w$, i.e. $\delta T^{\rho\cdots}_{\nu\cdots}=-w\Lambda_D T^{\rho\cdots}_{\nu\cdots}$, by $(\tilde\nabla_\mu+wb_\mu)T^{\rho\cdots}_{\nu\cdots}$. Here $\tilde\nabla_\mu$ contains a connection $\tilde\Gamma^\rho_{\mu\nu}$ that is invariant under local $\Lambda_D$ transformations obtained from the Levi--Civit\`a connection by replacing the ordinary derivative on the metric by the dilatation covariant one $(\partial_\mu-2b_\mu)g_{\nu\rho}$. This procedure makes the action invariant under a local $\Lambda_D$ transformation. The response of the action with respect to a variation of $b_\mu$ defines what is called the virial current $V^\mu$. For a relativistic theory we would thus have
\begin{eqnarray}
\delta_{\text{bg}}[\Lambda_D]S[g^{\mu\nu},b_\mu] & = & \int d^{d+1}x\sqrt{-g}\left(\frac{1}{2}T_{\mu\nu}\delta_{\Lambda_D} g^{\mu\nu}+V^\mu\delta_{\Lambda_D} b_\mu\right)\nonumber\\
& = & \int d^{d+1}x\sqrt{-g}\Lambda_D\left(T_{\mu\nu}g^{\mu\nu}-\frac{1}{\sqrt{-g}}\partial_\mu\left(\sqrt{-g}V^\mu\right)\right)\,,
\end{eqnarray}  
where we ignored a boundary term since we are only interested in on-shell identities. If we would also transform the matter fields we have a vanishing variation since the action is by construction invariant under local $\Lambda_D$ transformations. The variation of the matter fields contains a boundary term and a term that is proportional to the equations of motion. Hence on-shell we have
\begin{equation}\label{eq:relativisticscaleWI}
T_{\mu\nu}g^{\mu\nu}=\frac{1}{\sqrt{-g}}\partial_\mu\left(\sqrt{-g}V^\mu\right)\,.
\end{equation}
It is not automatic that the theory is also conformally invariant. There exist classical relativistic theories that are scale but not conformally invariant such as Maxwell theories in dimensions different from 4 \cite{ElShowk:2011gz}. Adding non-minimal coupling terms to the action leads to improvement transformations of both $T_{\mu\nu}$ and $V^\mu$.

Turning to scale invariant field theories on a TNC background, we note that the TNC analogue of the dilatation connection $b_\mu$ and the dilatation invariant connection $\Gamma^\rho_{\mu\nu}$ has been constructed in \cite{Bergshoeff:2014uea} (section 4.3). The dilatation connection reads
\begin{equation}\label{eq:dilatationconnection}
b_\mu=\frac{1}{z}\hat v^\rho\left(\partial_\rho\tau_\mu-\partial_\mu\tau_\rho\right)-\hat v^\rho b_\rho\tau_\mu\,.
\end{equation}
and this field transforms under \eqref{eq:trafosources} as
\begin{equation}
\delta b_\mu=\mathcal{L}_\xi b_\mu+\partial_\mu\Lambda_D\,,
\end{equation}
i.e. the same as in the relativistic case. However, an important difference with the relativistic case is that there $b_\mu$ is an independent field whereas here only the part $\hat v^\rho b_\rho$ is independent. The dilatation invariant affine connection is \cite{Bergshoeff:2014uea}
\begin{equation}\label{eq:tildeGammaTNC}
\tilde\Gamma^{\rho}_{\mu\nu} = -\hat v^\rho\left(\partial_\mu-zb_\mu\right)\tau_\nu+\frac{1}{2}h^{\rho\sigma}\left(\left(\partial_\mu-2b_\mu\right)\bar h_{\nu\sigma}+\left(\partial_\nu-2b_\nu\right) \bar h_{\mu\sigma}-\left(\partial_\sigma-2b_\sigma\right)\bar h_{\mu\nu}\right)\,.
\end{equation}

Because \eqref{eq:dilatationconnection} is partially a dependent gauge connection the details of the (anisotropic) Weyl gauging procedure are quite different. To get a flavor of what the differences are we consider a few scale invariant examples. 

\subsubsection{The Schr\"odinger model \label{sec:Schmodel}}

Consider the following action that we will refer to as the Schr\"odinger model for reasons that will become clear in section \ref{subsec:FTonflatNC}
\begin{equation}\label{eq:example1}
S=\int d^{d+1}xe\left(-i\phi^\star\hat v^\mu\partial_\mu\phi+i\phi\hat v^\mu\partial_\mu\phi^\star-h^{\mu\nu}\partial_\mu\phi\partial_\nu\phi^\star-2\tilde\Phi\phi\phi^\star-V_0(\phi\phi^*)^{\tfrac{d+2}{d}}\right)\,.
\end{equation}
This action is scale invariant under the $\Lambda_D$ transformations of the background fields as given in \eqref{eq:trafosources} and for $\delta\phi=-\tfrac{d}{2}\Lambda_D\phi$ with $z=2$ and $\Lambda_D$ constant. We can now apply the Weyl gauging method to this model, i.e. we replace $\partial_\mu\phi$ by $\left(\partial_\mu+\tfrac{d}{2}b_\mu\right)\phi$ where in \eqref{eq:dilatationconnection} we set $z=2$. This gives
\begin{eqnarray}
S & = & \int d^{d+1}xe\left(-i\phi^\star\hat v^\mu\partial_\mu\phi+i\phi\hat v^\mu\partial_\mu\phi^\star-h^{\mu\nu}\left(\partial_\mu+\frac{d}{2}b_\mu\right)\phi\left(\partial_\nu+\frac{d}{2}b_\nu\right)\phi^\star\right.\nonumber\\
&&\left.-2\tilde\Phi\phi\phi^\star-V_0(\phi\phi^*)^{\tfrac{d+2}{d}}\right)\,.\label{eq:example2}
\end{eqnarray}
Nothing happens with the $\hat v^\mu\partial_\mu$ derivatives because the $b_\mu$ drops out. The $b_\mu$ connection thus only enters via the part $h^{\mu\nu}b_\nu$ which is fully determined in terms of the invariants. We have thus managed to construct a local dilatation invariant action that only depends on the usual background fields $\hat v^\mu$, $h^{\mu\nu}$, $\tilde\Phi$ as well as the complex scalar $\phi$. Adding $b_\mu$ to the action has the effect of changing the energy momentum tensor $T^\mu{}_\nu$. Clearly if we vary \eqref{eq:example2} with respect to $\hat v^\mu$, $h^{\mu\nu}$, $\tilde\Phi$ we get a different answer for $T^\mu{}_\nu$ than if we vary these fields in \eqref{eq:example1}. Varying the background fields in \eqref{eq:example2} under a local $\Lambda_D$ transformation we get
\begin{eqnarray}
\delta_{\text{bg}}[\Lambda_D] S & = & \int d^{d+1}x e\Lambda_D\left[-z\tau_\nu\hat v^\mu T^\nu{}_\mu+\hat e^a_\nu e^{\mu a} T^\nu{}_\mu+2(z-1)\tau_\mu T^\mu\tilde\Phi\right]\,.
\end{eqnarray}
If we use the fact that the contribution to the total variation of \eqref{eq:example2} under a local $\Lambda_D$ transformation that comes from $\phi$ vanishes on-shell we get the $z=2$ version of the on-shell Ward identity \cite{Hartong:2014oma}
\begin{equation}\label{eq:scaleWI}
-z\tau_\nu\hat v^\mu T^\nu{}_\mu+\hat e^a_\nu e^{\mu a} T^\nu{}_\mu+2(z-1)\tau_\mu T^\mu\tilde\Phi=0\,.
\end{equation}
The $\Lambda_D$ transformation of the background fields is induced by diffeomorphisms in the form of conformal Killing vectors. This defines the notion of a conformal Killing vector as the solution $\xi^\mu=K^\mu$ and $\Lambda_D=\Omega$ to the equations
\begin{eqnarray}
\mathcal{L}_\xi\tau_\mu & = & -z\Lambda_D\tau_\mu\,,\label{eq:KE1}\\
\mathcal{L}_\xi\hat v^\mu & = & z\Lambda_D\hat  v^\mu\,,\label{eq:KE2}\\
\mathcal{L}_\xi\bar h_{\mu\nu} & = & -2\Lambda_D\bar h_{\mu\nu}\,,\label{eq:KE3}\\
\mathcal{L}_\xi h^{\mu\nu} & = & 2\Lambda_D h^{\mu\nu}\,,\label{eq:KE4}\\
\mathcal{L}_\xi\tilde\Phi & = & 2(z-1)\Lambda_D\tilde\Phi\,,\label{eq:KE5}
\end{eqnarray}
where here we take $z=2$.

For a Newton--Cartan background, i.e. $\partial_\mu\tau_\nu-\partial_\nu\tau_\mu=0$, the Ward identity \eqref{eq:scaleWI} where the $T^\nu{}_\mu$ is the one associated with \eqref{eq:example2} can be rewritten as follows. We do this by isolating the contributions from the variation of $b_\mu$\footnote{Even though $b_\mu$ vanishes on a Newton--Cartan background its variation evaluated on a NC background is nonzero and responsible for the occurrence of the virial current $V^\mu$.} to $T^\nu{}_\mu$ in \eqref{eq:scaleWI} and putting these terms on the right hand side. This gives
\begin{equation}\label{eq:alternativescaleWI}
-z\tau_\nu\hat v^\mu\tilde T^\nu{}_\mu+\hat e^a_\nu e^{\mu}_a\tilde T^\nu{}_\mu+2(z-1)\tau_\mu T^\mu\tilde\Phi=e^{-1}\partial_\mu\left(e V^\mu\right)\,,
\end{equation}
where $V^\mu$ is given by
\begin{equation}
V^\mu=\frac{d}{2}h^{\mu\nu}\partial_\nu\left(\phi\phi^\star\right)\,,
\end{equation}
and where $\tilde T^\nu{}_\mu$ in \eqref{eq:alternativescaleWI} is the one associated with \eqref{eq:example1}. Even though this scale Ward identity looks very similar to \eqref{eq:relativisticscaleWI} in the relativistic case, the way we get to it in the non-relativistic setting
is quite different. 

\subsubsection{Deformations of the Schr\"odinger model \label{sec:defSchmodel}}

If we set $\phi=\tfrac{1}{\sqrt{2}}\varphi e^{i\theta}$ the action \eqref{eq:example2} can be written as
\begin{eqnarray}\label{eq:example3}
S & = & \int d^{d+1}x e\left[\varphi^2\left(\hat v^\mu\partial_\mu\theta-\frac{1}{2}h^{\mu\nu}\partial_\mu\theta\partial_\nu\theta-\tilde\Phi\right)\right.\nonumber\\
&&\left.-\frac{1}{2}h^{\mu\nu}\left(\partial_\mu\varphi+\frac{d}{2}b_\mu\varphi\right)\left(\partial_\nu\varphi+\frac{d}{2}b_\nu\varphi\right)-V_0\varphi^{\tfrac{2(d+2)}{d}}\right]
\end{eqnarray}
We note that we can change the potential to a non-$U(1)$ invariant function with dilatation weight $d+2$ and all this would still be true, i.e. we can take e.g. $V=V_0\varphi^{\tfrac{2(d+2)}{d}}(1+b\theta^2)$. Another deformation of \eqref{eq:example3} that preserves local scale invariance is to add to \eqref{eq:example3} the term
\begin{equation}\label{eq:deformterm}
-a\int d^{d+1}xe\varphi^2 h^{\mu\nu}\tilde\nabla_\mu\partial_\nu\theta=2a\int d^{d+1}xe\varphi\left(\partial_\mu+\frac{d}{2}b_\mu\right)\varphi h^{\mu\nu}\partial_\nu\theta\,,
\end{equation}
where $\tilde\nabla_\mu$ contains the dilatation invariant connection \eqref{eq:tildeGammaTNC}. The field $\theta$ has dilatation weight zero so these terms are manifestly invariant under local scale transformations. In terms of the complex scalar $\phi$ this term is given by
\begin{equation}
-i\frac{a}{4}\int d^{d+1}x e\left(\frac{\phi^\star}{\phi}h^{\mu\nu}\left(\partial_\mu+\frac{d}{2}b_\mu\right)\phi\left(\partial_\nu+\frac{d}{2}b_\nu\right)\phi+\text{c.c.}\right)\,.
\end{equation}

In all these cases the scale Ward identity is of the form \eqref{eq:scaleWI}. Yet, we do not expect this to be the answer in general. The examples we have considered all have the property that the $\hat v^\mu b_\mu$ component, which is the independent component of $b_\mu$, drops out. This does not always need to happen and in those cases we expect modifications of \eqref{eq:scaleWI}, see for example \eqref{eq:example6} and just below \eqref{eq:Lifshitzmodel}.

\subsection{Local $U(1)$ transformations: promoting $M_\mu$ to a gauge connection}\label{subsec:toymodels}

So far we have looked at symmetries that relate to the invariant $T^\mu{}_\nu$. There is another such invariant which is $T^\mu$ that naturally appears when we vary with respect to $v^\mu$, $h^{\mu\nu}$ and $M_\mu$ as in \eqref{eq:variation4}.  As we have seen,  the scale symmetries come about by combining diffeomorphisms that act on the background fields $\hat v^\mu$, $h^{\mu\nu}$ and $\tilde\Phi$ with scale transformations that act on the fields living on the TNC background. We will now see that there is a second natural way in which symmetries can occur that relates to the presence of the background field $M_\mu$. We will show that it can happen that diffeomorphisms together with local boosts (and possibly local scale transformations) via \eqref{eq:trafosources} induce a transformation of the type 
\begin{equation}\label{eq:diffstype3}
\tilde N\;:\qquad\delta v^\mu =0\,,\qquad\delta h^{\mu\nu}=0\,,\qquad \delta M_\mu=\partial_\mu\tilde\sigma\,,
\end{equation}
with a specific $\tilde\sigma$ leaving the action invariant (due to diffeomorphism invariance of the theory). We denote this transformation by $\tilde N$. It is similar but not identical to the transformation denoted by $N$ in section \ref{subsec:localtrafosources}\footnote{At the beginning of section \ref{subsec:localtrafosources} we write $\tilde m_\mu=M_\mu+\partial_\mu\chi$ for the vector that transforms as a gauge connection under the $N$ transformation because $\delta_N\tilde m_\mu=\partial_\mu\sigma$. However $\tilde m_\mu$ does not transform nicely under dilatations. In \cite{Bergshoeff:2014uea} it is shown that it is rather the field $m_\mu$ defined as $m_\mu=M_\mu+\partial_\mu\chi-(2-z)\chi b_\mu$ with $b_\mu$ the dilatation connection, that is the natural $N$ gauge connection because this is how it appears in the gauging of the Schr\"odinger algebra. In this work we will have no need for $m_\mu$. We just mention it for the sake of completeness.}. The diffeomorphisms that lead to \eqref{eq:diffstype3} are of the form
\begin{eqnarray}
\mathcal{L}_\xi v^\mu & = & -\lambda^ae^\mu_a+z\Lambda_D v^\mu\,,\label{eq:type2symmetry-1}\\
\mathcal{L}_\xi h^{\mu\nu} & = & 2\Lambda_D h^{\mu\nu}\,,\label{eq:type2symmetry-2}\\
\mathcal{L}_\xi M_\mu & = & -e_\mu^a\lambda_a-(2-z)\Lambda_DM_\mu+\partial_\mu\alpha\,,\label{eq:type2symmetry-3}
\end{eqnarray}
whose solution we write as $\xi^\mu=L^\mu$, $\Lambda_D=\Omega$ and $\alpha=\tilde\sigma$. If the theory has a global $U(1)$ invariance that can be made local in which $M_\mu$ transforms as a gauge field the diffeomorphisms leading to \eqref{eq:diffstype3} can become global symmetries. The reason is that we can now do a sequence of two transformations that leaves the background fields invariant, namely first we perform a diffeomorphism of the type \eqref{eq:type2symmetry-1}--\eqref{eq:type2symmetry-3} and then we perform a compensating internal local $U(1)$ transformation. The combined effect of these two transformations leaves $M_\mu$ invariant and acts on the fields charged under the global $U(1)$. Since this symmetry crucially relies on the presence of the field $M_\mu$ it has no counterpart in a relativistic setting. We will see that the global $U(1)$ can be made local by carefully engineering the couplings to the TNC background such that the gauge connection becomes $M_\mu$. 

\subsubsection{Local $U(1)$ invariance of the deformed Schr\"odinger model \label{sec:locU}}

To make all this more explicit we consider the case of the $z=2$ scale invariant model \eqref{eq:example3} to which we add the deformation term \eqref{eq:deformterm}, i.e. we consider 
\begin{eqnarray}\label{eq:example4}
S & = & \int d^{d+1}x e\left[\varphi^2\left(\hat v^\mu\partial_\mu\theta-\frac{1}{2}h^{\mu\nu}\partial_\mu\theta\partial_\nu\theta-\tilde\Phi-ah^{\mu\nu}\tilde\nabla_\mu\partial_\nu\theta\right)\right.\nonumber\\
&&\left.-\frac{1}{2}h^{\mu\nu}\left(\partial_\mu\varphi+\frac{d}{2}b_\mu\varphi\right)\left(\partial_\nu\varphi+\frac{d}{2}b_\nu\varphi\right)-V_0\varphi^{\tfrac{2(d+2)}{d}}\right]\,.
\end{eqnarray}
To make the role of $M_\mu$ manifest we write it in terms of the $v^\mu$, $h^{\mu\nu}$ and $M_\mu$ background fields leading to
\begin{eqnarray}\label{eq:example5}
S & = & \int d^{d+1}x e\left[\varphi^2\left(v^\mu\left(\partial_\mu\theta+M_\mu\right)-\frac{1}{2}h^{\mu\nu}\left(\partial_\mu\theta+M_\mu\right)\left(\partial_\nu\theta+M_\mu\right)\right.\right.\nonumber\\
&&\left.\left.-ah^{\mu\nu}\tilde\nabla_\mu\left(\partial_\nu\theta+M_\nu\right)+ah^{\mu\nu}\tilde\nabla_\mu M_\nu\right)\right.\nonumber\\
&&\left.-\frac{1}{2}h^{\mu\nu}\left(\partial_\mu\varphi+\frac{d}{2}b_\mu\varphi\right)\left(\partial_\nu\varphi+\frac{d}{2}b_\nu\varphi\right)-V_0\varphi^{\tfrac{2(d+2)}{d}}\right]\,.
\end{eqnarray}
We see that this theory has a local symmetry $\delta M_\mu=\partial_\mu\alpha$, $\delta\theta=-\alpha$. However there is a term that spoils it. This is the $ah^{\mu\nu}\tilde\nabla_\mu M_\nu$ term on the second line. This problem can be cured by adding the following term to the action
\begin{eqnarray}
-a\int d^{d+1}xe\varphi^2 e^\mu_a\mathcal{D}_\mu M^a & = & -a\int d^{d+1}xe\varphi^2 \left(-e^{-1}\partial_\mu\left(e\hat v^\mu\right)+d\hat v^\mu b_\mu\right) \\
& = &  -2a\int d^{d+1}xe \varphi\hat v^\mu\left(\partial_\mu\varphi+\frac{d}{2}b_\mu\varphi\right)\nonumber\\
& = & -a\int d^{d+1}xe\left(2\varphi v^\mu\left(\partial_\mu\varphi+\frac{d}{2}b_\mu\varphi\right)+\varphi^2h^{\mu\nu}\tilde\nabla_\mu M_\nu\right)\,.\nonumber
\end{eqnarray}
The notation $\mathcal{D}_\mu M^a$ is borrowed from \cite{Bergshoeff:2014uea} and involves a dilatation covariant connection for local Galilean boosts to make the expression boost invariant. Since we have the identity \cite{Bergshoeff:2014uea}
\begin{equation}
e^\mu_a\mathcal{D}_\mu M^a=-e^{-1}\partial_\mu\left(e\hat v^\mu\right)+d\hat v^\mu b_\mu\,,
\end{equation}
it suffices for us to use the right hand side which is written in terms of quantities we already defined. The addition of this term can be compared to the introduction of the term $-\varphi^2\tilde\Phi$ in \eqref{eq:example1}. That term played no role until we started writing things in terms of $v^\mu$, $h^{\mu\nu}$ and $M_\mu$ and its purpose is to create the local symmetry $\delta M_\mu=\partial_\mu\alpha$, $\delta\theta=-\alpha$. 

We are thus led to consider the following model
\begin{eqnarray}\label{eq:example6}
S & = & \int d^{d+1}x e\left[\varphi^2\left(\hat v^\mu\partial_\mu\theta-\frac{1}{2}h^{\mu\nu}\partial_\mu\theta\partial_\nu\theta-ah^{\mu\nu}\tilde\nabla_\mu\partial_\nu\theta-\tilde\Phi-ae^\mu_a\mathcal{D}_\mu M^a\right)\right.\nonumber\\
&&\left.-\frac{1}{2}h^{\mu\nu}\left(\partial_\mu\varphi+\frac{d}{2}b_\mu\varphi\right)\left(\partial_\nu\varphi+\frac{d}{2}b_\nu\varphi\right)-V_0\varphi^{\tfrac{2(d+2)}{d}}\right]\nonumber\\
 & = & \int d^{d+1}x e\left[\varphi^2\left(v^\mu\left(\partial_\mu\theta+M_\mu\right)-\frac{1}{2}h^{\mu\nu}\left(\partial_\mu\theta+M_\mu\right)\left(\partial_\nu\theta+M_\mu\right)-ah^{\mu\nu}\tilde\nabla_\mu\left(\partial_\nu\theta+M_\nu\right)\right)\right.\nonumber\\
&&\left.-2a\varphi v^\mu\left(\partial_\mu\varphi+\frac{d}{2}b_\mu\varphi\right)-\frac{1}{2}h^{\mu\nu}\left(\partial_\mu\varphi+\frac{d}{2}b_\mu\varphi\right)\left(\partial_\nu\varphi+\frac{d}{2}b_\nu\varphi\right)-V_0\varphi^{\tfrac{2(d+2)}{d}}\right]\,.
\end{eqnarray}
Using that \eqref{eq:example6} has the local symmetry
\begin{equation}\label{eq:alphatrafo}
\delta M_\mu=\partial_\mu\alpha\,,\qquad \delta\theta=-\alpha\,,
\end{equation}
we obtain the on-shell Ward identity
\begin{equation}\label{eq:particlenumberconservation}
\partial_\mu\left(e T^\mu\right)=0\,,
\end{equation}
which is a way of writing the $\theta$ equation of motion. Hence diffeomorphisms of the type \eqref{eq:diffstype3} accompanied by a local shift of $\theta$ leave the action invariant leading to additional global space-time symmetries. 

This is quite analogous to what happened in the case of the scale transformations where diffeomorphisms $\xi^\mu=K^\mu$ generate a specific $\Lambda_D=\Omega$ transformation that is then compensated by a scale transformation of the scalar field, so also there it is the combined effect of diffeomorphism invariance plus a local scale transformation that leads to the existence of more global symmetries. In the case of the scale transformations we generalized the notion of Killing vectors to include the diffeomorphisms $\xi^\mu=K^\mu$, $\Lambda_D=\Omega$ that transform the background fields as a specific $\Lambda_D$ transformation and we called these Killing vectors conformal in analogy with Lorentzian geometry. One might consider to do the same for the case of the \eqref{eq:type2symmetry-1}--\eqref{eq:type2symmetry-3} diffeomorphisms that are generated by $\xi^\mu=L^\mu$ and $\Lambda_D=\Omega$. However these also involve specific local boost transformations ($\lambda_a$) that have no space-time counterpart and so we will refrain from calling them Killing vectors of some kind\footnote{This is in agreement with the fact Lifshitz space-times can accommodate Schr\"odinger invariant fields (see section \ref{subsec:probes}) but, as we will show in section \ref{subsec:flatNCKillingvectors}, its Killing vectors only realize the boundary TNC conformal Killing vectors that generate the Lifshitz algebra.}.

We also note that the objects $\mathcal{T}_\mu$ and $\mathcal{T}_{\mu\nu}$ appearing in \eqref{eq:variation4} are gauge invariant with respect to \eqref{eq:alphatrafo}. So we observe that $T^\mu{}_\nu$ is boost invariant whereas $h^{\nu\rho}\mathcal{T}_{\rho\mu}-v^\nu\mathcal{T}_\mu$ is not as follows from \eqref{eq:relationsEMTS3} while on the other hand $T^\mu{}_\nu$ is not gauge invariant under \eqref{eq:alphatrafo} whereas $h^{\nu\rho}\mathcal{T}_{\rho\mu}-v^\nu\mathcal{T}_\mu$ is as follows from \eqref{eq:variation4} (see also the example of \eqref{eq:example6}). Since we have $\delta_\alpha T^\nu{}_\mu=-T^\nu\partial_\mu\alpha$ using \eqref{eq:relationsEMTS3}, a gauge and boost invariant object for the model of \eqref{eq:example6} is $T^\nu{}_\mu-T^\nu\partial_\mu\theta$.

\subsection{Comments on the role of $M_\mu$ }\label{sec:comM}

In cases where we have the local symmetry \eqref{eq:alphatrafo} and we write $M_\mu=\tilde m_\mu-\partial_\mu\chi$ as we did in section \ref{subsec:localtrafosources}, we can gauge fix the $\alpha$ transformation to remove $\chi$ from the formalism and the new local symmetry becomes
\begin{equation}
\delta \tilde m_\mu=\partial_\mu\sigma\,,\qquad \delta\theta=-\sigma\,,
\end{equation}
where $\sigma$ is the parameter of local particle number $N$ transformations. When $\chi$ has been removed from the action, or what is the same, in the presence of the local symmetry \eqref{eq:alphatrafo}, the quantity $\langle O_\chi\rangle$ appearing in \eqref{eq:WI-N} vanishes. This is the situation discussed in \cite{Jensen:2014aia} and this can be reproduced by our formalism. The current $T^\mu$ thus corresponds to particle number and equation \eqref{eq:particlenumberconservation} expresses its conservation. This makes the boost Ward identity \eqref{eq:WI-G} or \eqref{eq:boostWI2} physical, i.e. not just an identity that has to be true due to a built-in structure of local symmetries, such as coming from the use of vielbeins and St\"uckelberg symmetries, but one that is the consequence of global space-time symmetries of the type \eqref{eq:type2symmetry-1}--\eqref{eq:type2symmetry-3}. 

We stress though that the notion of coupling a field theory to a TNC background that contains $\chi$ so that we work with $M_\mu$ rather than with $\tilde m_\mu$ does not require the presence of such local $U(1)$ transformations so that our formalism also works in more general settings. For example if we change the potential $V$ to a function of $\varphi$ and $\theta$ we break the $U(1)$ symmetry but we can still work with the general formalism of coupling to TNC geometries. In this case $T^\mu$ no longer corresponds to a conserved particle number current because $\langle O_\chi\rangle$ has becomes non-zero as a result of the potential depending on $\theta$. To compute $\langle O_\chi\rangle$ perform an $\alpha$ transformation such that $M_\mu$ becomes equal to $\tilde m_\mu$ by transforming $M_\mu=M'_\mu-\partial_\mu\chi$. Using subsequently that by definition $M_\mu=\tilde m_\mu-\partial_\mu\chi$, we get the desired result. This transformation acts on $\theta$ as $\theta=\theta'+\chi$ and introduces $\chi$ into the potential because now $V(\varphi,\theta)=V(\varphi,\theta'+\chi)$. If $\theta$ in the potential is the only source of particle number breaking we get $\langle O_\chi\rangle=-\partial_\chi V$. In such a theory the Ward identities \eqref{eq:WI-N} and \eqref{eq:WI-G} are just to be thought of as consequences of reparametrizations that have been built-in to the framework. The relevant quantities are now the background fields $\hat v^\mu$, $h^{\mu\nu}$ and $\tilde\Phi$ and the on-shell Ward identities for diffeomorphisms and local scale symmetries. This allows us to describe Lifshitz invariant field theories using TNC geometries. The response to varying $\tilde\Phi$ can be called the mass density, $\tau_\mu T^\mu$, but it does not correspond to the component of some conserved mass current. It simply can appear in the diffeomorphism and local scale Ward identities. 

A note on our terminology: in cases where we couple to $\tilde\Phi$ we call $T^\mu$ the mass current regardless of whether or not $T^\mu$ is conserved, i.e. we call $T^\mu$ the mass current (or particle number current) regardless of whether or not we have a local $U(1)$ symmetry whose gauge connection is $M_\mu$. We do this because we can either isolate the terms responsible for the explicit breaking of the conservation of $T^\mu$ by introducing $\chi$ in the manner just described in the text, i.e. because we can compute $\langle O_\chi\rangle=-\partial_\chi V$ or because we can show that $\langle O_\chi\rangle=e^{-1}\partial_\mu\left(e J^\mu\right)$, so that the current $T^\mu-J^\mu$ is the equation for particle number conservation. This latter option occurs in cases where we couple to a TNC geometry in a manner such that there is no local $U(1)$. In the case of the model \eqref{eq:example6} we can spoil the local $U(1)$ symmetry by removing the terms $\tilde\Phi$ and $e^\mu_a\mathcal{D}_\mu M^a$ in the first line of \eqref{eq:example6}. At the end of section \ref{subsec:FTonflatNC} and in appendix \ref{subsec:comments} we explain in detail how one can show that $\langle O_\chi\rangle=e^{-1}\partial_\mu\left(e J^\mu\right)$ and thus find the particle number current for the case of a flat NC background. 

In section \ref{subsec:FTonflatNC} we will study the complex scalar field theory mentioned above on a flat Newton--Cartan background and show that we can have various degrees of space-time symmetries such as scale invariance with or without conformal invariance and/or Galilean boosts. Put another way TNC geometries can accommodate Schr\"odinger invariant field theories just as easily as Lifshitz invariant ones. We will see that the action \eqref{eq:example6} on flat space corresponds to a Schr\"odinger invariant theory for $a=0$ but that if we change the potential to break the $U(1)$ symmetry while retaining scale invariance this gets reduced to Lifshitz symmetries. The case of \eqref{eq:example6} with $a\neq 0$ breaks special conformal symmetry while retaining Galilean boost invariance. In general the generic space-time symmetries for a scale invariant theory are given by the Lifshitz algebra which we will show are the conventional symmetries originating from the TNC conformal Killing vectors. The enhancement to larger algebras can be realized by the aforementioned mechanism relating to the local $U(1)$ symmetry. However before we can discuss these models we need to know a bit more about Newton--Cartan space-times in particular flat NC space-times and its conformal Killing vectors.

\subsection{No coupling to $\tilde\Phi$}\label{subsec:notildePhicoupling}

We may also consider the situation in which we do not couple to $\tilde\Phi$ and we only have $\hat v^\mu$ and $h^{\mu\nu}$ as for example in the case of the $z=2$ Lifshitz model
\begin{equation}\label{eq:Lifshitzmodel}
S=\int d^{d+1}xe\left[\frac{1}{2}\left(\hat v^\mu\partial_\mu\phi\right)^2-\frac{\lambda}{2}\left(h^{\mu\nu}\nabla_\mu\partial_\nu\phi\right)^2\right]\,.
\end{equation}
In such cases we have no need for $T^\mu$ as we can define everything in terms of the response to varying $\hat v^\mu$ and $h^{\mu\nu}$, i.e. $T^\nu{}_\mu$. Here $\phi$ is a real scalar with dilatation weight $(d-2)/2$. We note that we could make the model invariant under local scale transformations by replacing $\partial_\mu$ by $\partial_\mu+\tfrac{d-2}{2}b_\mu$.

When we do not couple to $\tilde\Phi$ it is more convenient to change the affine connection to 
\begin{equation}\label{eq:hatGammaTNC}
\hat\Gamma^{\rho}_{\mu\nu} = -\hat v^\rho\partial_\mu\tau_\nu+\frac{1}{2}h^{\rho\sigma}\left(\partial_\mu\hat h_{\nu\sigma}+\partial_\nu \hat h_{\mu\sigma}-\partial_\sigma\hat h_{\mu\nu}\right)\,,
\end{equation}
where $\hat h_{\mu\nu}=\delta_{ab}\hat e^a_\mu\hat e^b_\nu=\bar h_{\mu\nu}+2\tilde\Phi\tau_\mu\tau_\nu$. This connection is also a $G$, $J$, $N$ invariant that is metric compatible in the sense that $\hat\nabla_\mu\tau_\nu=0=\hat\nabla_\mu h^{\nu\rho}$ in which $\hat\nabla_\mu$ contains $\hat\Gamma^{\rho}_{\mu\nu}$. The existence of this connection is explained in footnote \ref{footnote:newconnection}. Hence when we do not couple to $\tilde\Phi$, we couple to $\hat h_{\mu\nu}$, $\tau_\mu$, $\hat v^\mu$ and $h^{\mu\nu}$ using the affine connection $\hat\Gamma^{\rho}_{\mu\nu}$.

\subsubsection{Lorentz invariants}

It is interesting to note that 
the invariants $\tau_\mu$, $h^{\mu\nu}$, $\hat v^\mu$, $\bar h_{\mu\nu}$ and $\tilde\Phi$ of section \ref{subsec:TNCgeometry} satisfying the relations \eqref{eq:relations} can be used to build non-degenerate symmetric rank 2 tensors with Lorentzian signature $g_{\mu\nu}$ that in the case of a relativistic theory we would refer to as a Lorentzian metric. The metric $g_{\mu\nu}$ and its inverse $g^{\mu\nu}$ are given by
\begin{eqnarray}
g_{\mu\nu} & = & -\tau_\mu\tau_\nu+\bar h_{\mu\nu}+2\tilde\Phi\tau_\mu\tau_\nu=-\tau_\mu\tau_\nu+\hat h_{\mu\nu}\,,\\
g^{\mu\nu} & = & -\hat v^\mu\hat v^\nu+h^{\mu\nu}\,,
\end{eqnarray}
for which we have
\begin{eqnarray}
g_{\mu\nu}\hat v^\mu & = & \tau_\nu\,,\\
g_{\mu\nu} e^\mu_a & = & \hat e_{\nu a}\,.
\end{eqnarray}
We just discussed a subclass of field theories coupled to TNC geometries that do not couple to $\tilde\Phi$, i.e. actions of the form $S=S[\hat v^\mu, h^{\mu\nu}]$. We can thus equally write this as $S=S[\hat v^\mu, g^{\mu\nu}]$. This is the situation discussed in \cite{Hoyos:2013qna}. We refrain from calling $g_{\mu\nu}$ a Lorentzian metric except in cases where we do not separately couple to $\hat v^\mu$ and we simply have $S=S[g^{\mu\nu}]$. When we are dealing with an action of the form $S=S[g^{\mu\nu}]$ it is of course best to use the Christoffel connection. All the connections used here, i.e. 
$\Gamma^{\rho}_{\mu\nu}$ of \eqref{eq:GammaTNC}, $\hat\Gamma^{\rho}_{\mu\nu}$ of \eqref{eq:hatGammaTNC} and the Christoffel connection are related by redefinitions such that any two of these connections differ by a tensor. Hence any one is as good as any other one, or put another way they are all affine connections. The difference resides from demanding different notions of metric compatibility conditions and dependence on $\tilde\Phi$ or $M_\mu$.

\section{Flat Newton--Cartan space-time \label{sec:flatspace}}

Building on the general results of the previous section, we now turn to the symmetry properties of non-relativistic field theories on flat NC space-time. 
To this end, we first define our notion of flat NC space-time, after which we determine the residual coordinate transformations
that leave this invariant. These ingredients will then be used to discuss scale invariant field theories on flat NC backgrounds,
including the particular toy models introduced in the previous section. 

\subsection{Definition}\label{subsec:flatNC}

We first need to define what it means for a  Newton--Cartan space-time to be flat. This is a relevant question as often we are interested in field theories on flat space-time. For us the main reason is that this will turn out to be the dual boundary geometry of a Lifshitz space-time in a certain class of coordinates. We are not aware of a covariant definition of such a concept and we will define it in what will be referred to as global inertial coordinates (see also \cite{Andringa:2013mma,Bergshoeff:2014uea}). We will give expressions for the variables $\tau_\mu$, $e^a_\mu$ and $M_\mu$. We start with the vielbeins. For suitably chosen coordinates $(t,x^i)$ they are
\begin{equation}
\tau_\mu = \delta^t_\mu\,,\qquad e^a_\mu = \delta_\mu^i\delta^a_i\,.
\end{equation}
This implies that we have
\begin{equation}\label{eq:flatNC}
\begin{array}{rclccrcl}
\tau_\mu & = & \delta^t_\mu\,,&&&&&\\
h^{tt} & = & h^{ti}=0\,,&&& h^{ij} & = & \delta^{ij}\,,\\
v^\mu & = &  -\delta^\mu_t\,,&&&&&\\
h_{tt} & = & h_{ti}=0\,,&&& h_{ij} & = & \delta_{ij}\,.
\end{array}
\end{equation}

So far we did not specify yet what we should take for  $M_\mu$. In our setup the space-time is not dynamical, but we would like things to be such that if we probe the geometry with a standard non-relativistic particle of mass $m$ with quadratic dispersion relation it obeys Newton's second law. Since we have set $\tau_\mu = \delta^t_\mu$ there is no torsion in the affine connection \eqref{eq:GammaTNC} and so we are within the context of ordinary Newton--Cartan geometry. 

The motion of a non-relativistic particle of mass $m$ on a NC background is governed by the following action \cite{Kuchar:1980tw,Bergshoeff:2014gja}
\begin{equation}\label{eq:particleaction}
S=\int d\lambda L=\frac{m}{2}\int d\lambda\frac{\bar h_{\mu\nu}\dot x^\mu \dot x^\nu}{\tau_\rho\dot x^\rho}\,,
\end{equation}
where the dot denotes differentiation with respect to $\lambda$. This action has a world-line reparametrization symmetry of the form $\delta\lambda=\xi(\lambda)$ and $\delta x^\mu=\xi(\lambda)\dot x^\mu$. Using this to fix a gauge in which $\tau_\mu\dot x^\mu=1$ it can be shown that the equations of motion are given by the geodesic equation
\begin{equation}\label{eq:TNCgeodesic}
\frac{d^2 x^\mu}{d\lambda^2}+\Gamma^\mu_{\nu\rho}\frac{dx^\nu}{d\lambda}\frac{dx^\rho}{d\lambda}=0\,,
\end{equation}
where $\Gamma^\rho_{\mu\nu}$ is given by \eqref{eq:GammaTNC}. We expect this to be the relevant geodesic equation for any TNC geometry, however the equation of motion obtained by varying the action \eqref{eq:particleaction} only gives rise to \eqref{eq:TNCgeodesic} when the background is NC. In our coordinates the components of $\Gamma^\rho_{\mu\nu}$ are
\begin{equation}
\begin{array}{rcl}
\Gamma^t_{\mu\nu} & = & 0\,,\\
\Gamma^i_{tt} & = &  -\delta^{ij}\left(\partial_t M_j-\partial_j M_t\right)\,,\\
\Gamma^i_{tk} & = &  -\frac{1}{2}\delta^{ij}\left( \partial_k M_j-\partial_j M_k\right)\,,\\
\Gamma^i_{kl} & = & 0\,.
\end{array}
\end{equation}
Hence in order that we obtain Newton's second law we must choose
\begin{eqnarray}
M_t & = & \partial_t M+\Phi\,,\label{eq:NPPhi}\\
M_i & = & \partial_i M\,,\label{eq:NPPhi2} 
\end{eqnarray}
so that
\begin{equation}
\frac{d^2 x^i}{dt^2}+\delta^{ij}\partial_j \Phi=0\,,
\end{equation}
where $\Phi$ is the Newton potential\footnote{Calling $\Phi$ the Newton potential is only justified for particle motion governed by \eqref{eq:TNCgeodesic}. In general depending on the dispersion relation one may need to consider more general geodesic equations, see e.g. \cite{Kiritsis:2009rx} in the context of Ho\v rava--Lifshitz gravity.}. Hence in a flat space we expect straight line motion in a suitable coordinate system, which here means that we need to take $\Phi=0$. Consequently, 
 our coordinate dependent specification of flat space entails the statement
\begin{equation}\label{eq:flatNC-V}
\Gamma^\rho_{\mu\nu}=0\rightarrow M_\mu=\partial_\mu M\,.
\end{equation}

Returning to our discussion of flat NC space-time, we have thus imposed \eqref{eq:flatNC} and \eqref{eq:flatNC-V} leaving us with a function $M$. We now address the significance of this function $M$.  So far the description of flat space-time is universal. Certainly flat space should include the case $M=\text{cst}$. However we will show in section \ref{subsec:FTonflatNC} that sometimes we can allow for more general functions $M$ because they are identical to $M$ by local symmetries of the theory. The set of $M$'s that are identical to each other by local transformations of the theory is what we will call the orbit of $M$. 

Before discussing the orbits of $M$ for the various scalar field theory models mentioned in section \ref{subsec:toymodels} we study in the next subsection the most general diffeomorphisms that preserve our choices \eqref{eq:flatNC} and \eqref{eq:flatNC-V} under the local TNC transformations \eqref{eq:trafosources}. We will initially set up the computation more generally including a Newton potential as this is interesting and not more difficult than taking $\Phi=0$, i.e. we start with \eqref{eq:NPPhi} and \eqref{eq:NPPhi2}.

\subsubsection{Energy-momentum tensors for non-relativistic particles}

We pause our discussion of flat NC space-time briefly to use the opportunity to study the various notions of energy-momentum tensors defined in section \ref{subsec:EMTs} for the case of the point particle \eqref{eq:particleaction}. By varying \eqref{eq:particleaction} with respect to $\hat v^\mu$, $h^{\mu\nu}$ and $\tilde\Phi$ using \eqref{eq:variationidentity1} and \eqref{eq:variationidentity2} we obtain for $T^\mu{}_\nu$ the result
\begin{equation}
T^\mu{}_\nu = -P_\nu\dot x^\mu\,,
\end{equation}
where we used the completeness relation given at the end of \eqref{eq:relations} after we computed the components of $T^\mu{}_\nu$ with the help of \eqref{eq:variation3}. In this expression $P_\mu$ is the generalized momentum defined by 
\begin{equation}
P_\mu=\frac{\partial L}{\partial\dot x^\mu}=-\frac{m}{2}\tau_\mu\frac{h_{\rho\sigma}\dot x^\rho \dot x^\sigma}{(\tau_\alpha\dot x^\alpha)^2}+m\frac{h_{\mu\nu}\dot x^\nu}{\tau_\alpha\dot x^\alpha}-m M_\mu\equiv p_\mu-mM_\mu\,,
\end{equation}
where $L$ is given in \eqref{eq:particleaction} and where we also defined the linear momentum $p_\mu$. Next we compute the objects $\mathcal{T}_\mu$ and $\mathcal{T}_{\mu\nu}$ using either \eqref{eq:relationEMTS1} and \eqref{eq:relationEMTS2} or by directly varying with respect to $v^\mu$, $h^{\mu\nu}$ and $M_\mu$ using \eqref{eq:variation4}. The result is 
\begin{eqnarray}
\mathcal{T}_\mu & = & -\tau_\alpha\dot x^\alpha p_\mu\,,\\
\mathcal{T}_{\mu\nu} & = & 2\tau_{(\mu}h_{\nu)\rho}\dot x^\rho v^\sigma p_\sigma-h^\rho{}_{(\mu}h_{\nu)\sigma}\dot x^\sigma p_\rho\,.
\end{eqnarray}
As before we can fix $X$ (see eq.~\eqref{eq:relationEMTS2}) by demanding that $v^\mu v^\nu \mathcal{T}_{\mu\nu}=0$. Further we can choose $\tau_\alpha\dot x^\alpha=1$ by fixing the world-line reparametrization freedom. The current $T^\mu$ is simply given by
\begin{equation}
T^\mu=-m\dot x^\mu\,.
\end{equation}
We see that the difference between $T^\mu{}_\nu$ and $\mathcal{T}_\mu$, $\mathcal{T}_{\mu\nu}$ is that $T^\mu{}_\nu$ depends on $M_\mu$ whereas $\mathcal{T}_\mu$, $\mathcal{T}_{\mu\nu}$ do not. In other words on a flat NC space-time with a Newton potential $\Phi$, i.e. assuming \eqref{eq:flatNC}, \eqref{eq:NPPhi} and \eqref{eq:NPPhi2}, the energy momentum tensor $T^\mu{}_\nu$ depends on $\Phi$ while $\mathcal{T}_\mu$, $\mathcal{T}_{\mu\nu}$ only depend on the properties of the particle. For example $v^\mu\mathcal{T}_\mu$ gives the kinetic energy of the particle whereas $-v^\mu\tau_\nu T^\nu{}_\mu$ gives the kinetic plus gravitational potential energy of the particle.

\subsection{Residual coordinate transformations of flat NC space-time}\label{sec:residualtrafos}

As said we start by asking what are the transformations among \eqref{eq:trafosources} that leave \eqref{eq:flatNC}, \eqref{eq:NPPhi} and \eqref{eq:NPPhi2} invariant. Substituting  \eqref{eq:flatNC}, \eqref{eq:NPPhi} and \eqref{eq:NPPhi2} into \eqref{eq:trafosources} (where the transformation of $\chi$ is irrelevant as we work here with $M_\mu$ and not with $\tilde m_\mu$) and demanding that we get zero gives
\begin{equation}
\begin{array}{rcl}
0=\delta\tau_t & = &  \partial_t\xi^t+z\Lambda_D\,,\\
0=\delta\tau_i & = & \partial_i\xi^t\,,\\
0=\delta v^i & = & \partial_t\xi^i+\lambda^i\,,\\
0=\delta v^t & = & \partial_t\xi^t+z\Lambda_D\,,\\
0=\delta h^{tt} & = & -2h^{\rho t}\partial_\rho\xi^t\,,\\
0=\delta h^{ti} & = & -\delta^{ij}\partial_j\xi^t\,,\\
0=\delta h^{ij} & = & -\delta^{jk}\partial_k\xi^i-\delta^{ik}\partial_k\xi^j-2\Lambda_D \delta^{ij}\,.
\end{array}
\end{equation}
This leads to
\begin{eqnarray}
\Lambda_D & = & -\frac{1}{z}\partial_t\xi^t\qquad\text{with $\xi^t=\xi^t(t)$}\,,\label{eq:LambdaD}\\
\lambda^i & = & -\partial_t\xi^i\,,\label{eq:lambda-i}\\
0 & = & \partial_i\xi_j+\partial_j\xi_i+2\Lambda_D\delta_{ij}\,.\label{eq:partialxi}
\end{eqnarray}

Continuing with the conditions for $M_\mu$ we first consider
\begin{equation}\label{eq:partialideltaM}
\delta M_i=\partial_i\delta M=\partial_i\left( \xi^t\partial_t M+\xi^j\partial_j M+(2-z)\Lambda_D M\right)+\lambda_i\,,
\end{equation}
which follows from $\delta M_\mu$ given in \eqref{eq:trafosources} together with \eqref{eq:LambdaD}--\eqref{eq:partialxi}. We conclude from this that we need
\begin{equation}\label{eq:partialiF2}
\lambda_i=\partial_iF\,,
\end{equation} 
so that \eqref{eq:lambda-i} implies that
\begin{equation}\label{eq:partialiF}
\partial_iF= -\partial_t\xi_i\,,
\end{equation}
leading to $\partial_i\partial_jF= -\partial_t\partial_j\xi_i$ so that
\begin{equation}
\partial_i\partial_t\xi_j-\partial_j\partial_t\xi_i=0\,.
\end{equation}
Differentiating \eqref{eq:partialxi} with respect to $t$ and using \eqref{eq:partialiF} we get
\begin{equation}
\partial_i\partial_j F = \partial_t\Lambda_D\delta_{ij}\,,
\end{equation}
which can be integrated to 
\begin{equation}\label{eq:F}
F=A(t)+B_i(t)x^i+\frac{1}{2}\partial_t\Lambda_D x^ix^i\,,
\end{equation}
where $A$ and $B_i$ are arbitrary functions of $t$. Equations \eqref{eq:partialiF2} and \eqref{eq:partialiF} become
\begin{equation}
B_i(t)+\partial_t\Lambda_D x_i= -\partial_t\xi_i=\lambda_i\,.
\end{equation}
By integration over $t$ we obtain for $\xi^i$ the expression
\begin{equation}
\xi^i=-\int^t dt' B^i(t')-\Lambda_D(t) x^i+A^i(x)\,,
\end{equation}
where \eqref{eq:partialxi} implies
\begin{equation}
\partial_iA_j+\partial_jA_i=0\,,
\end{equation}
so that
\begin{equation}
A_i=a_i+\lambda_{ij} x^j\,,
\end{equation}
with $a_i$ and $\lambda_{ij}=-\lambda_{ji}$ constants.

We thus have now for the local parameters $\Lambda_D$, $\lambda^i$ and $\xi^\mu$ the following conditions
\begin{eqnarray}
\Lambda_D & = & -\frac{1}{z}\partial_t\xi^t\,,\label{eq:LambdaD(t)}\\
\lambda^i & = & -\partial_t\xi^i\,,\label{eq:lambdi-xi}\\
\xi^t & = & \xi^t(t)\,,\\
\xi^i & = & -\int^t dt' B^i(t')+ a^i+\lambda^i{}_j x^j-\Lambda_D(t)x^i\,.\label{eq:aeg}
\end{eqnarray}
Further from equations \eqref{eq:partialideltaM}, \eqref{eq:partialiF2} and \eqref{eq:F} we find for $\delta M$ 
\begin{equation}\label{eq:deltaM-ABLambdaD}
\delta M = \xi^t\partial_t M+\xi^j\partial_j M+(2-z)\Lambda_D M+A(t)+B_i(t)x^i+\frac{1}{2}\partial_t\Lambda_D x^ix^i\,.
\end{equation}
We still have the condition $M_t=\Phi+\partial_tM$. Using $\delta M_\mu$ given in \eqref{eq:trafosources} together with \eqref{eq:LambdaD}--\eqref{eq:partialxi} and \eqref{eq:deltaM-ABLambdaD} we find that the Newton potential transforms as
\begin{equation}\label{eq:deltaPhi-ABLambdaD}
\delta \Phi = \xi^\rho\partial_\rho \Phi+2(1-z)\Lambda_D\Phi-\partial_tA(t)-\partial_tB_i(t)x^i-\frac{1}{2}\partial^2_t\Lambda_D x^ix^i+(z-2)(\partial_t\Lambda_D)M\,,
\end{equation}
where $\xi^\mu$ and $\Lambda_D$ are given in \eqref{eq:LambdaD(t)}--\eqref{eq:aeg}. This includes the acceleration extended Galilei symmetries (see e.g. \cite{Andringa:2013mma}) but also transformations under dilatations and special conformal transformations when $z=2$ (which correspond to a non-constant time dependent $\Lambda_D$). These transformations are also contained in \cite{Bergshoeff:2014uea} (sections 2.2.3 and 2.3.2) but they were not made explicit there because of different gauge fixing conditions\footnote{We thank Eric Bergshoeff for useful discussions on this point.}. When $(z-2)\partial_t\Lambda_D\neq 0$ there is an additional term in \eqref{eq:deltaPhi-ABLambdaD}. The relevance of this term will be discussed in the next subsection.

Going back to our notion of a flat boundary as defined in the previous subsection we set $\Phi$ equal to zero. In order that this choice is respected by the transformations of our holographic setup we must demand that $\delta\Phi=0$ leading for $(z-2)\partial_t\Lambda_D=0$ to the conditions
\begin{equation}\label{eq:ABLambdaD}
A=-C\,,\qquad B_i=-v_i\,,\qquad\partial^2_t\Lambda_D=0\,,
\end{equation}
where $C$ and $v^i$ are constants and for $(z-2)\partial_t\Lambda_D\neq 0$ to the condition
\begin{equation}\label{eq:deltaPhi-ABLambdaD-II}
\partial_tA(t)+\partial_tB_i(t)x^i+\frac{1}{2}\partial^2_t\Lambda_D x^ix^i=(z-2)(\partial_t\Lambda_D)M\,.
\end{equation}
We will now summarize the results regarding the residual coordinate transformations of flat NC space-time.

\subsubsection{Summary}

Consider first the case $(z-2)\partial_t\Lambda_D=0$. Using \eqref{eq:LambdaD(t)}--\eqref{eq:deltaM-ABLambdaD} as well as \eqref{eq:ABLambdaD} we conclude that the conditions \eqref{eq:flatNC} and \eqref{eq:flatNC-V}, which are necessary for a flat NC space-time, are preserved by the following local transformations of our holographic model
\begin{eqnarray}
\Lambda_D & = & -\lambda-\delta_{z,2}c t\,,\label{eq:LambdaDagain}\\
\xi^t & = & a+z\lambda t+\delta_{z,2}c t^2\,,\label{eq:xi-t}\\
\xi^i & = & v^it+a^i+\lambda^i{}_j x^j+\lambda x^i+\delta_{z,2}c t x^i\,,\label{eq:xi-i}\\
\lambda^i & = & -v^i-\delta_{z,2}c x^i\,,\label{eq:lambdai}
\end{eqnarray}
with $M$ transforming as
\begin{equation}\label{eq:deltaM}
\delta M=\xi^t\partial_t M+\xi^i\partial_i M-(2-z)\lambda M-C-v^i x^i-\frac{1}{2}\delta_{z,2}c x^ix^i\,.
\end{equation}

The finite versions of these transformations are
\begin{equation}\label{eq:finiteresidualtrafos}
\begin{array}{rclcrclcrcl}
M'(x) & = & M(x)+C &&&&&&&&\\
t' & = & t+a &&&&&& M'(x') & = & M(x)\\
x'^i & = & x^i+a^i &&&&&& M'(x') & = & M(x)\\
x'^i & = & R^i{}_j x^j &&&&&& M'(x') & = & M(x)\\
t' & = & \lambda^z t && x'^i & = & \lambda x^i && M'(x') & = & \lambda^{2-z}M(x)\\
x'^i & = & x^i+v^i t && t' & = & t  && M'(x') & = & M(x)-\frac{1}{2}v^iv^it+v^i x^i
\end{array}
\end{equation}
where $R^i{}_j R^j{}_k=\delta^i_k$. For $z=2$ we also have the special conformal $(K)$ transformation
\begin{equation}\label{eq:Ktrafo}
t' =\frac{t}{1-ct}\,,\qquad x'^i=\frac{x^i}{1-ct}\,,\qquad M'(x')=M(x)+\frac{c}{2}\frac{x^ix^i}{1-ct}\,.
\end{equation}
To go back to the infinitesimal versions note that we use $x'^\mu-x^\mu=\xi^\mu$ and $M(x)-M'(x)=\delta M$. For some parameters we use the same symbol for the finite and infinitesimal transformations.

When $(z-2)\partial_t\Lambda_D\neq 0$ we conclude that the residual transformations are \eqref{eq:LambdaD(t)}--\eqref{eq:deltaM-ABLambdaD} where the functions $A(t)$, $B_i(t)$ and $\Lambda_D$ must obey \eqref{eq:deltaPhi-ABLambdaD-II}.

In order to get a feeling of the role of the $M$-changing residual coordinate transformations we will now study the toy models of section \ref{subsec:toymodels} on flat NC backgrounds.

\subsection{Scale invariant field theories on flat NC backgrounds}\label{subsec:FTonflatNC}

In section \ref{sec:FT} we studied field theories on general TNC geometries. In this section we will take a closer look at the case of ($z=2$) scale invariant field theories on a flat NC space-time and study in particular the role played by $M$. To this end we consider the models \eqref{eq:example6} and \eqref{eq:Lifshitzmodel}.

If we specify our background to a flat NC space-time as given in \eqref{eq:flatNC} and \eqref{eq:flatNC-V} the action \eqref{eq:example6} becomes
\begin{eqnarray}
S & = & \int d^{d+1}x\left(-\varphi^2\left[\partial_t\left(\theta+M\right)+\frac{1}{2}\partial_i\left(\theta+M\right)\partial^i\left(\theta+M\right)+a\partial_i\partial^i\left(\theta+M\right)\right]\right.\nonumber\\
&&\left.-\frac{1}{2}\partial_i\varphi\partial^i\varphi-V_0\varphi^{\tfrac{2(d+2)}{d}}\left(1+b\theta^2\right)\right)\,,\label{eq:Schmodel}
\end{eqnarray}
where we discarded the term $2a\int d^{d+1}x\varphi\partial_t\varphi$ (coming from the first term in the last line of \eqref{eq:example6}) since it is a boundary term and where we have added a $\theta$ shift symmetry breaking term to the potential. Further we will also consider the Lifshitz model \eqref{eq:Lifshitzmodel} which upon substituting \eqref{eq:flatNC} and \eqref{eq:flatNC-V} reads
\begin{equation}\label{eq:Lifmodel}
S=\int d^{d+1}x\left[\frac{1}{2}\left(\partial_t\phi+\partial^iM\partial_i\phi\right)^2-\frac{\lambda}{2}\left(\partial_i\partial^i\phi\right)^2\right]\,.
\end{equation}

We now address the question which of the residual transformations \eqref{eq:finiteresidualtrafos} and \eqref{eq:Ktrafo} leave \eqref{eq:Schmodel} and \eqref{eq:Lifmodel} form invariant. For $a=0$ the action \eqref{eq:Schmodel} is invariant under the $K$ transformations \eqref{eq:Ktrafo} with $\varphi$ transforming as $\varphi=(1+ct')^{d/2}\varphi'$. However when $a\neq 0$ in the action \eqref{eq:Schmodel} or when we consider the action \eqref{eq:Lifmodel} the $K$ transformations \eqref{eq:Ktrafo} are no longer local symmetries. With respect to the transformations \eqref{eq:finiteresidualtrafos} both models \eqref{eq:Schmodel} (with arbitrary $a$ and $b$) and \eqref{eq:Lifmodel} remain form invariant. The field $\theta$ transforms as a scalar with zero dilatation weight under these residual coordinate transformations while $\varphi$ in \eqref{eq:Schmodel} and $\phi$ in \eqref{eq:Lifmodel} transform as $\varphi=\lambda^{d/2}\varphi'$ and $\phi=\lambda^{(d-2)/2}\phi'$, respectively\footnote{The models \eqref{eq:Schmodel} with $b\neq 0$ and \eqref{eq:Lifmodel} both correspond to Lifshitz invariant field theories, but note the different scaling dimensions of the scalars $\varphi$ and $\phi$. It is much easier to construct interacting Lifshitz invariant theories that are of the type \eqref{eq:Schmodel} with $b\neq 0$, which in fact is an example of an interacting Lifshitz theory, than it is for \eqref{eq:Lifmodel}. The model \eqref{eq:Schmodel} has the property that when the interactions are turned off, i.e. $a=b=0$, it becomes Schr\"odinger invariant.}. We will only speak of global symmetries  once we have removed $M$ from the action as we will do shortly.

If the models \eqref{eq:Schmodel} and \eqref{eq:Lifmodel} really correspond to flat space we should be able remove $M$ somehow since we defined flat NC space-time as corresponding to $M=\text{cst}$ together with all other $M$'s that give actions that are identical to the one
with $M=\text{cst}$  by local transformations, i.e. field redefinitions. When $b=0$ in \eqref{eq:Schmodel} this can be done by defining $\tilde\theta=\theta+M$ which is gauge invariant under the local $\alpha$ transformations discussed in section \ref{subsec:toymodels}. If we make this redefinition with $b=0$ we get
\begin{equation}\label{eq:physicalaction}
S = \int d^{d+1}x\left(-\varphi^2\left[\partial_t\tilde\theta+\frac{1}{2}\partial_i\tilde\theta\partial^i\tilde\theta+a\partial^2\tilde\theta\right]-\frac{1}{2}\partial_i\varphi\partial^i\varphi-V_0\varphi^{\tfrac{2(d+2)}{d}}\right)\,.
\end{equation}
It is straightforward to check that this theory has Lifshitz symmetries. Further it also has Galilean boost invariance because
\begin{eqnarray}
t & = & t'\,,\qquad x^i = x'^i-v^it'\,,\\
\tilde\theta & = & \tilde\theta'+\frac{1}{2}v^iv^i t'-v^ix'^i\,,
\end{eqnarray}
leaves the action invariant. However, the special conformal transformation $K$ is broken by $a\neq 0$.

The fact that the model \eqref{eq:Schmodel} with $b=0$ has a local $U(1)$ (whose local parameter is $\alpha$) is what enables us to remove $M$ from the action entirely so that we have no more background fields and we can just work with fields and their transformations. It is the combined effect of the local $U(1)$ ($\alpha$) symmetry and the residual coordinate transformations \eqref{eq:finiteresidualtrafos} and \eqref{eq:Ktrafo} that makes the model whose `physical' field is $\tilde\theta=\theta+M$ Schr\"odinger invariant for $a=0$ and Lifshitz plus Galilean boost invariant for $a\neq 0$. When we speak of flat NC space-time in these models we mean any $M$ that can be generated by \eqref{eq:finiteresidualtrafos} and for $a=0$ even including \eqref{eq:Ktrafo}.

If we consider the model \eqref{eq:Schmodel} with $b\neq 0$ or the Lifshitz model \eqref{eq:Lifmodel} there is no local symmetry that allows us to perform a field redefinition that removes $M$ from the action. Hence in this case flat space corresponds to setting $M=\text{cst}$ and the only residual coordinate transformations are those that preserve this choice of $M$. From \eqref{eq:finiteresidualtrafos} it is clear that these form the Lifshitz group\footnote{We thank Jan de Boer for useful discussions on the roles of Lifshitz symmetries and global $U(1)$ transformations. For example an interesting question is the following. It is clear that the existence of a global $U(1)$ symmetry is a necessary condition for the occurrence of Galilean boost symmetries. Making this a local symmetry relies on how we choose the couplings to the TNC geometry in section \ref{subsec:toymodels} such that $M_\mu$ becomes a gauge field. 
An important question is then how general this mechanism is.  In other words given a global $U(1)$,  can we always promote it to a local $U(1)$ using $M_\mu$ or are there restrictions.  Put yet another way, when does Lifshitz plus a global $U(1)$ imply Galilean boosts?}. In the potential of \eqref{eq:Schmodel} we can also consider $V=V_0\varphi^{\tfrac{2}{d}(2+d)}(1+b\cos^2(c\theta))$ so that we keep a nontrivial discrete shift symmetry while breaking Galilean boosts.

The energy momentum tensor $\mathcal{T}_\mu$, $\mathcal{T}_{\mu\nu}$ of section \ref{subsec:EMTs} for the case of the model \eqref{eq:Schmodel} with $b=0$ is gauge invariant (as shown in \ref{subsec:toymodels}) under the $\alpha$ transformations and thus depends on the field $\tilde\theta$ while the $M$ dependent object $T^\nu{}_\mu$ is better suited for the study of conserved currents (see sections \ref{subsec:diffsandKVs} and \ref{subsec:scaleandCKVs}). When $b\neq 0$ in \eqref{eq:Schmodel} or when we are dealing with \eqref{eq:Lifmodel} on flat NC space-time we have $M_\mu=0$ so that $\mathcal{T}_\mu$, $\mathcal{T}_{\mu\nu}$ and $T^\nu{}_\mu$ become the same object as follows from the relation \eqref{eq:relationsEMTS3}.

The model \eqref{eq:Schmodel} was obtained by putting the action \eqref{eq:example6} on a flat NC background. The action \eqref{eq:example6} has a built-in local $U(1)$ symmetry with gauge connection $M_\mu$ which was very convenient for us to deduce that there must a exist a conserved current $T^\mu$ related to particle number. The terms that are responsible for the extra local $U(1)$ symmetry are the last two terms on the first line of \eqref{eq:example6} involving $\tilde\Phi$ and $e^\mu_a\mathcal{D}_\mu M^a$. 
Since one might wonder how crucial these terms are, we have included appendix \ref{subsec:comments} which address this issue.

\subsection{Orbits of $M$}\label{subsec:orbitsM}

We first consider the case $z=2$. We have seen in the previous subsection that the orbits of $M$, i.e. all $M$ that are equivalent to $M=\text{cst}$ depend on the theory. What is model independent is that for sure the $M$'s in an orbit are related by the transformations \eqref{eq:finiteresidualtrafos} and \eqref{eq:Ktrafo} and that $M=\text{cst}$ is among them. If this is all we assume we obtain the largest possible orbit of $M$. As we have seen in the previous section this is the orbit that underlies the Schr\"odinger invariant theories (see the $a=b=0$ version of \eqref{eq:Schmodel}). If we take $a\neq 0$ the orbit shrinks because \eqref{eq:Ktrafo} or no longer allowed and when $b\neq 0$ it collapses to the point $M=\text{cst}$. In this section we will study the largest orbit, in more detail. The results will prove useful later when we look at Lifshitz space-times.

Using the finite transformations \eqref{eq:finiteresidualtrafos} and \eqref{eq:Ktrafo} we can by starting with $M=\text{cst}$ generate a function that is at most quadratic in $x^i$ where the quadratic piece is a trace by which we mean a term of the form $x^ix^i$, i.e. $M$ will be of the form
\begin{equation}\label{eq:formM}
M=a(t)+b_i(t)x^i+c(t)x^ix^i\,.
\end{equation}
However the time dependence of the coefficients is not arbitrary. The time dependence is fixed by the following observation. The solution $M=\text{cst}$ and all other $M$ obtained from this by performing the residual transformations \eqref{eq:xi-t}, \eqref{eq:xi-i} and \eqref{eq:deltaM} are solutions of the equation
\begin{equation}\label{eq:noNP}
\tilde\Phi=\partial_tM+\frac{1}{2}\partial_iM\partial^iM=0\,.
\end{equation}
Put another way, acting on any solution of \eqref{eq:noNP} with a Schr\"odinger transformation of the form \eqref{eq:xi-t}, \eqref{eq:xi-i} and \eqref{eq:deltaM} leads to another solution of equation\footnote{We thank Matthias Blau for pointing this out.} \eqref{eq:noNP}. Equation \eqref{eq:noNP} allows us to fix the time dependence of the form of $M$ given in \eqref{eq:formM} to be either
\begin{equation}\label{eq:quadratic}
M=C+\frac{(x^i-x_0^i)(x^i-x_0^i)}{2(t-t_0)}\,,
\end{equation}
when $c$ in \eqref{eq:formM} is nonzero or
\begin{equation}\label{eq:linear}
M=C-\frac{1}{2}V^iV^it+V^ix^i\,,
\end{equation}
when $c$ in \eqref{eq:formM} is zero. In these expressions $t_0$ and $x_0^i$ are arbitrary constants. There are thus three families of solutions given by i). $M=\text{cst}$,  ii). $M$ is linear in $x^i$ as in \eqref{eq:linear} and iii). $M$ is (trace) quadratic in $x^i$ as in \eqref{eq:quadratic}. Equation \eqref{eq:formM} is equivalent to the following differential equations for $M$
\begin{eqnarray}
0 & = & \partial_i\partial_j\partial^j M\,,\label{eq:condition1}\\
0 & = & \partial_i\partial_j M-\frac{1}{d}\delta_{ij}\partial_k\partial^k M\,.\label{eq:condition2}
\end{eqnarray}
We conclude that a complete specification in terms of differential equations of the functions $M$ that are related to $M=\text{cst}$ via the residual transformations \eqref{eq:finiteresidualtrafos} and \eqref{eq:Ktrafo} is given by \eqref{eq:noNP}, \eqref{eq:condition1} and \eqref{eq:condition2}. These differential equations allow us to treat all three cases of functions $M$ in a uniform manner. 

As a curiosity we mention that when \eqref{eq:condition1} (but not necessarily \eqref{eq:condition2}) holds there is a map from solutions to \eqref{eq:noNP} to solutions of the Schr\"odinger equation. This follows from the fact that\footnote{By zero we mean up to possible delta functions on the right hand side as for example the function $\exp[\frac{i}{2}M-\frac{1}{2}\int^t dt'\partial^2M]$ for $M$ given by \eqref{eq:quadratic} is the Green's function of the free particle Schr\"odinger equation, see e.g. \cite{Henkel:1993sg,Henkel:2003pu}.}
\begin{equation}
\left(i\partial_t+\partial^2\right)\exp[\frac{i}{2}M-\frac{1}{2}\int^t dt'\partial^2M]=0\,,
\end{equation}
for any $M$ satisfying \eqref{eq:noNP} and \eqref{eq:condition1} where $\partial^2=\partial_i\partial^i$. This includes solutions to \eqref{eq:noNP} and \eqref{eq:condition1} that are not in the $M=\text{cst}$ orbit because they do not solve \eqref{eq:condition2}. An example of such a function $M$ is
\begin{equation}\label{eq:anotherorbit1}
M = \frac{x^2}{2t}\,,
\end{equation}
where $x$ denotes a single coordinate and not the vector $x^i$. This solution does not obey \eqref{eq:condition2}. To see how the Schr\"odinger transformations \eqref{eq:xi-t}, \eqref{eq:xi-i} and \eqref{eq:deltaM} generate new solutions to \eqref{eq:noNP} and \eqref{eq:condition1} one can check that there is another solution to \eqref{eq:noNP} and \eqref{eq:condition1} that is in the same orbit as \eqref{eq:anotherorbit1} given by
\begin{equation}\label{eq:anotherorbit2}
M = \frac{x^2}{2t}-\frac{c}{2}\frac{y^2}{1-ct}\,,
\end{equation}
where we took $d=2$. This is obtained by acting on \eqref{eq:anotherorbit1} with the $K$ transformation \eqref{eq:Ktrafo}. Since the functions \eqref{eq:anotherorbit1} and \eqref{eq:anotherorbit2} are not in the $M=\text{cst}$ orbit they do not correspond to a flat NC space-time. 

We now turn to  the orbits of $M$ when $z\neq 2$. In section \ref{sec:residualtrafos} we showed that for $\partial_t\Lambda_D=0$ the residual transformations are given by \eqref{eq:finiteresidualtrafos} and that when $\partial_t\Lambda_D\neq 0$ they are given by \eqref{eq:LambdaD(t)}--\eqref{eq:deltaM-ABLambdaD} subject to \eqref{eq:deltaPhi-ABLambdaD-II}. The latter requirement tells us that again $M$ can be at most quadratic in $x^i$ so that it is of the form \eqref{eq:formM}. We conclude that \eqref{eq:condition1} and \eqref{eq:condition2} also apply to the case $z\neq 2$. 

In the $z=2$ case we were able to conclude, by using the $z=2$ Schr\"odinger transformations, that $\tilde\Phi$ must vanish and that therefore \eqref{eq:noNP} has to be obeyed. However we could have derived it in another way as well which uses an argument that is valid for all $z$ and that goes as follows. The function $\tilde\Phi$ is a scalar with dilatation weight $2(z-1)$ under all local transformations of our model namely \eqref{eq:trafosources}, i.e. it transforms as
\begin{equation}
\delta\tilde\Phi=\xi^\rho\partial_\rho\tilde\Phi+2(1-z)\Lambda_D\tilde\Phi\,,
\end{equation}
which follows from its definition \eqref{eq:tildePhi} and \eqref{eq:trafosources}. For the $M=\text{cst}$ orbit $\tilde\Phi$ is zero because $M_\mu=0$ so that $\delta\tilde\Phi=0$. Hence $\tilde\Phi$ vanishes for all solutions of the $M=\text{cst}$ orbit because these are generated by the $\delta\tilde\Phi$. We conclude that for any $M$ in our $M=\text{cst}$ orbit it must be that $\tilde\Phi=0$. Hence also when $z\neq 2$ we must obey \eqref{eq:noNP}. Therefore for $z\neq 2$ the function $M$ must obey the same set of equations, namely \eqref{eq:noNP}, \eqref{eq:condition1} and \eqref{eq:condition2} as for $z=2$. 

It is crucial that for each $M$ which solves these three equations we can find a residual transformation that makes it equivalent to $M=\text{cst}$. Going between $M=\text{cst}$ and a linear $M$ of the form \eqref{eq:linear} is achieved by \eqref{eq:finiteresidualtrafos} with $z\neq 2$. For transformations from $M=\text{cst}$ or the linear $M$ of \eqref{eq:linear} to the trace quadratic $M$ of \eqref{eq:quadratic} we need to use \eqref{eq:LambdaD(t)}--\eqref{eq:deltaM-ABLambdaD} subject to \eqref{eq:deltaPhi-ABLambdaD-II}. We can solve the latter equation separately for the three families of $M$. For example if we take the quadratic $M$ of \eqref{eq:quadratic} we get\footnote{We will not explicitly write the other solutions of \eqref{eq:deltaPhi-ABLambdaD-II} for $M=\text{cst}$ and a linear $M$ and the corresponding residual transformations \eqref{eq:LambdaD(t)}--\eqref{eq:deltaM-ABLambdaD}.}
\begin{eqnarray}
A & = & -\frac{1}{2}c x_0^ix_0^i(t-t_0)^{z-2}\,,\\
B^i & = & c x_0^i (t-t_0)^{z-2}\,,\\
\Lambda_D & = & -\frac{c}{z-1}(t-t_0)^{z-1}\,,
\end{eqnarray} 
in which case we obtain the residual transformation
\begin{eqnarray}
\xi^t & = & \frac{c}{z-1}(t-t_0)^z\,,\label{eq:xitzneq2}\\
\xi^i & = &  \frac{c}{z-1}(x^i-x_0^i)(t-t_0)^{z-1}\,,\label{eq:aegzneq2}\\
\delta M & = & \xi^t\partial_t M+\xi^j\partial_j M+(2-z)\Lambda_D M-\frac{1}{2}c(x^i-x_0^i)(x^i-x_0^i)(t-t_0)^{z-2}\,.\label{eq:deltaM-ABLambdaDzneq2}
\end{eqnarray}
These transformations with parameter $c$ look like a $z\neq 2$ version of a special conformal transformation. However we will see in the next subsection that the generic $\xi^\mu$ in \eqref{eq:xi-t} and \eqref{eq:xi-i} does not combine with the $\xi^\mu$ given in \eqref{eq:xitzneq2} and \eqref{eq:aegzneq2} under the action of the Lie bracket to give a residual transformation. This only works if we set $a=a^i=0$. Hence the residual transformations with $a=a^i=0$ form a group with the $c\neq 0$ transformations (that will be shown to be isomorphic to the Lifshitz algebra in the next subsection) and the residual transformations with $c=0$ form the $z\neq 2$ Schr\"odinger algebra without particle number.

\subsection{Conformal Killing vectors of flat NC space-time}\label{subsec:flatNCKillingvectors}

Now that we have the residual transformations of flat NC space-time at our disposal, namely \eqref{eq:LambdaDagain}--\eqref{eq:deltaM} and \eqref{eq:xitzneq2}--\eqref{eq:deltaM-ABLambdaDzneq2}, we can ask which of these transformations correspond to conformal Killing vectors. Since $M$ is the only field left that is still transforming we simply need to set $\delta M=0$. In this section we will show that we can get the same answer by solving the TNC conformal Killing equations \eqref{eq:KE1}--\eqref{eq:KE5}. To this end we substitute \eqref{eq:flatNC} and \eqref{eq:flatNC-V} into the TNC conformal Killing equations.

Substituting the above choices into \eqref{eq:KE1} we get
\begin{eqnarray}
\partial_t K^t & = & -z\Omega\,,\label{eq:solutionKE1}\\
\partial_i K^t & = & 0\,.
\end{eqnarray}
It follows that 
\begin{equation}
\partial_i\Omega=0\,.
\end{equation}
Doing the same with equation \eqref{eq:KE4} we find
\begin{equation}\label{eq:solutionKE2}
\partial_i K_j+\partial_j K_i=-2\Omega\delta_{ij}\,.
\end{equation}
The $t$ component of equation \eqref{eq:KE4} is equivalent to \eqref{eq:solutionKE1} while the $i$ component leads to 
\begin{equation}\label{eq:solutionKE3}
\partial_t K_i=\partial_i\left(\mathcal{L}_K M-(z-2)\Omega M\right)\,.
\end{equation}
The most general solution to \eqref{eq:solutionKE2} can be written as
\begin{equation}\label{eq:generalformKi}
K^i=A^i(t)+\lambda^i{}_j(t)x^j-\Omega(t)x^i\,,
\end{equation}
where $\lambda^i{}_j(t)$ is antisymmetric so that 
\begin{equation}\label{eq:solutionKE4}
\partial_t K^i=\partial_t A^i(t)+\partial_t\lambda^i{}_j(t)x^j-\partial_t\Omega x^i\,,
\end{equation}
Differentiating \eqref{eq:solutionKE4} with respect to $x^j$ and using \eqref{eq:solutionKE3} to establish symmetry in $i$ and $j$ we get
\begin{equation}\label{eq:cstrotations}
\partial_t\lambda^i{}_{j}=0\,.
\end{equation}
Using \eqref{eq:solutionKE4} with \eqref{eq:cstrotations} on the left hand side of equation \eqref{eq:solutionKE3} we can integrate the right hand side of \eqref{eq:solutionKE3} to obtain
\begin{equation}\label{eq:solutionKE5}
\mathcal{L}_{K}M = x^i\partial_t A^i-\frac{1}{2}x^ix^i\partial_t\Omega+(z-2)\Omega M+C(t)\,,
\end{equation}
where $C(t)$ is an arbitrary function of $t$. The $ti$ and $ij$ components of equation \eqref{eq:KE3} give nothing new but the $tt$ component tells us that
\begin{equation}\label{eq:solutionKE6}
\partial_t\left(\mathcal{L}_{K}M\right)=(z-2)\Omega\partial_t M\,.
\end{equation}
Substituting \eqref{eq:solutionKE5} into \eqref{eq:solutionKE6} we find
\begin{equation}\label{eq:solutionKE7}
x^i\partial^2_t A^i-\frac{1}{2}x^ix^i\partial^2_t\Omega+(z-2)\partial_t\Omega M+C'(t)=0\,.
\end{equation}
Equation \eqref{eq:KE2} gives nothing new.

When $(z-2)\partial_t\Omega=0$ equation \eqref{eq:solutionKE7} is solved by 
\begin{eqnarray}
A^i & = & a^i+v^it\,,\\
\Omega & = & -\lambda-\delta_{z,2}c t\,,\\
C & = & \text{cst}\,,
\end{eqnarray}
and equation \eqref{eq:solutionKE5} becomes
\begin{equation}\label{eq:solutionKE8}
\mathcal{L}_{K}M = v^ix^i+\frac{1}{2}\delta_{z,2}c x^ix^i+(2-z)\lambda M+C\,,
\end{equation}
and from \eqref{eq:solutionKE1} and \eqref{eq:generalformKi} we see that the Killing vectors become
\begin{eqnarray}
K^t & = & a+z\lambda t+\delta_{z,2}c t^2\,,\label{eq:KV1}\\
K^i & = & a^i+v^i t+\lambda^i{}_j x^j+\lambda x^i+\delta_{z,2}c tx^i\,,\label{eq:KV2}
\end{eqnarray}
provided we can solve \eqref{eq:solutionKE8}. We thus see that the conformal Killing vectors $K^\mu$ agree with the residual diffeomorphisms $\xi^\mu$ given in \eqref{eq:xi-t} and \eqref{eq:xi-i} whenever $M$ is such that \eqref{eq:deltaM} vanishes.

Next we consider the case $(z-2)\partial_t\Omega\neq 0$. As we saw in the previous subsection there are three families of functions $M$. If we take $M=\text{cst}$ and a linear $M$ and we substitute this into
\eqref{eq:solutionKE5} and \eqref{eq:solutionKE7} it follows that $\partial_t\Omega=0$ and so these cases have already been covered. However if we take the quadratic $M$ of \eqref{eq:quadratic} we find a new solution to \eqref{eq:solutionKE5} and \eqref{eq:solutionKE7} which reads
\begin{eqnarray}
\Omega & = & -(t-t_0)^{z-1}\,,\\
K^t & = & (t-t_0)^z\,,\\
K^i &= & (x^i-x_0^i)(t-t_0)^{z-1}\,.
\end{eqnarray}
This can also be found by setting $\delta M=0$ in equation \eqref{eq:deltaM-ABLambdaDzneq2} with $M$ given by \eqref{eq:quadratic}.

What we find is that for each of the three families of $M$ the Killing vectors that obey \eqref{eq:solutionKE5} and \eqref{eq:solutionKE7} always form the Lifshitz algebra. 
\begin{eqnarray}
M & = & \text{cst}\hspace{4.9cm}H, D, P_i, J_{ij}\,,\label{eq:constantM}\\
M & = & \frac{x^2+y^2}{2t}\hspace{4cm}K, D, G_i, J_{ij}\,,\label{eq:quadraticM}\\
M & = & -\frac{1}{2}V^iV^it+V^ix^i\hspace{2.4cm}H, D, P_i, J_{ij}\,,
\end{eqnarray}
where $V^i$ is some constant velocity and where we set $t_0=0=x^i_0$ in \eqref{eq:quadratic}. The last one requires
\begin{eqnarray}
C & = & -\frac{1}{2}V^iV_ia+V^ia_i\,,\\
v_j & = & V^i\lambda_{ij}+(z-1)\lambda V_j\,,
\end{eqnarray}
in \eqref{eq:solutionKE8} for a KV of the form
\begin{equation}
aH+a^iP_i+\frac{1}{2}\lambda^{ij}J_{ij}+\lambda D\,.
\end{equation}
The Killing vectors $H$, $P_i$, $G_i$, $J_{ij}$, $D$, $K$ are given by
\begin{equation}\label{eq:HDGJDK-KV}
\begin{array}{rclccrcl}
H & = & \partial_t\,,&&& P_i & = & \partial_i\,,\\
G_i & = & t\partial_i\,,&&& J_{ij} & = & x_i\partial_j-x_j\partial_i\,,\\
D & = & zt\partial_t+x^i\partial_i\,,&&& K & = &  t^z\partial_t+t^{z-1}x^i\partial_i\,,
\end{array}
\end{equation}
where the latter requires $\Omega=- t^{z-1}$. For $z\neq 2$ this has the property $\partial_t\Omega\neq 0$. 

\subsubsection{A local realization of the Schr\"odinger algebra on $M$}

We can use these vectors to generate the maximal orbit studied in section \ref{subsec:orbitsM}. To this end we denote by $\tilde N$ the operator that shifts $M$ by a constant (see equation \eqref{eq:diffstype3}). We can write down the following Schr\"odinger algebra of vectors
\begin{equation}\label{eq:Schgens}
\begin{array}{rclccrcl}
H & = & \partial_t\,,&&& P_i & = & \partial_i\,,\\
G_i & = & t\partial_i+x_i\tilde N\,,&&& J_{ij} & = & x_i\partial_j-x_j\partial_i\,,\\
D & = & zt\partial_t+x^i\partial_i\,,&&&&&
\end{array}
\end{equation}
where for $z=2$ we also have
\begin{equation}\label{eq:Kgen}
K =  t^2\partial_t+tx^i\partial_i+\frac{1}{2}x^ix^i\tilde N\,.
\end{equation}
These generate the transformation of $M$ as given in \eqref{eq:deltaM}. Solving $\delta M=0$ for a given $M$ always leads to a Lifshitz subgroup of the Schr\"odinger algebra. The generators that do not leave $M$ invariant were denoted by $L^\mu$ below \eqref{eq:type2symmetry-3}. In order that these orbit generators become global symmetries we need to consider couplings to TNC geometries for which $M_\mu$ becomes a gauge connection as explained in sections \ref{subsec:toymodels} and \ref{subsec:FTonflatNC}.

We have not studied any field theory models with $z\neq 2$, so it is more difficult to say what happens in that case. Again, when $z\neq 2$ we get a Lifshitz algebra of conformal Killing vectors for every choice of $M$. It would be interesting to extend the analysis of section \ref{subsec:FTonflatNC} to the $z\neq 2$ cases and see how the Galilean boosts can be added. In particular it would be interesting to understand the status of the $K$ transformation which for $z\neq 2$ cannot be added to the Schr\"odinger algebra.

\section{The Lifshitz vacuum}\label{sec:Lifspace}

In section \ref{sec:holoLif} we have shown that the sources in Lifshitz holography transform under a local action of the Schr\"odinger algebra. Here we will derive this for the case of an exact Lifshitz space-time, i.e. the sources that describe the Lifshitz vacuum transform under a local Schr\"odinger group consisting entirely of bulk PBH transformations. The Killing symmetries are always given by a Lifshitz subalgebra of the Schr\"odinger algebra spanned by the PBH generators. In a suitable set of bulk coordinates and seen from the boundary point of view this corresponds to a flat Newton--Cartan geometry whose conformal Killing vectors span the Lifshitz algebra with the Schr\"odinger symmetries being realized locally on the Newton--Cartan vector $M_\mu=\partial_\mu M$.

We have shown by studying field theory on Newton--Cartan geometries in sections \ref{subsec:toymodels} and \ref{subsec:FTonflatNC} that this is the natural way in which field theories realize global Schr\"odinger invariance through a mechanism in which the fields eat up the background field $M$ in such a way that $M$ disappears from the theory. This involves an $M$-dependent field redefinition. The resulting field theory has a global Schr\"odinger symmetry in which always those generators outside a Lifshitz subalgebra are realized as projective transformations. In the example of section \ref{subsec:FTonflatNC} it is the field $\phi=\frac{1}{\sqrt{2}}\varphi e^{i\tilde\theta}$ that transforms projectively under the non-centrally extended Schr\"odinger group, i.e. the Schr\"odinger group without the particle number generator. These are obtained from unitary representations of the centrally extended Schr\"odinger group \cite{Niederer:1972zz,Perroud:1977qh}. This is based on the fact that the unitary irreducible representations of the Galilei group require the central extension to the Bargmann algebra \cite{Bargmann:1954gh}. Here the central element corresponds to shifting $M$ which is not a space-time coordinate. Hence the representations become projective. This is what we see in the case of the toy models of section \ref{subsec:FTonflatNC}. These projective realizations of space-time symmetries cannot be predicted by only looking at Killing vectors. To this end we study probe fields on a $z=2$ Lifshitz background in section \ref{subsec:probes} and show that we can construct probe actions that are invariant under the entire $z=2$ Schr\"odinger algebra. We take this to suggest that holographic realizations of Schr\"odinger invariant field theories involve dynamics on Lifshitz geometries in the bulk. The role of particle number is tied to the manner in which the fields couple to the Newton--Cartan vector $M_\mu=\partial_\mu M$. Before we get to those results we start by explaining how the function $M$ appears in the Lifshitz metric.

\subsection{One Lifshitz metric for all $M$}\label{subsec:examples}

It is well-known that the Lifshitz metric can be written in Poincar\'e type coordinates as
\begin{equation}
ds^2=\frac{dr^2}{r^2}-\frac{dt^2}{r^{2z}}+\frac{1}{r^2}dx^idx^i\,.
\end{equation}
The Killing vectors of this metric agree with \eqref{eq:constantM} where for the dilatations we need to add an obvious $r\partial_r$ to the conformal Killing vector $D$ in \eqref{eq:HDGJDK-KV}. It is thus tempting to suggest that this form of the metric corresponds to $M=\text{cst}$. 
Another possibility is to consider a  trace quadratic $M$. From \eqref{eq:quadraticM} we read off that in that case the boundary conformal Killing vectors are given by $G$, $J$, $D$ and $K$ in \eqref{eq:HDGJDK-KV}. These form a Lifshitz algebra, and we now address
the question how these can be realized in the bulk. We make a naive suggestion which is to add $r$ to the boundary conformal Killing vectors that are not also Killing vectors, i.e. $D$ and $K$ in \eqref{eq:HDGJDK-KV} as if it were another $x^i$ coordinate. That is, following
\cite{Hartong:2014pma} we try
\begin{eqnarray}
G_i & = & t\partial_i\,,\label{eq:G-KVbulk}\\
J_{ij} & = & x_i\partial_j-x_j\partial_i\,,\label{eq:J-KVbulk}\\
D & = & zt\partial_t+x^i\partial_i+r\partial_r,,\label{eq:D-KVbulk}\\
K & = &  t^z\partial_t+t^{z-1}\left(x^i\partial_i+r\partial_r\right)\, . \label{eq:K-KVbulk}
\end{eqnarray}
Imposing that these are the Killing vectors of a metric leads to the following expression
\begin{equation}\label{eq:metricletter2}
ds^2=\left(\frac{dr}{r}-\frac{dt}{t}\right)^2-\frac{dt^2}{r^{2z}}+\frac{1}{r^2}\left(dx^i-\frac{x^i}{t}dt\right)^2\,.
\end{equation}
To see that this is indeed a Lifshitz metric one can use the transformation (for $z=2$)
\begin{equation}\label{eq:nonradialtoradial}
t=-\frac{1}{t'}\,,\qquad r=-\frac{r'}{t'}\,,\qquad x^i=-\frac{x'^i}{t'}\,,
\end{equation}
which brings the metric to the usual form. The general $z$ transformation will be given below.

The metric \eqref{eq:metricletter2} depends on boundary coordinates and it is suggestive to rewrite this in terms of $M=\tfrac{x^ix^i}{2t}$ via $\partial_i M=x^i/t$ and $\partial_i\partial^i M=d/t$. We never need to use time derivatives of $M$ as these are determined via \eqref{eq:noNP} in terms of spatial derivatives. Doing so we get
\begin{equation}\label{eq:LifshitzallM}
ds^2=\left(\frac{dr}{r}-\frac{1}{d}\partial_i\partial^iMdt\right)^2-\frac{dt^2}{r^{2z}}+\frac{1}{r^2}\left(dx^i-\partial^i Mdt\right)^2\,.
\end{equation}
In section \ref{subsec:orbitsM} we have shown that the orbit of $M$ relevant for flat NC space-time contains only three cases: constant, linear and trace quadratic $M$ functions. Hence it may well be that \eqref{eq:LifshitzallM} is indeed a Lifshitz metric for any $M$ in the maximal orbit of section \ref{eq:LifshitzallM}. We will now show this to be the case.

Define
\begin{equation}
e^i=\exp[-\frac{1}{d}\int^t dt'\partial^2M]\left(dx^i-\partial^i Mdt\right)\,.
\end{equation}
One can show that
\begin{equation}
de^i=0
\end{equation}
provided that \eqref{eq:condition1} and \eqref{eq:condition2} hold. Hence we can write
\begin{equation}
dx^i-\partial^i Mdt=\exp[\frac{1}{d}\int^t dt'\partial^2M]dx'^i\,,
\end{equation}
where $x'^i$ are some new coordinates. In order that
\begin{equation}
\frac{1}{r}\left(dx^i-\partial^i Mdt\right)=\frac{1}{r'}dx'^i
\end{equation}
we define
\begin{equation}
r'=r\exp[-\frac{1}{d}\int^t dt'\partial^2M]\,.
\end{equation}
This also turns $\frac{dr}{r}-\frac{1}{d}\partial_i\partial^iMdt$ into $\frac{dr'}{r'}$. Finally in order that $r^{-z}dt=r'^{-z}dt'$ we define
\begin{equation}
dt'=\exp[-\frac{z}{d}\int^t dt'\partial^2M]dt\,.
\end{equation}
We conclude from this that the metric \eqref{eq:LifshitzallM} is pure Lifshitz for any $M$ satisfying \eqref{eq:condition1} and \eqref{eq:condition2} but that \eqref{eq:noNP} is not needed. Further the massive vector field is for any metric of the form \eqref{eq:LifshitzallM} always simply $B=\tfrac{dt}{r^z}$.

In section \ref{sec:holoLif} we defined the sources for asymptotically locally Lifshitz space-times in a (conformally) radial gauge \eqref{eq:gaugemetric}. For an exact Lifshitz space-time $R=1$ and we are in radial gauge. However \eqref{eq:LifshitzallM} is not in radial gauge. Suppose that somehow the off-diagonal $drdt$ term in \eqref{eq:LifshitzallM} was not there. Then we can use the dictionary of section \ref{sec:holoLif}
to read off that the sources are 
\begin{equation}\label{eq:flatNCagain}
\begin{array}{rclccrcl}
\tau_\mu & = & \delta^t_\mu\,,&&&&&\\
h^{tt} & = & h^{ti}=0\,,&&& h^{ij} & = & \delta^{ij}\,,\\
v^\mu & = & -\delta^\mu_t\,,&&&&&\\
h_{tt} & = & h_{ti}=0\,,&&& h_{ij} & = & \delta_{ij}\,,\\
M_\mu & = & \partial_\mu M\,,&&&&&\\
\tilde\Phi & = & 0\,.&&&&&
\end{array}
\end{equation}
where we used \eqref{eq:noNP} to conclude that $M_t=\partial_t M$ and where $M$ obeys \eqref{eq:condition1}--\eqref{eq:condition2}. In the next section we will show that there always exists a coordinate transformation that brings \eqref{eq:LifshitzallM} to radial gauge without modifying the sources. We remind the reader that this is exactly the two step way of viewing a PBH transformation as explained in section \ref{subsec:localtrafosources}. First we perform a boundary dependent rescaling of the radial coordinate possibly together with a boundary diffeomorphism as in \eqref{eq:nonradialtoradial} (corresponding to choose $\Lambda_D$ and $\xi^\mu$ in \eqref{eq:PBH1} and \eqref{eq:PBH2} and then we perform a second transformation which is subleading in that it does not act on the sources that brings the metric back to radial gauge. In the next subsection we construct this transformation for the case $M=\tfrac{x^ix^i}{2t}$. Once we have established it for that case it is straightforward to generalize it to any trace quadratic $M$ as in \eqref{eq:quadratic}. For linear $M$ the metric \eqref{eq:LifshitzallM} is already in radial gauge so there is nothing to do.

\subsection{Coordinate transformation to radial gauge}

Consider the metric \eqref{eq:metricletter2} for $M=x^ix^i/2t$ and $z=2$ with the massive vector given by
\begin{equation}
B=\frac{dt}{r^2}\,.
\end{equation}
We know how to transform this to the standard Lifshitz metric. This goes via the transformation \eqref{eq:nonradialtoradial} leading to
\begin{equation}\label{eq:standardLifshitz}
ds^2=-\frac{dT^2}{R^4}+\frac{dR^2}{R^2}+\frac{1}{R^2}dX^idX^i\,,
\end{equation}
where the massive vector is
\begin{equation}\label{eq:massivevectorstandard}
B=\frac{dT}{R^2}\,,
\end{equation}
where we replaced primed coordinates by capitalized coordinates.

Next perform the following coordinate transformation\footnote{The $(T,R)$ to $(t,r)$ coordinate transformation is an isometry of the AdS$_2$ metric $-\frac{dT^2}{R^4}+\frac{dR^2}{R^2}$.}
\begin{eqnarray}
T & = & -\frac{1}{t}\frac{1}{1-\frac{1}{4}\frac{r^4}{t^2}}\,,\label{eq:coordinatetrafo1}\\
R & = & -\frac{r}{t}\frac{1}{\left(1-\frac{1}{4}\frac{r^4}{t^2}\right)^{1/2}}\,,\\
X^i & = & -\frac{x^i}{t}\,.\label{eq:coordinatetrafo3}
\end{eqnarray}
This leads to the following radial gauge metric
\begin{equation}\label{eq:radialquadraticM}
ds^2=\frac{dr^2}{r^2}-\frac{dt^2}{r^4}+\frac{1}{r^2}\delta_{ij}\left(1-\frac{1}{4}\frac{r^4}{t^2}\right)\left(dx^i-\frac{x^i}{t}dt\right)\left(dx^j-\frac{x^j}{t}dt\right)
\end{equation}
and massive vector
\begin{equation}\label{eq:BforquadraticMradialgaugemetric}
B=\frac{1+\frac{1}{4}\frac{r^4}{t^2}}{1-\frac{1}{4}\frac{r^4}{t^2}}\frac{dt}{r^2}-\frac{\frac{r^2}{t}}{1-\frac{1}{4}\frac{r^4}{t^2}}\frac{dr}{r}\,.
\end{equation}
We have thus obtained a radial gauge metric with the sources corresponding to a flat NC space-time with $M=x^ix^i/2t$. 

We see that close to the boundary at $r=0$ the coordinate transformation \eqref{eq:coordinatetrafo1}--\eqref{eq:coordinatetrafo3} becomes the inverse of \eqref{eq:nonradialtoradial}. In fact the transformation \eqref{eq:nonradialtoradial} is of the form of a $(t,x^i)$-dependent rescaling of the radial coordinate $r$ accompanied by a boundary diffeomorphism which is precisely what a PBH transformation is at leading order (see section \ref{subsec:localtrafosources}). What a PBH transformation does on top of this is that it ensures that the radial gauge form of the metric is preserved. In other words for every $\Lambda_D$ and $\xi^\mu$ that constitute the leading order part of a PBH transformation \eqref{eq:PBH1} and \eqref{eq:PBH2} there exists a trivial bulk diffeomorphism that brings it back to radial gauge. By a trivial bulk diffeomorphism we mean those coordinate transformations that do not act on the sources which therefore are of order $r^2$ and higher in \eqref{eq:PBH1} and \eqref{eq:PBH2}. This is precisely what happens in \eqref{eq:coordinatetrafo1}--\eqref{eq:coordinatetrafo3}; it is a combination of the inverse of \eqref{eq:nonradialtoradial} followed by a trivial bulk diffeomorphism which are subleading in $r$ to maintain the radial gauge form of the metric. Hence the residual coordinate transformations of \eqref{eq:radialquadraticM} act in exactly the same manner on the sources as those of \eqref{eq:metricletter2}\footnote{Since we are dealing with the Lifshitz vacuum there are no vevs turned on. If one defines the vevs via certain coefficients in the near boundary expansion in the gauge \eqref{eq:gaugemetric} it is important to maintain the conformally radial gauge of \eqref{eq:gaugemetric} at least up to orders where the vevs appear in order to find out how they transform under a PBH transformation.}.

The way in which we obtain the coordinate transformation \eqref{eq:coordinatetrafo1}--\eqref{eq:coordinatetrafo3} is as follows. The metric \eqref{eq:metricletter2} has manifest $K$, $G_a$, $D$ and $J_{ab}$ Killing vectors. In radial gauge we have to drop manifest $K$ invariance. We thus make an ansatz for the most general metric with manifest $G_a$, $J_{ab}$ and $D$ Killing vectors. This ansatz is of the form
\begin{equation}\label{eq:radialquadraticMnew}
ds^2=\frac{dr^2}{r^2}-F_1\frac{dt^2}{r^4}+\frac{1}{r^2}\delta_{ij}F_2\left(dx^i-\frac{x^i}{t}dt\right)\left(dx^j-\frac{x^j}{t}dt\right)
\end{equation}
and massive vector
\begin{equation}\label{eq:BforquadraticMradialgaugemetricnew}
B=H_1\frac{dt}{r^2}+H_2\frac{dr}{r}\,,
\end{equation}
with $F_1$, $F_2$, $H_1$ and $H_2$ arbitrary functions of $\tfrac{r^2}{t}$ (as follows from $G_a$, $J_{ab}$, $D$ invariance). The equations \eqref{eq:pureLif-I}--\eqref{eq:pureLif-VI} provide us with a coordinate independent definition of a Lifshitz space-time. We solve equations \eqref{eq:pureLif-I}--\eqref{eq:pureLif-VI} with the boundary condition that $F_1$ and $F_2$ go to unity as $r$ goes to zero. The solution we obtain is \eqref{eq:radialquadraticM}. By comparing \eqref{eq:radialquadraticM} and \eqref{eq:massivevectorstandard} with \eqref{eq:standardLifshitz} and \eqref{eq:BforquadraticMradialgaugemetric} we obtain \eqref{eq:coordinatetrafo1}--\eqref{eq:coordinatetrafo3}. 

One can perform a similar calculation for $z\neq 2$ and the structure of the PBH transformations guarantees that a transformation to radial gauge should exist, so we leave the explicit construction of this transformation for general $z$ for future work.

\subsection{Symmetries of the Lifshitz space-time}\label{subsec:symmetriesLifspace}

In section \ref{sec:residualtrafos} we derived the residual coordinate transformations that preserve the gauge choice in which we defined flat NC space-time. These transformations are \eqref{eq:LambdaDagain}--\eqref{eq:deltaM}. 
We now want to understand what these correspond to from a bulk perspective. The transformations used to derive the residual transformations \eqref{eq:LambdaDagain}--\eqref{eq:deltaM} were \eqref{eq:trafosources} which have been shown in section \ref{subsec:localtrafosources} to correspond to the local bulk transformations that preserve the boundary conditions. Since we can bring \eqref{eq:LifshitzallM} to radial gauge without changing the sources, the bulk duals of the transformations are \eqref{eq:LambdaDagain}--\eqref{eq:deltaM} must be the bulk diffeomorphisms that preserve the form of the metric \eqref{eq:LifshitzallM}. As a check of this statement we will show that this is the case for $z=2$.

The residual bulk diffs are generated by a $\zeta^M$ that obeys
\begin{equation}\label{eq:bulkdiff1}
\begin{array}{rcl}
\delta g_{rr} & = & \mathcal{L}_\zeta g_{rr}=0\,,\\
\delta g_{rt} & = & \mathcal{L}_\zeta g_{rt}=-\frac{1}{d}\frac{1}{r}\partial^2\delta M\,,\\
\delta g_{ri} & = & \mathcal{L}_\zeta g_{ri}=0\,,\\
\delta g_{it} & = & \mathcal{L}_\zeta g_{it}=-\frac{1}{r^2}\partial_i\delta M\,,\\
\delta g_{ij} & = & \mathcal{L}_\zeta g_{ij}=0\,,\\
\delta g_{tt} & = & \mathcal{L}_\zeta g_{tt}=\frac{2}{r^2}\partial^iM\partial_i\delta M+\frac{2}{d^2_s}\partial^2M\partial^2\delta M\,.
\end{array}
\end{equation}
Further we need to demand that the conditions \eqref{eq:condition1} and \eqref{eq:condition2} that make the metric Lifshitz are preserved, meaning we impose
\begin{eqnarray}
\partial_i\partial^2\delta M & = & 0\,,\label{eq:constrainedvar1}\\
\partial_i\partial_j\delta M-\frac{1}{d}\delta_{ij}\partial^2\delta M & = & 0\,.\label{eq:constrainedvar2}
\end{eqnarray} 
Finally, on the boundary we have imposed the conditions $\Gamma^\rho_{\mu\nu}=0$ and $\tilde\Phi=0$. This means that we need to preserve \eqref{eq:noNP} as well which means
\begin{equation}\label{eq:bulkdiff9}
\partial_t\delta M+\partial^i M\partial_i\delta M=0\,.
\end{equation}
Solving \eqref{eq:bulkdiff1}--\eqref{eq:bulkdiff9} leads to
\begin{eqnarray}
\zeta^r & = & -r\Lambda_D(t)\,,\qquad\zeta^\mu=\xi^\mu\,,\label{eq:xi^r}\\
\xi^t & = & \xi^t(t)\qquad\text{such that $\partial_t\xi^t=-2\Lambda_D$}\,,\\
\partial_i\xi_j+\partial_j\xi_i & = & -2\delta_{ij}\Lambda_D\,,\\
\partial_t\xi_i & = & -\partial_iF\,,\\
\delta M & = & \xi^\mu\partial_\mu M+F\qquad\text{such that $\partial_tF=0$}\,,\label{eq:deltaM-bulk}
\end{eqnarray}
where $F$ is defined in section \ref{sec:residualtrafos}, see around equation \eqref{eq:partialiF}. The combination $\xi^\mu\partial_\mu M+F$ was called $\tilde\sigma$ in \eqref{eq:diffstype3}. The solution to these equations is exactly given by \eqref{eq:LambdaDagain}--\eqref{eq:xi-i} and \eqref{eq:deltaM}. In order to obtain the result \eqref{eq:lambdai} one must demand that the local Galilean boosts only affect $M_\mu$ and not $h_{\mu\nu}$, i.e. impose $\delta h_{\mu\nu}=0$ using the transformations \eqref{eq:trafosources}.

All bulk residual coordinate transformations \eqref{eq:xi^r}--\eqref{eq:deltaM-bulk} are nothing other than ordinary PBH transformations. Hence they are local symmetries of the on-shell action. Therefore to find the symmetries of the space-time we solve
\begin{equation}
\delta g_{MN}=0=\delta B_M
\end{equation}
which using \eqref{eq:bulkdiff1}--\eqref{eq:bulkdiff9} implies 
\begin{equation}
\delta M=0
\end{equation}
and the resulting set of symmetries are none other than \eqref{eq:solutionKE8}--\eqref{eq:KV2}. For every $M$ that lies in the $M=\text{cst}$ orbit the solution to $\delta M=0$ provides us with a set of Lifshitz Killing vectors. The condition $\delta B_M=0$ with $B=\tfrac{dt}{r^2}$ gives nothing new as it is an invariant under the residual coordinate transformations. 

The $\delta M$ transformations are generated by \eqref{eq:Schgens} that form the Schr\"odinger algebra. In other words the generators of the PBH transformations that preserve the boundary conditions \eqref{eq:flatNCagain} span the Schr\"odinger algebra. In section \ref{subsec:probes} we will see how this structure can give rise to global Schr\"odinger invariance of certain probe fields on a Lifshitz space-time. 

\subsection{The particle number current}\label{subsec:particlenumbercurrent}

The local transformations of the source $M$ will lead to a Ward identity for $\partial_\mu T^\mu$ in much the same way as we derived in appendix \ref{subsec:comments}. Any solution of the bulk equations of motion of our bulk theory \eqref{eq:action} with boundary conditions such that the boundary geometry is described by flat NC space-time, i.e. with sources fixed to be as in \eqref{eq:flatNCagain}, will have a local Schr\"odinger algebra realized on $M$. Since the transformations acting on $M$ are induced by bulk diffeomorphisms we have the result that
\begin{equation}\label{eq:varosaction}
\delta S^{\text{ren}}_{\text{on-shell}}[M]=-\int d^{d+1}x\partial_\mu T^\mu\delta M\,,
\end{equation}
where $\delta S^{\text{ren}}_{\text{on-shell}}[M]$ is the variation of the on-shell action obtained after performing holographic renormalization for sources given by \eqref{eq:flatNCagain}. This action only depends on $M$ which is the only source left unfixed. For details about the holographic renormalization see \cite{Hartong:2014}. The precise form of the counterterms is not relevant for the discussion here. For variations $\delta M$ that obey \eqref{eq:constrainedvar1}--\eqref{eq:bulkdiff9} the variation \eqref{eq:varosaction} vanishes. Hence we obtain
\begin{eqnarray}
\partial_\mu T^\mu & = & -\partial_t\lambda_1-\partial_i(\lambda_1\partial^iM)-\partial_i\partial_j\partial^j\lambda^i+\left(\partial_i\partial_j-\frac{1}{d}\delta_{ij}\partial_k\partial^k\right)\lambda^{ij}\nonumber\\
&=&  -\partial_t\lambda_1-\partial_i(\lambda_1\partial^iM)+\left(\partial_i\partial_j\Lambda^{ij}-\frac{1}{d}\partial_i\partial^i\Lambda^k{}_k\right)\equiv \partial_\mu J^\mu\,,\label{eq:J}
\end{eqnarray}
where $\Lambda^{ij}=\lambda^{ij}-\frac{d}{d-1}\partial^{(i}\lambda^{j)}$ and where we defined a current $J^\mu$. In appendix \ref{subsec:comments} we find a similar result using the method of Lagrange multipliers. Here we argue as follows. Consider the case $M=\text{cst}$ and let us restrict to the case $z=2$. We then have $\delta M=-C-v^ix^i-\tfrac{1}{2}c x^ix^i$ as follows from \eqref{eq:deltaM}. This tells us that 
\begin{equation}
\int d^{d+1}x\partial_\mu T^\mu\left(v^i x^i+\frac{1}{2}c x^i x^i\right)=0\,.
\end{equation}
Performing a partial integration we get 
\begin{equation}
\int d^{d+1}x J^i\left(v^i+c x^i\right)=0\,,
\end{equation}
where $T^\mu=\tilde T^\mu+J^\mu$ with $\tilde T^\mu$ a conserved current. Consider first the case $v^i=0$. It must be that\footnote{One way to show this goes as follows. Define $\hat F(k)=\int d^dx e^{i\vec k\cdot\vec x}F(x)$, i.e. $\hat F(k)$ is the Fourier transform of $F(x)$. We then have $\hat F(0)=\int d^dx F(x)$. Suppose the function $F$ is such that $\int d^dx F(x)=0$, which is the case we are dealing with if we take $F=J^i x^i$, then we get $\hat F(0)=0$. By Taylor expanding $F(k)$ around $k=0$ we see that $\hat F=k_i \hat F^i$, so that when we do the inverse Fourier transform we obtain $F=\partial_i F^i$.}
\begin{equation}
x^i J^i=\partial_i F^i\,.
\end{equation}
This in turn can be written as 
\begin{equation}
J^i=\partial_j\left(F^{ij}-\frac{1}{d}\delta^{ij}F^k{}_k\right)\,,
\end{equation}
where $x^i\left(F^{ij}-\frac{1}{d}\delta^{ij}F^k{}_k\right)=F^j$. This form for $J^i$ is also compatible with $v^i\neq 0$. We do not find any constraint on $J^t$ since 
\begin{equation}
\int d^{d+1}x\partial_t J^t\left(v^i x^i+\frac{1}{2}c x^i x^i\right)=0\,.
\end{equation}
We have thus reproduced the expression for $\partial_\mu J^\mu$ in equation \eqref{eq:J}. By making the time derivative $\partial_t$ Galilean boost invariant by replacing it by $\partial_t+\partial^i M\partial_i$ we can reproduce the result for $\partial_\mu J^\mu$ for the case that $\partial_iM$ is constant or in other words for a linear $M$ of the form \eqref{eq:linear}. The case of a quadratic $M$ as in \eqref{eq:quadratic} can be dealt with by observing that $\partial_t+\partial^i M\partial_i$ transforms homogeneously under \eqref{eq:Ktrafo} and that $\partial_t\lambda_1+\partial_i(\lambda_1\partial^iM)$ can be written as $\partial_{t'}\lambda'_1+\partial^iM\partial_i\lambda'_1$ by making a redefinition of $t$ and $\lambda_1$ of the form $\partial_{t'}=\exp[-\int^t dt'\partial_i\partial^iM]\partial_t$ and  $\lambda'_1=\exp[\int^t dt'\partial_i\partial^iM]\lambda_1$.

We thus conclude that the local Schr\"odinger invariance of the on-shell action with flat NC boundary conditions {\it{can}} lead to a conserved current of the form
\begin{equation}
\partial_\mu\left(T^\mu-J^\mu\right)=0\,.
\end{equation}
We emphasize `can' because there is the possibility that $T^\mu=J^\mu$ plus terms that are trivially conserved in which case there is no non-trivial conserved current. In order to see that we can in fact have particle number conservation as well as e.g. Galilean boost invariance we need to add matter fields just like in TNC geometries: Galilean boosts are never a symmetry of the space-time only, they require matter (see section \ref{subsec:FTonflatNC}). In the next section we show that one can write down simple probe actions on a Lifshitz space-time that are invariant under the full Schr\"odinger group.

It is interesting to observe that the transformation properties of $M_\mu$ (here $M$) are tied to the boundary conditions. In our case the $\delta M$ transformations result from the PBH transformations. This means that the existence of a conserved particle number current is in part tied to the choice of boundary conditions. This is a pretty uncommon feature and is due to the fact that $M_\mu$ plays kind of a dual role: it is on the one hand part of the geometry and on the other hand coupling to a current.

We have thus established that the field theory dual to Lifshitz space-times with flat NC boundaries have global Lifshitz symmetries for every $M$ in the $M=\text{cst}$ orbit that is generated by the Schr\"odinger algebra and that there can be a conserved particle number current associated with the local shifts in $M$. The local shifts in $M$ are generated by Galilean ($v^i$) and special conformal transformations ($c$) (see eq.~\eqref{eq:deltaM}).

\subsection{Schr\"odinger invariant probe actions}\label{subsec:probes}

In this section we set the number of spatial dimensions $d=2$. The question we wish to address is what a natural  
probe field for a Lifshitz space-time looks like. A probe action that has been considered frequently in the literature is to take a real Klein--Gordon field on a Lifshitz background. With our new perspective on Lifshitz symmetries we will take a fresh look at the problem of constructing probe actions and find some interesting results. The main question connected to a probe is of course what one one precisely wants to probe. Here we wish to write down a probe action that is Schr\"odinger invariant. In order to gain some intuition about what kind of action to take,
we consider the following probe action (inspired by section 2.2 of \cite{Christensen:2013rfa})
\begin{equation}\label{eq:probeaction1}
S=\int d^4x\sqrt{-g}\left(D_M\phi^\star D^M\phi-m^2\phi^\star\phi\right)\,,
\end{equation}
where $D_M=\partial_M-iqA_M$. This seems to have some good ingredients such as a complex scalar which is crucial for Schr\"odinger symmetries and it has a local gauge symmetry $\phi=e^{iq\Lambda}\phi'$ and $A_M=A'_M+\partial_M\Lambda$ where $A_M$ is the field appearing in the St\"uckelberg decomposition $B_M=A_M-\partial_M\Xi$. We can thus by a local gauge transformation replace $A_M$ by $B_M$, and from now on we will use  $B_M$. 

The equation of motion, using the metric \eqref{eq:LifshitzallM}, is
\begin{equation}\label{eq:almostSch}
-r^2D_t\left(r^2D_t\phi\right)+r^2\partial_i\partial^i\phi+2iqr^2D_t\phi+r^2\partial_r^2\phi-3r\partial_r\phi-(m^2-q^2)\phi=0\,,
\end{equation}
where we used that for a $z=2$ Lifshitz background $B^2=-1$ and that we always have that $\nabla_M B^M=0$ and where we denote by $D_t$ the following operator
\begin{equation}
D_t=\partial_t+\partial^iM\partial_i+\frac{1}{2}\partial^2M r\partial_r\,,
\end{equation}
which is covariant under the residual coordinate transformations of \eqref{eq:LifshitzallM}. The equation \eqref{eq:almostSch} looks almost like a Schr\"odinger equation. The term that spoils it is the first one containing two time derivatives. 

In order to determine whether it makes sense to drop this term, we recall from appendix \ref{app:coordinateindependentLif} that every Lifshitz metric can be written as 
\begin{equation}
ds^2=\left(-B_MB_N+\gamma_{MN}\right)dx^Mdx^N\,,
\end{equation}
where $B_M$ is the massive vector field for which $B^2=-1$ and $\gamma_{MN}$ is orthogonal to $B^M$. In this language we can rewrite \eqref{eq:probeaction1} as follows
\begin{eqnarray}
S & = & \int d^4x\sqrt{-g}\left(\gamma^{MN}\partial_M\phi^\star \partial_N\phi+iq\phi^\star B^M\partial_M\phi-iq\phi B^M\partial_M\phi^\star\right.\nonumber\\
&&\left.-B^M\partial_M\phi^\star B^N\partial_N\phi-(m^2-q^2)\phi^\star\phi\right)\,.\label{eq:probeaction2}
\end{eqnarray}
The first term in \eqref{eq:almostSch} comes from the $-B^M\partial_M\phi^\star B^N\partial_N\phi$ term in the probe action, 
so that it is natural to drop this term. This gives rise to the following probe action 
\begin{equation}\label{eq:probeaction3}
S = \int d^4x\sqrt{-g}\left(\gamma^{MN}\partial_M\phi^\star \partial_N\phi+iq\phi^\star B^M\partial_M\phi-iq\phi B^M\partial_M\phi^\star-(m^2-q^2)\phi^\star\phi\right)\,,
\end{equation}
where $\gamma^{MN}=g^{MN}+B^MB^N$. This is a Schr\"odinger invariant probe action on a Lifshitz space-time whose equation of motion, in the coordinates of \eqref{eq:LifshitzallM}, is 
\begin{equation}\label{eq:Schinvbulkeom}
r^2\left(\partial_i\partial^i\phi+2iqD_t\phi\right)+r^2\partial_r^2\phi-3r\partial_r\phi-(m^2-q^2)\phi=0\,.
\end{equation}

We will next study how the Schr\"odinger invariance comes about and how this is tied to the role of $M$ in the Lifshitz metric \eqref{eq:LifshitzallM}. By construction \eqref{eq:Schinvbulkeom} is form invariant under the residual bulk transformations \eqref{eq:xi^r}--\eqref{eq:deltaM-bulk}. Further we can remove $M$ from the equation of motion by the following field redefinition
\begin{equation}\label{eq:fieldredef}
\phi=\exp[-iqM-\frac{i}{4}qr^2\partial^2M]\tilde\phi\,,
\end{equation}
so that $\tilde\phi$ satisfies
\begin{equation}\label{eq:Scheq}
r^2\left(\partial_i\partial^i\tilde\phi+2iq\partial_t\tilde\phi\right)+r^2\partial_r^2\tilde\phi-3r\partial_r\tilde\phi-(m^2-q^2)\tilde\phi=0\,.
\end{equation}
This requires using all the properties of $M$, i.e. equations \eqref{eq:condition1}, \eqref{eq:condition2} and \eqref{eq:noNP}. The redefinition implies that there is a local symmetry $M=M'+\alpha$ and $\phi=\exp[iq\alpha+\frac{i}{4}qr^2\partial^2\alpha]\phi'$ which is the analogue of the $\alpha$ symmetry of section \ref{subsec:FTonflatNC} and is the reason we can promote the flat NC residual transformations \eqref{eq:finiteresidualtrafos} and \eqref{eq:Ktrafo} to global symmetries.

We note that equation \eqref{eq:Scheq} is exactly the same equation of motion as that of  a complex Klein--Gordon scalar on a $z=2$ Schr\"odinger space-time with null momentum equal to $q$ \cite{Son:2008ye,Balasubramanian:2008dm}. As an interesting consequence\footnote{We thank Cindy Keeler for pointing this out to us.}, this means that these probes evade the bulk reconstruction issues  \cite{Keeler:2013msa} that are present for minimally coupled scalars in Lifshitz space-times. Indeed, this is what one could have expected from the fact that our probe actions are invariant under the full Schr\"odinger symmetry, thus constraining the Green functions. 

On a $z=2$ Schr\"odinger space-time we can perform a coordinate transformation to global Schr\"odinger coordinates \cite{Blau:2009gd}. It would be interesting to see if we can reproduce the equation of a complex scalar in global Schr\"odinger coordinates \cite{Blau:2010fh} on a Lifshitz space-time. From the Schr\"odinger space-time point of view the global coordinates appear as if a Newton potential has been turned on \cite{Blau:2009gd} (in the sense that the time-time component of the bulk metric near the Schr\"odinger boundary has a term that looks like a potential). It therefore might be an idea to use the equations \eqref{eq:pureLif-I}--\eqref{eq:pureLif-VI} to find Lifshitz space-times that are dual to flat NC boundaries with a Newton potential turned on and to consider the probe action \eqref{eq:probeaction3} in those Lifshitz coordinates.

The equation \eqref{eq:almostSch} was inspired by the work \cite{Christensen:2013rfa} which is a case in which we obtained the Lifshitz space-time by Scherk--Schwarz reduction along a circle that is everywhere spacelike in the bulk of an asymptotically AdS$_5$ space-time. The resulting 4-dimensional theory is of the same type as we discussed in this paper. From the boundary point of view the reduction is along a null circle of $\mathcal{N}=4$ super Yang--Mills with a theta angle turned on that is uniformly distributed along the null circle which is expected to give rise to a $z=2$ Lifshitz Chern--Simons theory \cite{Mulligan:2010wj}. This is a simple way of understanding that the Lifshitz boundary geometry is described by Newton--Cartan geometry with torsion as this is the result of reducing Lorentzian geometry along a null circle\footnote{We refer to \cite{Kuenzle:1972zw,Duval:1990hj,Julia:1994bs,Bekaert:2013fta} for more details about null reductions of pp-waves and space-times with hypersurface orthogonal null Killing vectors and torsionless Newton--Cartan geometry and to \cite{Christensen:2013rfa} for generalizations to more general null reductions of any space-time with a null Killing vector and the importance of including of torsion once the higher dimensional space-time is no longer a pp-wave.}. Furthermore since we are reducing a CFT on a null circle we expect the dual field theory to be Schr\"odinger invariant in the UV. The equation of motion of the probe \eqref{eq:almostSch} was obtained by reducing the equation of a real Klein--Gordon scalar on the 5-dimensional asymptotically AdS space-time (which is a $z=0$ Schr\"odinger space-time \cite{Balasubramanian:2010uk,Donos:2010tu,Costa:2010cn,Cassani:2011sv,Chemissany:2011mb}) that upon reduction gives a $z=2$ Lifshitz space-time. We see that \eqref{eq:almostSch} close to the boundary becomes equal to \eqref{eq:Schinvbulkeom} in agreement with our expectation that the dual field theory has a Schr\"odinger invariant UV fixed point. For large $r$ the probe \eqref{eq:almostSch} sees Lifshitz symmetries, so it seems that there is a flow to a Lifshitz IR.

\section{Outlook}

We have shown that the Lifshitz vacuum dual to a flat Newton-Cartan space-time has a local action of a Schr\"odinger group acting on the one remaining source which is $M$, a subgroup of which is described by Killing vectors that generate the Lifshitz algebra. 
Moreover,  the boundary theory can have a conserved current related to conservation of particle number. We have exhibited
that this is precisely the same manner in which a field theory on Newton-Cartan space-time sees Schr\"odinger symmetries. 
Furthermore, in order to show that the theory is invariant under global Schr\"odinger symmetries one needs to 
know what type of matter fields live on the space and how they they are coupled to the geometry.  As important evidence that this is possible 
in the holographic setup, we have shown that  one can construct scalar probes on a bulk Lifshitz background that are invariant under a global Schr\"odinger group.

There are a number of interesting future research directions that we hereby briefly discuss.

The holographic models that have led us to consider TNC geometries have been derived using a bulk theory containing Einstein gravity coupled to massive vector fields. In 4 dimensions there are two alternative bulk theories known that admit Lifshitz solutions. The first is a model introduced in \cite{Taylor:2008tg} that can be thought of as setting $W=0$ in our bulk action. This is commonly known as the Einstein--Maxwell-dilaton model (EMD). In this case the Lifshitz geometries are supported by a Maxwell gauge field and a logarithmically running dilaton. Allowing for a logarithmically running dilaton is also possible when $W\neq 0$ and in general leads to a second exponent $\alpha$ related to Lifshitz scaling violating due to the matter fields \cite{Gouteraux:2012yr,Gath:2012pg}. It would be very interesting to understand the role of this additional exponent from the dual field theory perspective (see e.g. \cite{Khveshchenko:2014nka,Karch:2014mba,Hartnoll:2015sea} in this context).

Further one could wonder how the TNC geometry comes about in that model and what the role of the extra local $U(1)$ is in this case. Once one understands holography for general exponents $z$ and $\alpha$ one can include hyperscaling violation by considering non-Einstein frames as in \cite{Chemissany:2014xpa,Chemissany:2014xsa} in which the theory has only two exponents $z$ and $\alpha$. The hyperscaling exponent $\theta$ then comes about by transforming to the Einstein frame. The other 4-dimensional model that allows for Lifshitz solutions are of the Ho\v rava--Lifshitz type \cite{Horava:2009uw,Griffin:2012qx,Janiszewski:2012nb}. It would be interesting to see if in the context of bulk Ho\v rava--Lifshitz models \cite{Griffin:2012qx,Janiszewski:2012nb} one can similarly speak of boundaries described by TNC geometries. 

Our results for the holographic description of Lifshitz space-times also suggest a new perspective on existing results, notably the computation of doing perturbations around a Lifshitz vacuum and adding temperature by looking at Lifshitz black branes. 

Considering first the subject of perturbations around the Lifshitz vacuum \cite{Ross:2009ar,Danielsson:2009gi,Cheng:2009df,Ross:2011gu,Baggio:2011cp,Baggio,Holsheimer:2013ula}. The way this is normally done is to consider the Lifshitz metric with $M=\text{cst}$ and to perturb around it using radial perturbations. This is then divided in terms of pairs of modes that form source/vev pairs that are then used as the basis for constructing asymptotic expansions of full non-linear asymptotically Lifshitz space-times. Although the last step is rarely carried out (see however \cite{Chemissany:2014xpa,Chemissany:2014xsa}). Following this approach one finds scaling dimensions of the sources and vevs that are in general rather complicated functions of $z$ and possibly parameters in the potential $V$. In particular the source we denote by $\tilde\Phi$ was not seen by the linearized perturbations around the $M=\text{cst}$ Lifshitz metric. Instead another scalar source appears in the spectrum that is denoted by $\psi$ in \cite{Ross:2011gu,Chemissany:2014xsa} whose scaling dimension differs from that of $\tilde\Phi$. This seems at odds with our general non-linear analysis of the sources of section \ref{subsec:bdryconditions} which do include $\tilde\Phi$ and whose scaling dimensions have a rather simple dependence on $z$ with no dependence on the potential (with the exception of $\Delta$ in \eqref{eq:bdryPhi}). One potential explanation is that there is a relation between $\tilde\Phi$ and $\psi$. It would be interesting to understand better what precisely is going on. It might also be interesting to perform perturbations around Lifshitz for general $M$.

In appendix \ref{app:coordinateindependentLif} we have constructed a coordinate independent definition of a Lifshitz space-time. In view of the above discussion and in relation to finding the analogue of a complex scalar on global Schr\"odinger space-time by consider a Schr\"odinger invariant probe on a Lifshitz metric (see the discussion at the end of section \ref{subsec:probes}) it would be interesting to use the results of appendix \ref{app:coordinateindependentLif} to find the most general Lifshitz metric with a flat NC boundary but with a nonzero Newton potential, i.e. with $M_t=\partial_t M+\Phi$ and $M_i=\partial_i M$. This might also be interesting for the study of more general Lifshitz black branes that asymptote to such a boundary geometry.

Regarding the subject of Lifshitz black branes our analysis suggests that they should be dual to Galilean invariant fluids. It would be interesting to consider Lifshitz black branes, and to see if they can be boosted in such a way that the dual energy-momentum tensor is that of a Galilean perfect fluid at leading order in some hydrodynamic expansion, comparing with the work of \cite{Jensen:2014ama}. We hope to report on such an analysis in the near future. Along similar lines it would be interesting to use our machinery of coupling fields theories to TNC geometries to study hydrodynamics of both Galilean and Lifshitz invariant theories and to compare with \cite{Jensen:2014ama,Hoyos:2013eza,Hoyos:2013qna}. More generally, in parallel to the renewed development of relativistic fluid and superfluid dynamics  that was initiated and inspired by the fluid/gravity correspondence \cite{Policastro:2001yc,Bhattacharyya:2008jc}, we expect that our holographic approach to Lifshitz space-times will lead to further novel insights into the dynamics and hydrodynamics of
non-relativistic field theories.

Finally, especially for applications to condensed matter physics it would be interesting to add charge into the game both in the context of field theory coupled to TNC geometries by adding more background fields such as a vector potential but also from the holographic point of view by adding a Maxwell type vector field.

\section*{Acknowledgments}
\label{ACKNOWL}

We would like to thank Jay Armas, Eric Bergshoeff, Matthias Blau, Jan de Boer, Jean--Pierre Derendinger, Troels Harmark, Kristan Jensen,
Cindy Keeler and Jan Rosseel for many valuable discussions.
The work of JH is supported by the advanced ERC grant `Symmetries and Dualities in Gravity and M-theory' of Marc Henneaux. The work of EK was supported in part by European Union's Seventh Framework Programme under grant agreements (FP7-REGPOT-2012-2013-1) no 316165, PIF-GA-2011-300984, the EU program ``Thales'' MIS 375734, by the European Commission under the ERC Advanced Grant BSMOXFORD 228169 and was also co-financed by the European Union (European Social Fund, ESF) and Greek national funds through the Operational Program ``Education and Lifelong Learning'' of the National Strategic Reference Framework (NSRF) under ``Funding of proposals that have received a positive evaluation in the 3rd and 4th Call of ERC Grant Schemes''. The work of NO is supported in part by  Danish National Research Foundation project ``New horizons in particle and condensed matter physics from black holes".

\appendix

\section{Coordinate independent definition of Lifshitz space-times}\label{app:coordinateindependentLif}

For the purpose of finding Lifshitz metrics in different coordinate systems using an ansatz based on symmetries it is useful to have a coordinate independent definition, i.e. a set of tensor equations for the metric and the massive vector field whose only solution is a Lifshitz space-time locally. In other words we look for the equivalent of the well known result that all solutions to 
\begin{equation}
R_{MNPQ}=-\left(g_{MP}g_{NQ}-g_{MQ}g_{NP}\right)
\end{equation}
are locally AdS. Such a definition will be provided in this appendix.

Consider the equations of motion
\begin{eqnarray}
\hspace{-1cm}\frac{1}{\sqrt{-g}}\partial_M\left(\sqrt{-g}ZF^{MN}\right) & = & WB^N\,,\\
\hspace{-1cm}\square\Phi & = & \frac{1}{4}Z'F^2+\frac{1}{2}W'B^2+V'\,,\\
\hspace{-1cm}R_{MN} & = & \frac{1}{2}Vg_{MN}+\frac{1}{2}Z\left(F_{MP}F_N{}^P-\frac{1}{4}F^2g_{MN}\right)+\frac{1}{2}WB_M B_N\,.
\end{eqnarray}
The Einstein equation is compatible with the following statement for the Riemann tensor
\begin{eqnarray}
R_{MNPQ} & = & C_{MNPQ}+\left(\frac{1}{6}V+\alpha_1 ZF^2-\frac{1}{12}WB^2\right)\left(g_{MP}g_{NQ}-g_{MQ}g_{NP}\right)+\nonumber\\
&&+\frac{1}{4}W\left(B_M B_P g_{NQ}-B_N B_P g_{MQ}-B_M B_Q g_{NP}+B_N B_Q g_{MP}\right)+\nonumber\\
&&+\alpha_2 Z\left(2F_{MN}F_{PQ}+F_{MP}F_{NQ}+F_{MQ}F_{PN}\right)+\\
&&+\alpha_3 Z\left(F_{MR}F_P{}^R g_{NQ}-F_{NR}F_P{}^R g_{MQ}-F_{MR}F_Q{}^R g_{NP}+F_{NR}F_Q{}^R g_{MP}\right)\,,\nonumber
\end{eqnarray}
where $C_{MNPQ}$ is the Weyl tensor and where 
\begin{eqnarray}
\alpha_1 & = & -\frac{1}{24}-\frac{1}{3}\alpha_3\\
\alpha_2 & = & \frac{1}{6}-\frac{2}{3}\alpha_3\,.
\end{eqnarray}
The term 
\begin{equation}
2F_{MN}F_{PQ}+F_{MP}F_{NQ}+F_{MQ}F_{PN}=3F_{MN}F_{PQ}-3F_{[MN}F_{PQ]}
\end{equation}
has the same index structure as the Riemann tensor. We will now choose the Weyl tensor such that 
\begin{eqnarray}
R_{MNPQ} & = & \left(\frac{1}{6}V-\frac{1}{24}ZF^2-\frac{1}{12}WB^2\right)\left(g_{MP}g_{NQ}-g_{MQ}g_{NP}\right)+\nonumber\\
&&+\frac{1}{4}W\left(B_M B_P g_{NQ}-B_N B_P g_{MQ}-B_M B_Q g_{NP}+B_N B_Q g_{MP}\right)+\nonumber\\
&&+\frac{1}{6} Z\left(2F_{MN}F_{PQ}+F_{MP}F_{NQ}+F_{MQ}F_{PN}\right)\,.
\end{eqnarray}
It can be checked that a pure Lifshitz space-time satisfies this equation. This expression for the Riemann tensor is strikingly similar to the expression obtained in appendix A of \cite{Hartong:2013cba} for the case of a pure Schr\"odinger space-time. In fact the analysis in section 3 and appendix A of \cite{Hartong:2013cba} have been the inspiration for the coordinate independent definition of a Lifshitz space-time that we will get to now.

For a pure Lifshitz space-time $\Phi$ is a constant and provided we choose functions $Z$, $W$ and $V$ such that the scalar equation is satisfied the remaining equations become
\begin{eqnarray}
\frac{1}{\sqrt{-g}}\partial_M\left(\sqrt{-g}\bar F^{MN}\right) & = & 2z\bar B^N\label{eq:vectoreom}\\
R_{MN} & = & -\frac{1}{2}\left(z^2+z+4\right)g_{MN}+\frac{z-1}{z}\left(\bar F_{MP}\bar F_N{}^P-\frac{1}{4}\bar F^2g_{MN}\right)\nonumber\\
&&+2(z-1)\bar B_M \bar B_N
\end{eqnarray}
where $B_M=A_0\bar B_M$ and $F_{MN}=A_0\bar F_{MN}$. We used here that $W_0=2zZ_0$, $V_0=-(z^2+z+4)$ and $A_0^2=\tfrac{2(z-1)}{zZ_0}$. With these choices the Riemann tensor \eqref{eq:Riemann} can be written as\footnote{The factor of $-1$ in front of the metric part is what motivated the choice made earlier for the Weyl tensor.}
\begin{eqnarray}
R_{MNPQ} & = & -\left(g_{MP}g_{NQ}-g_{MQ}g_{NP}\right)+\nonumber\\
&&+(z-1)\left(\bar B_M\bar  B_P g_{NQ}-\bar B_N\bar  B_P g_{MQ}-\bar B_M \bar B_Q g_{NP}+\bar B_N \bar B_Q g_{MP}\right)+\nonumber\\
&&+\frac{z-1}{z}\left(\bar F_{MN}\bar F_{PQ}-\bar F_{[MN}\bar F_{PQ]}\right)\,.\label{eq:Riemann}
\end{eqnarray}
Further one can check that for a pure Lifshitz space-time we have
\begin{equation}
\bar F_{[MN}\bar F_{PQ]}=0\,.
\end{equation}
Further we also have for a pure Lifshitz space-time that
\begin{equation}
\bar B^2=-1\qquad\bar F^2=-2z^2\,.
\end{equation}
We define $X_N$ as
\begin{equation}
X_N=\frac{1}{z}\bar B^M\bar F_{MN}\,.
\end{equation}
We then furthermore have
\begin{equation}
X^2=1\,,\qquad X\cdot B=0\,,
\end{equation}
and
\begin{equation}
\frac{1}{z}\bar F_{MN}=X_M\bar B_N-X_N\bar B_M\,.
\end{equation}
We define the projector $\gamma_M{}^N$ as
\begin{equation}
\gamma_M{}^N=\delta_M{}^N+\bar B_M\bar B^N\,.
\end{equation}

Let us consider the vector equation of motion \eqref{eq:vectoreom}. Taking the covariant derivative we get
\begin{equation}
\partial_M\left(\sqrt{-g}\bar B^M\right)=0\,.
\end{equation}
Contracting it with $\bar B_N$ gives
\begin{equation}
\frac{1}{\sqrt{-g}}\partial_M\left(\sqrt{-g}\bar X^M\right)=z+2\,.
\end{equation}
Using $\frac{1}{z}\bar F_{MN}=X_M\bar B_N-X_N\bar B_M$ together with the divergences of $\bar B^M$ and $X^M$ leads to
\begin{equation}
X^M\partial_M\bar B^N-\bar B^M\partial_M X^N=-z\bar B^N\,.
\end{equation}
These last three equations and therefore the vector equation of motion are solved if we have
\begin{eqnarray}
\nabla_M \bar B_N & = & -z\bar B_M X_N\,,\label{eq:pureLif1}\\
\nabla_M X_N & = & \gamma_{MN}-X_M X_N-z\bar B_M \bar B_N\,,\label{eq:pureLif2}
\end{eqnarray}
together with
\begin{equation}\label{eq:pureLif3}
\bar B^2=-1\,,\qquad X^2=1\,,\qquad X\cdot B=0\,.
\end{equation}
It can be checked that equations \eqref{eq:pureLif1}--\eqref{eq:pureLif3} are satisfied for a pure Lifshitz space-time. From equation  \eqref{eq:pureLif1} it follows that the extrinsic curvature $K_{MN}=\gamma_M{}^P\nabla_P \bar B_N=0$. Equation \eqref{eq:Riemann} implies that
\begin{equation}\label{eq:pureLif4}
\gamma_A{}^M \gamma_B{}^N \gamma_C{}^P \gamma_D{}^Q R_{MNPQ}=-\left(\gamma_{AC}\gamma_{BD}-\gamma_{AD}\gamma_{BC}\right)\,.
\end{equation}
Since the extrinsic curvature vanishes the Gauss--Codazzi equations imply that the Riemann tensor of the metric $\gamma_{MN}$ is locally AdS. One can also show that given \eqref{eq:pureLif1}--\eqref{eq:pureLif4} the rest of the Riemann tensor \eqref{eq:Riemann} follows. We have checked that Lifshitz solves \eqref{eq:pureLif1}--\eqref{eq:pureLif4}. Now we will show the converse, namely all solutions of \eqref{eq:pureLif1}--\eqref{eq:pureLif4} are locally Lifshitz with metric given by
\begin{equation}
ds^2=\left(-\bar B_M \bar B_N+\gamma_{MN}\right)dx^M dx^N\,.
\end{equation}

From equation \eqref{eq:pureLif2} we conclude that $\partial_M X_N-\partial_N X_M=0$ so that
\begin{equation}
X_M=\partial_M\Omega
\end{equation}
locally for some $\Omega$. Equation \eqref{eq:pureLif1} then implies that there exists a function $f(\Omega)$ such that $H_M=f\bar B_M$ is a Killing vector. The function turns out to be $f=e^{z\Omega}$. More precisely for $\bar B_M=e^{-z\Omega}H_M$ equation \eqref{eq:pureLif1} becomes
\begin{eqnarray}
 0 & = & \mathcal{L}_H g_{MN}\,,\\
 0 & = & \partial_M\left(e^{-2z\Omega}H_N\right)-\partial_N\left(e^{-2z\Omega}H_M\right)
\end{eqnarray}
for the symmetric and anti-symmetric parts respectively. The latter equation implies that
\begin{equation}
\bar B_M=e^{z\Omega}\partial_M T\,.
\end{equation}
Equations \eqref{eq:pureLif3} then imply 
\begin{equation}\label{eq:XandB}
\mathcal{L}_H T=-1\,,\qquad\mathcal{L}_{H}\Omega=0\,,\qquad\mathcal{L}_XT=0\,,\qquad\mathcal{L}_X\Omega=1\,.
\end{equation}
Next using that
\begin{equation}
\mathcal{L}_X\bar B_M=z\bar B_M
\end{equation}
we can show that the symmetric part of \eqref{eq:pureLif2} is equivalent to
\begin{equation}
\mathcal{L}_X\gamma_{MN}=2\gamma_{MN}-2X_M X_N\,.
\end{equation}
By contraction with $X^N$ this implies that $\mathcal{L}_XX_M=0$ (which also follows from $\partial_M X_N-\partial_N X_M=0$ and $X^2=1$ and is hence not a new condition), so that we can also write
\begin{equation}
\mathcal{L}_X\bar \gamma_{MN}=2\bar \gamma_{MN}\,,
\end{equation}
where
\begin{equation}
\bar \gamma_{MN}=\gamma_{MN}-X_M X_N\,.
\end{equation}
The metric $\bar \gamma_{MN}$ is the projection of the metric $\gamma_{MN}$ onto the space orthogonal to $X^M$. Hence we have
\begin{equation}
\bar \gamma_M{}^N\partial_N\Omega=0\,,
\end{equation}
i.e. $\Omega$ is constant on the $d$ dimensional space that $\bar \gamma_M{}^N$ projects onto. The last relation of \eqref{eq:XandB} implies that we can write
\begin{equation}\label{eq:gammaX}
\mathcal{L}_X\sigma_{MN}=0\,.
\end{equation}
where we defined
\begin{equation}
\sigma_{MN}=e^{-2\Omega}\bar \gamma_{MN}
\end{equation}
we can write for the metric
\begin{equation}\label{eq:Lifshitz}
ds^2=\left(-e^{2z\Omega}\partial_M T\partial_N T+\partial_M\Omega\partial_N\Omega+e^{2\Omega}\sigma_{MN}\right)dx^M dx^N\,.
\end{equation}
We will finally show that $\bar \gamma_{MN}$ is a flat metric and since $\Omega$ is constant on the $d$ dimensional space that $\bar \gamma_M{}^N$ projects onto this implies that $\sigma_{MN}$ is a flat metric.
We have earlier argued that the Riemann tensor of the metric $\gamma_{MN}$ satisfies
\begin{equation}\label{eq:Riemannh}
R^{(\gamma)}_{ABCD}=-\left(\gamma_{AC}\gamma_{BD}-\gamma_{AD}\gamma_{BC}\right)\,.
\end{equation}
The $\gamma$-covariant derivative $\nabla^{(\gamma)}_M$ of $X_M$ (which is orthogonal to $\bar B^M$) is defined as
\begin{equation}
\nabla^{(\gamma)}_A X_B=\gamma_A{}^C\gamma_B{}^D\nabla_C X_D=\bar \gamma_{AB}
\end{equation}
where we used \eqref{eq:pureLif2}. The extrinsic curvature of the co-dimension one space (inside the space orthogonal to $\bar B^M$) orthogonal to $X^M$ is given by
\begin{equation}\label{eq:extrinsiccurvbarh}
K^{(\bar \gamma)}_{AB}=\bar \gamma_A{}^C\nabla^{(\gamma)}_C X_B=\bar \gamma_{AB}\,.
\end{equation}
Using the Gauss--Codazzi equations
\begin{equation}
R^{(\gamma)}_{ABCD}=R^{(\bar \gamma)}_{ABCD}-K^{(\bar \gamma)}_{AC}K^{(\bar \gamma)}_{BD}+K^{(\bar \gamma)}_{AD}K^{(\bar \gamma)}_{BC}
\end{equation}
with equations \eqref{eq:Riemannh} and \eqref{eq:extrinsiccurvbarh} we obtain 
\begin{equation}
R^{(\bar \gamma)}_{ABCD}=0
\end{equation}
so that $\bar \gamma_{MN}$ and thus $\sigma_{MN}$ are flat Euclidean metrics of dimensionality $d$. This together with \eqref{eq:gammaX} and all the properties of $\Omega$, i.e. $\mathcal{L}_{H}\Omega=0=\bar \gamma_M{}^N\partial_N\Omega$ and $\mathcal{L}_X\Omega=1$, makes \eqref{eq:Lifshitz} a Lifshitz metric.

This analysis also shows that the equations \eqref{eq:pureLif1}--\eqref{eq:pureLif3} are equivalent to
\begin{eqnarray}
\partial_M\bar B_N-\partial_N\bar B_M & = & z\left(X_M\bar B_N-X_N\bar B_M\right)\,,\label{eq:pureLif-I}\\
\partial_M X_N-\partial_N X_M & = & 0\,,\\
\mathcal{L}_{\bar B}g_{MN} & = & -z\left(X_M\bar B_N+X_N\bar B_M\right)\,,\\
\mathcal{L}_{X}\bar \gamma_{MN} & = & 2\bar \gamma_{MN}\,,\\
\bar B^2 & = & -1\,,\qquad X^2=1\,,\label{eq:pureLif-V}\\
R^{(\bar\gamma)}_{MNPQ} & = & 0\,.\label{eq:pureLif-VI}
\end{eqnarray}
It is in this form that we will solve equations \eqref{eq:pureLif1}--\eqref{eq:pureLif3}. Contracting the first of these equations with $\bar B^M$ and using $\bar B^2=-1$ we see that the vector $X$ is determined in terms of $\bar B$ via
\begin{equation}\label{eq:X}
X_M = \frac{1}{z}\mathcal{L}_{\bar B}\bar B_M\,,
\end{equation}
so that we automatically have $\bar B\cdot X=0$.

\section{Comments on $T^\mu$ and demanding $M_\mu$ to become a gauge field}\label{subsec:comments}

In this appendix we study the question of defining the particle number current in cases where we couple to a TNC geometry in a manner that there is no local $U(1)$ symmetry whose gauge connection is $M_\mu$. Looking at the model \eqref{eq:example6} we see that the terms responsible for the gauge invariance are those with $\tilde\Phi$ and $e^\mu_a\mathcal{D}_\mu M^a$.  We now consider what
happens when we remove these terms. If we put the resulting action on a flat NC background we obtain instead of \eqref{eq:Schmodel} the action
\begin{eqnarray}
S & = & \int d^{d+1}x\left(-\varphi^2\left[\partial_t\theta+\partial^iM\partial_i\theta+\frac{1}{2}\partial_i\theta\partial^i\theta+a\partial_i\partial^i\theta\right]\right.\nonumber\\
&&\left.-\frac{1}{2}\partial_i\varphi\partial^i\varphi-V_0\varphi^{\tfrac{2(d+2)}{d}}\right)\,,\label{eq:Schmodeldifferent}
\end{eqnarray}
where we put $b=0$ since we are not interested in explicit breaking of the $\theta$ shift symmetry here. This action can also be written as
\begin{equation}\label{eq:Schmodeldifferent2}
S = S_{\text{$U(1)$}}+\int d^{d+1}x\varphi^2\left(\partial_tM+\frac{1}{2}\partial_iM\partial^iM+a\partial_i\partial^iM\right)\,,
\end{equation}
where by $S_{\text{$U(1)$}}$ we denote the action \eqref{eq:Schmodel} with a local $U(1)$ invariance. Flat NC means that we take $M=\text{cst}$ together with all other $M$ that give identical actions. Clearly all $M$ satisfying 
\begin{eqnarray}
0 & = & \partial_tM+\frac{1}{2}\partial_iM\partial^iM\,,\label{eq:conditionsM-1}\\
0 & = & \partial_i\partial^iM\,,\label{eq:conditionsM-2}
\end{eqnarray}
lead to the same action \eqref{eq:physicalaction} with $\tilde\theta=\theta+M$. This gives the strong suspicion that demanding there to be a local $U(1)$ symmetry whose gauge field is $M_\mu$ is convenient but not strictly necessary. For example if we vary $M$ in \eqref{eq:Schmodel} we get
\begin{equation}
\delta_{\text{bg}}S=-\int d^{d+1}x\partial_\mu T^\mu \delta M\,,
\end{equation}
from which we can conclude that on-shell 
\begin{equation}
-\partial_\mu T^\mu=\partial_t\varphi^2+\partial_i\left(\varphi^2\partial^i(\theta+M)-a\partial^i\varphi^2\right)=0
\end{equation}
where the conservation follows from the fact that $\delta M=\alpha$, $\delta\theta=-\alpha$ is a local symmetry. 

If we vary $M$ in \eqref{eq:Schmodeldifferent} we obtain
\begin{equation}\label{eq:varM}
\delta_{\text{bg}}S=-\int d^{d+1}x\partial_\mu \tilde T^\mu \delta M\,,
\end{equation}
where $-\partial_\mu \tilde T^\mu=\partial_i\left(\varphi^2\partial^i\theta\right)$. The action \eqref{eq:Schmodeldifferent2} still has some local symmetry namely $\delta M=\tilde\alpha$, $\delta\theta=-\tilde\alpha$ where $\tilde\alpha$ obeys
\begin{equation}\label{eq:tildealpha}
\partial_t\tilde\alpha+\partial^iM\partial_i\tilde\alpha=0\,,\qquad\partial_i\partial^i\tilde\alpha=0\,.
\end{equation}
This follows from demanding that $\partial_tM+\frac{1}{2}\partial_iM\partial^iM$ and $\partial_i\partial^iM$ remain invariant under shifting $M$. Demanding that \eqref{eq:varM} is zero for $\tilde\alpha$ satisfying \eqref{eq:tildealpha} leads to an equation of the form
\begin{equation}\label{eq:newconservation}
-\partial_\mu \tilde T^\mu-\partial_t\lambda_1-\partial_i(\lambda_1\partial^iM)+\partial_i\partial^i\lambda_2=0\,,
\end{equation}
for some undetermined functions $\lambda_1$ and $\lambda_2$. To prove this we add the following terms to the action \eqref{eq:Schmodeldifferent} or \eqref{eq:Schmodeldifferent2}
\begin{equation}\label{eq:Lagrangeterms}
\int d^{d+1}x\left[\lambda_1\left(\partial_tM+\frac{1}{2}\partial_iM\partial^iM\right)+\lambda_2\partial_i\partial^iM\right]\,,
\end{equation}
where $\lambda_1$ and $\lambda_2$ are Lagrange multipliers. We can assign transformations to $\lambda_1$ and $\lambda_2$ such that the action \eqref{eq:Schmodeldifferent} plus \eqref{eq:Lagrangeterms} is gauge invariant under any $\tilde\alpha$, i.e. without any constraints. Varying this new action with respect to $\delta M=\tilde\alpha$ we find that off-shell the term obtained by varying $M$ with respect to $\tilde\alpha$ is proportional to the equations of motion of the Lagrange multipliers and $\theta$, so that we get the on-shell equation \eqref{eq:newconservation}. We conclude that in the model without the local $U(1)$ invariance the current $\tilde T^\mu$ is not quite the particle number current but according to \eqref{eq:newconservation} it can be improved to become equal to $T^\mu$.

\renewcommand{\theequation}{\thesection.\arabic{equation}}

\providecommand{\href}[2]{#2}\begingroup\raggedright\endgroup

\end{document}